\documentclass[twocolumn,english,aps,prb,amsmath,floatfix]{revtex4}
\usepackage{amssymb}
\usepackage[dvipdfm]{graphicx,color}

\newcommand{\brac}[3]{\left#1 #2 \right#3}

\newcommand{\nn}[1]{\left< #1 \right>}

\newcommand{\sigu}{\uparrow}
\newcommand{\sigd}{\downarrow}

\newcommand{\etal}{\emph{et al}}
\newcommand{\hspA}{\hspace{-1.25cm}}
\newcommand{\hspB}{\hspace{-.75cm}}
\newcommand{\Ds}{\left<\Delta_{s}\right>}
\newcommand{\Dp}{\left<\Delta_{p}\right>}
\newcommand{\Dd}{\left<\Psi_{d}\right>}
\newcommand{\Dex}{\left<E_{1}-E_{0}\right>}
\newcommand{\PDP}{P\left(\Delta_{p}^{k}>0\right)}
\newcommand{\PSS}{P\left(S=1\right)}
\newcommand{\ssp}{\sigma_{p}}

\begin{document}

\title{Disorder effects on superconducting tendencies in the checkerboard Hubbard model}

\author{Peter M. Smith and Malcolm P. Kennett}

\affiliation{Physics Department, Simon Fraser University, 8888 University Drive,
Burnaby, BC, Canada, V5A 1S6}

\date{\today}

\begin{abstract}
The question of whether spatially inhomogeneous hopping in the two dimensional Hubbard 
model can lead to enhancement of superconductivity has been tackled by a number of authors in the 
context of the checkerboard Hubbard model (CHM).  We address the effects of disorder on
superconducting properties of the CHM by using exact diagonalization calculations
for both potential and hopping disorder.  We characterize the 
superconducting tendencies of the model by focusing on the pair binding energy, 
the spin gap, and $d$-wave pairing order parameter.   We find that superconducting tendencies, 
particularly the pair binding energy, are more robust to disorder 
when there is inhomogeneous hopping than for the uniform  Hubbard model.  We also 
study all possible staggered potentials for an eight site CHM cluster and relate
the behaviour of these configurations to the disordered system.
\end{abstract}

\maketitle

\section{Introduction}
\label{sec:Intro}

The problem of high temperature superconductivity (HTS) in cuprate materials has been at the 
forefront of condensed matter research since its discovery over twenty five
years ago.\cite{BednorzMuller} Despite the overwhelming theoretical and experimental 
efforts focused towards this topic, the microscopic origin of HTS remains elusive. 
From the theoretical side, the doped two dimensional Hubbard model has been 
central to many attempts to understand HTS.\cite{Anderson1987,dopedHubbard}  
Recent numerical simulations of this model appear to confirm that it can 
support $d$-wave superconductivity.\cite{Maier}  Nevertheless, variants of this 
model are appealing to study as they may 
allow for further insights into the two dimensional Hubbard model. 
One such variant that has received much recent attention is the
Checkerboard Hubbard model (CHM)\cite{TsaiKivelson1,TsaiKivelson2,YaoKivelson1,YaoKivelson2,Doluweera,Tremblay,BaruchOrgad}
in which hopping on the two dimensional lattice is spatially modulated in a checkerboard pattern.\cite{footnote01}  This modulation
can be tuned to interpolate between the limit of isolated plaquettes and the limit of the uniform
two dimensional Hubbard model.  The isolated plaquette limit is exactly solvable since it is possible to write down the wavefunctions and energies
as a function of electron number, Hubbard $U$, and intraplaquette hopping parameter $t$  for a four site Hubbard model.\cite{Schumann} As interplaquette hopping $t'$ is turned on but $t'/t\ll1$, one can view the CHM as weakly coupled plaquettes and a perturbative approach can be developed.\cite{TsaiKivelson1}
Additional motivation for studying Hubbard models with modulated hopping comes from the evidence for spatial modulations of 
electronic properties in underdoped cuprate materials\cite{Stripes,cupratecheck,Smectic} 
and proposals for realizing checkerboard fermionic Hubbard models in cold atom systems.\cite{Peterson,Demler,AMRey,Kuns}

Particularly in the context of cuprates, there has been much discussion as to whether inhomogeneity in hopping
can enhance superconductivity or not.  Kivelson and collaborators\cite{TsaiKivelson1,TsaiKivelson2,YaoKivelson1,Karakonstantakis,Cho}
have argued that there is an optimal inhomogeneity in the hopping
for superconductivity in the Hubbard model based on analytic calculations and exact diagonalization studies.
Work using contractor renormalization methods supports this claim. \cite{BaruchOrgad}
However, calculations using quantum Monte Carlo (QMC) by Doluweera {\it et al.}\cite{Doluweera}
and cluster dynamical mean field theory (DMFT) by Chakraborty {\it et al.}\cite{Tremblay} 
suggest that hopping inhomogeneity may enhance superconductivity for some interaction strengths and dopings, 
but at others it does not.  

Previous studies of the checkerboard Hubbard model have not included disorder, and in this paper
we study the effect of disorder on the superconducting properties of the CHM using exact 
diagonalization calculations.  We study the effects of both weak and strong potential and hopping disorder 
on three proxies for superconducting order:  the pair binding energy (PBE), $\Delta_{p}$;
the spin gap, $\Delta_{s}$; and a $d$-wave order parameter, $\Psi_{d}$.
 The pair binding energy  
is a measure of the tendency towards pairing of hole excitations. The spin gap is the gap 
between the lowest energy $S=0$ state and the lowest energy $S=1$ state. Numerical studies of the 
homogeneous two-dimensional Hubbard model \cite{Maier,Kemper,Maier2, Maier3} using the dynamical 
cluster approximation (DCA) and quantum Monte Carlo (QMC) simulations have suggested that $S$=1 
particle-hole spin  fluctuations act to mediate $d$-wave pairing. However, neither the PBE nor the spin 
gap give information about the symmetry of the ground state. Hence, we also calculate a $d$-wave 
order parameter, which has been investigated by several authors, to obtain insight into the ground 
state symmetry of the CHM.\cite{TsaiKivelson2,Tremblay}
Although potential disorder is a pair-breaking perturbation for $d$-wave 
superconductors,\cite{Garg} a recent numerical study of the uniform Hubbard model suggested that very weak 
potential disorder can enhance antiferromagnetic spin correlations and lead to a small increase in the 
critical temperature.\cite{Kemper} 

We study eight- and twelve-site systems at dopings $x=1/8$ and $x=1/12$ respectively, for both potential 
and hopping disorder over a wide range of disorder strengths.  Our results in the weak disorder limit
are similar to previous exact diagonalization studies: we find that superconducting tendencies are
enhanced for intermediate hopping inhomogeneity, with the tendency most pronounced in the PBE and the 
spin gap.  We also find that with increasing disorder, superconducting tendencies are most robust to 
disorder in the region of intermediate hopping inhomogeneity, which is our main result.  We note
that potential disorder can be considered as a random linear combination of specific 
staggered potentials although the resulting electronic properties are not a simple linear combination 
of the properties for each potential configuration. For eight-site clusters, it is straightforward to 
enumerate all inequivalent staggered potentials and we study the effects of each of these potentials 
on the pair binding energy, the spin gap, and the $d$-wave order parameter.  

This paper is organized as follows. In Sec.~\ref{Sec:Model}, we introduce the disordered checkerboard Hubbard model and 
define the quantities we calculate.  In Sec.~\ref{Sec:Results}, we show the results of our finite diagonalization studies 
of eight and twelve-site clusters and discuss how disorder averaging affects the properties of the PBE (Sec.~\ref{Sec:PBE}), the spin gap  (Sec.~\ref{Sec:SpinGap}), and the $d$-wave pairing order parameter  (Sec.~\ref{Sec:Dwave}).  In Sec.~\ref{Sec:Staggered} we discuss the effects of introducing  staggered potential disorder on these quantities for an eight-site system. We conclude and discuss our results in Sec.~\ref{Sec:Discussion}.

\section{Model and Quantities Calculated}
\label{Sec:Model}
In this section, we define the checkerboard Hubbard model, specify the different types of disorder we consider, and define the quantities we calculate: 
the pair binding energy, the spin gap, and the $d$-wave order parameter.  The disordered CHM consists of 
$N$ electrons on an $M$ site lattice with a Hubbard-Anderson Hamiltonian 

\begin{eqnarray}
\label{eq:HubbardH}
H 	& = 	& 	- \sum_{ij} t_{ij}\brac{(}{\hat{c}_{i\sigma}^\dagger \hat{c}_{j\sigma} + \mathrm{H. c.}}{)} 
        	  		+ \sum_{i}U_i \hat{n}_{i\sigu}\hat{n}_{i\sigd}\nonumber\\
	&   	& 	+ \sum_{i\sigma}W_i \hat{n}_{i\sigma},
\end{eqnarray}
where $ \hat{c}^{\dagger}_{i\sigma} $ is a fermionic creation operator for a spin-$\sigma$ electron on site $i$, 
$ \hat{n}_{i\sigma} = \hat{c}^{\dagger}_{i\sigma}  \hat{c}_{i\sigma} $ is the number operator,
 $t_{ij}$ is a hopping amplitude, $U_{i}$ is an on-site Hubbard interaction, and 
$W_{i}$ is an on-site disorder potential. We choose $U_{i}=U$ on all sites of the lattice and  $W_i = W\delta_i$,  
where $\delta_i$ is drawn with uniform probability from  $[-0.5,0.5]$. 
\begin{figure}
\includegraphics[scale=0.75]{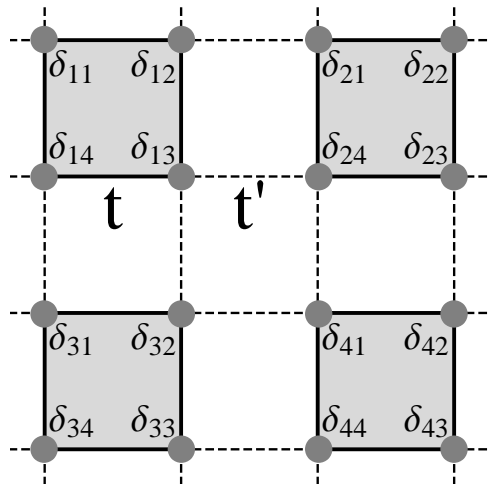}
\caption{
\label{fig:SiteDisorder} Illustration of the inhomogeneous hopping terms $t$ and $t'$ and the distribution of disorder in the checkerboard Hubbard model with potential disorder. $\delta_{Ab}$ is the disorder potential on site $b$ of plaquette $A$.}
\end{figure}
We focus on the situation in which the hopping parameters define a 
checkerboard,\cite{TsaiKivelson1,TsaiKivelson2} as illustrated in Fig.~\ref{fig:SiteDisorder}; 
\begin{eqnarray*}
t_{ij} & = & \left\{
              \begin{array}{cc}
               t,  & \nn{ij} \in I,\\ 
               t', & \nn{ij}, i \in I,j\in J, I \neq J,\\
               0,  & {\rm otherwise.}
              \end{array} \right. ,
\end{eqnarray*}
Lowercase letters ($ij$) label sites on the lattice, uppercase letters ($IJ$) label
plaquettes, and $\nn{ij}$ indicates that sites $i$ and $j$ are
nearest neighbours on the lattice. Thus, $t$ gives the magnitude of the
hopping term on a single plaquette, while $t'$ gives the magnitude of
inter-plaquette hopping. Without loss of generality, we choose $t'\le t$. 

We consider two types of disorder: on-site potential disorder, as discussed above, and  hopping disorder, 
corresponding to disorder in the intra-plaquette ($t_{ij}=t$) nearest neighbour hopping terms. 
For sites $i$ and $j$ on plaquette $I$, we choose $t_{ij} = t\left[1 + \eta_{ij}\left({W}/{t}\right)\right]$, 
where $\eta_{ij}$ is drawn with uniform probability from $[-0.5,0.5]$.  For random hopping we always consider
$W < 2$, so that the $t_{ij}$ are always positive.

\subsection{Pair binding energy}
\label{Sec:PBE}

Let $E_m$ be the ground state energy of an $M$ 
site cluster with $m$ holes, where $m\ge1$ and is measured from half-filling. The pair binding energy (PBE) for that $M$-site cluster is 
\begin{eqnarray}
\label{eq:PBE}
\Delta_{p} & = & 2 E_m - (E_{m+1}+E_{m-1}).
\end{eqnarray}
The PBE can interpreted physically in the following way: for a system of two identical clusters 
with an average of $m$ holes per cluster, $\Delta_{p}>0$ indicates that it is energetically favourable 
to place $m+1$ holes on one cluster and $m-1$ holes on the other rather than $m$ holes on both clusters.

To study the role of disorder on the pair binding properties of the inhomogeneous Hubbard model, 
we study both the disorder averaged pair binding energy and the distribution of
pair binding energies as functions of $U/t$, $t^\prime/t$
and $W/t$.  For a specific disorder configuration $k$, we define the associated PBE to be
$$\Delta_{p}^k = 2 E_m^k - (E_{m+1}^k + E_{m-1}^k),$$ 
where $E_m^k$ is the ground state energy of disorder configuration $k$ when it has $m$ holes. 
The disorder-averaged PBE is
\begin{eqnarray}
\label{eq:DAPBE}
\left<\Delta_{p}\right> & = & \frac{1}{K}\sum_{k=1}^{K}\Delta_{p}^k,
\end{eqnarray}
where the angle brackets $\left<...\right>$ indicate an average over $K$ disorder configurations 
and it is understood that $\Dp$ is a function of $U/t$, $W/t$, and $t'/t$. 
We also calculate the probability of measuring a 
positive PBE for any given disorder configuration by averaging over $K$ configurations at fixed $W$. We denote this 
quantity by $P(\Delta_{p}^k>0)$.

\subsection{Spin gap}
Let $ E_{0}(m=2,S=0)$ and $E_{0}(m=2,S=1)$ denote the energies of the lowest-lying $S=0$ state and the 
lowest-lying $S=1$ state of 2 holes, respectively. The spin gap,
\begin{eqnarray}
\Delta_{s} 	& 	= 	& E_{0}(m=2,S=1)-E_{0}(m=2,S=0), 
\end{eqnarray}
also provides a measure of the energy scale towards pairing in the CHM. In BCS theory, in the thermodynamic limit,  $$\lim_{N\to\infty}\Delta_{s}=\lim_{N\to\infty}\Delta_{p}=2\Delta_{0},$$ where $N$ is the system size and $\Delta_{0}$ is the superconducting gap.\cite{Karakonstantakis} We calculate the disorder averaged spin gap,
\begin{eqnarray}
\left<\Delta_{s}\right> 	& 	= 	& \frac{1}{K}\sum_{k=1}^{K}\Delta_{s}^{k}, 
\end{eqnarray} 
and study how it behaves as a function of disorder strength.

In the absence of disorder, the total spin eigenvalue of the $m=2$ ground state is $S=0$ for all $t'/t$ and $U/t$ not too large ($U/t\lesssim20$). The introduction of disorder alters the energy spectrum, which may lead to level crossings between $S=0$ states and $S=1$ states depending on the strength of interactions, intra-plaquette hopping, and disorder strength. Hence, we also calculate the probability of finding $S=1$ in the ground state of the $m=2$ system for each cluster size as a function of $U/t$, $t'/t$, and $W/t$. 

\subsection{d-wave pairing order parameter}
The PBE and the spin gap provide measures of the tendency towards superconductivity in the CHM. However, 
neither of these quantities give information about the symmetry of the ground state in the region of parameter 
space where these quantities are positive. It is 
expected\cite{TsaiKivelson1,TsaiKivelson2,YaoKivelson1,YaoKivelson2,BaruchOrgad,Doluweera,Tremblay,Karakonstantakis} 
that superconductivity in the CHM has $d$-wave symmetry, hence we calculate a 
$d$-wave order parameter of a standard form.\cite{Garg} Let $\hat{D}$ be the singlet operator acting on the bonds, defined by
\begin{eqnarray}
\hat{D}	& = & \sum_{\left<ij\right>}D_{ij} c_{i\sigu}c_{j\sigd},
\end{eqnarray}
where $D_{ij}$ is equal to +1 on bonds oriented along the $x$ direction and -1 for bonds along the $y$ direction. 
The disorder averaged $d$-wave order parameter is then the matrix element between ground states with $m$ and $m-2$ holes: 
\begin{eqnarray}
\left<\Psi_d\right> & = & \frac{1}{K}\sum_{k=1}^{K}\left<S=0,m;k\right|\hat{D}\left|S=0,m-2;k\right> .
\label{eq:DWOP}
\end{eqnarray}

\section{Exact Diagonalization Results}
\label{Sec:Results}
\begin{figure}[h]
\begin{center}
\includegraphics[angle=0,height=2cm]{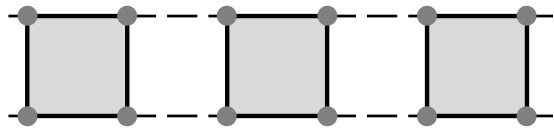} 
\end{center}
\caption{
\label{fig:Geometries}
Lattice geometry studied in this paper, shown for 12 sites.}
\end{figure}

We now discuss the results obtained from exact diagonalization of the disordered CHM in the presence of 
on-site and hopping disorder on eight- and twelve-site clusters. We consider the ladder geometries shown in 
Fig.~\ref{fig:Geometries} for eight- and twelve-site clusters. 
The boundary conditions are periodic along the direction of the length of the ladder for ladder geometries.
We study the dopings $x=1/8$ for eight-site clusters and $x=1/12$ for twelve-site clusters. We calculate all 
quantities discussed in Sec.~\ref{Sec:Model}  as a function of 
increasing disorder strength and compare our results to the clean system. We measure all energies in units of 
$t$. For each type of disorder, we average over 256 disorder realizations for eight-site clusters 
and 64 disorder realizations for twelve-site clusters. 

The data presented in this paper focused primarily on exact diagonalization of ladder systems. 
Tsai \etal.\cite{TsaiKivelson1,TsaiKivelson2} performed exact diagonalization calculations for 
the sixteen-site lattice at doping $x=1/16$ and $3/16$ and found evidence for an optimal value of inhomogeneity 
in the CHM. Tsai {\it et al.} were
able to make use of a number of symmetries to simplify their exact diagonalization calculations.  As soon as
disorder is introduced, all real-space symmetries of the lattice are lost immediately, which means that it is
harder to solve sixteen site systems, and makes calculating the disorder averaged spin gap and $d$-wave order parameter 
much more difficult.  We have hence mainly focused on obtaining disorder averaged results for eight and twelve site 
lattices.

\subsection{Pair-binding energy}
\label{Sec:PBE}
Numerical studies of the CHM employing dynamical cluster QMC\cite{Doluweera} and cluster 
DMFT\cite{Tremblay} suggest that $d$-wave superconductivity is generally suppressed relative to the homogeneous
 case by introducing inhomogeneity in hopping. These studies suggest that the maximum values of either 
$T_{c}$ or the $d$-wave order parameter generally do not exceed that of the homogeneous system. 
On the other hand, results from exact diagonalization studies of the CHM on 
4$\times$4 lattices\cite{TsaiKivelson1,TsaiKivelson2} and DMRG studies of the 
CHM on ladders\cite{Karakonstantakis} suggest that at low doping the CHM can have 
enhanced $d$-wave pairing compared to the uniform Hubbard model. In 
Refs.~\onlinecite{TsaiKivelson1,TsaiKivelson2,Karakonstantakis}, the optimal parameters that 
maximize the PBE are $t'=0.5\,t$ and $U=8\,t$ for a 4$\times$4 system at doping $x=1/16$ and  
$t'\approx0.6-0.8\,t$ and $U=6\,t$ in a ladder system at doping $x=1/8$. Moreover, 
these results indicate that other quantities relevant to superconductivity, 
such as the spin gap, $d$-wave pairing operator, and pair-field correlations, 
are also optimized in the region where the PBE is maximal. Thus, the PBE should be a reasonable 
measure for predicting where in parameter space the tendency towards superconductivity in the CHM 
may be strongest. 

\begin{figure}
$
\begin{array}{cc}
\hspA(a)	& 
\hspB(b)\vspace{-0.5cm}  	 \\

\hspA	\includegraphics[angle=-90,width=5cm]{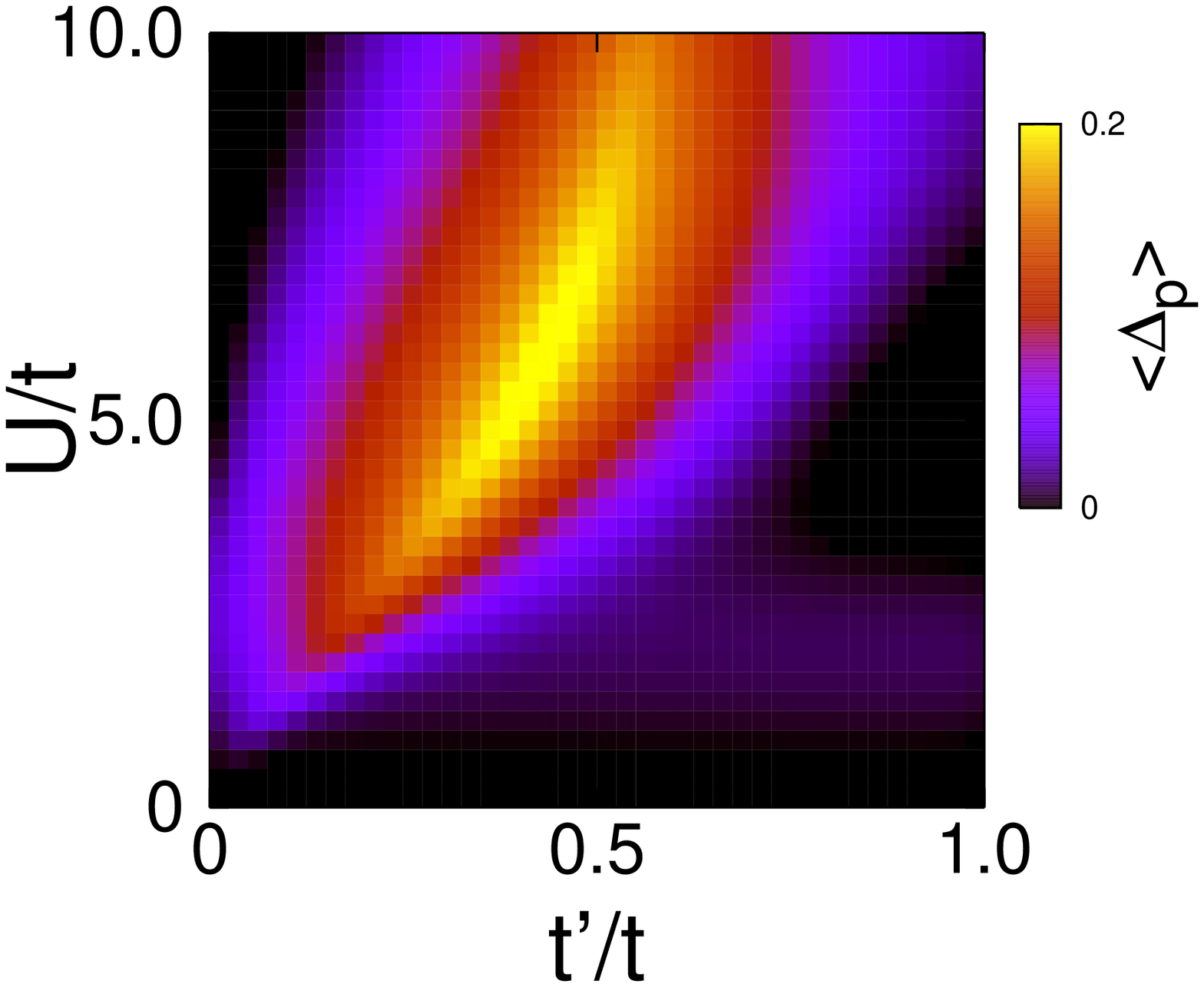} 	& 
\hspB  	\includegraphics[angle=-90,width=5cm]{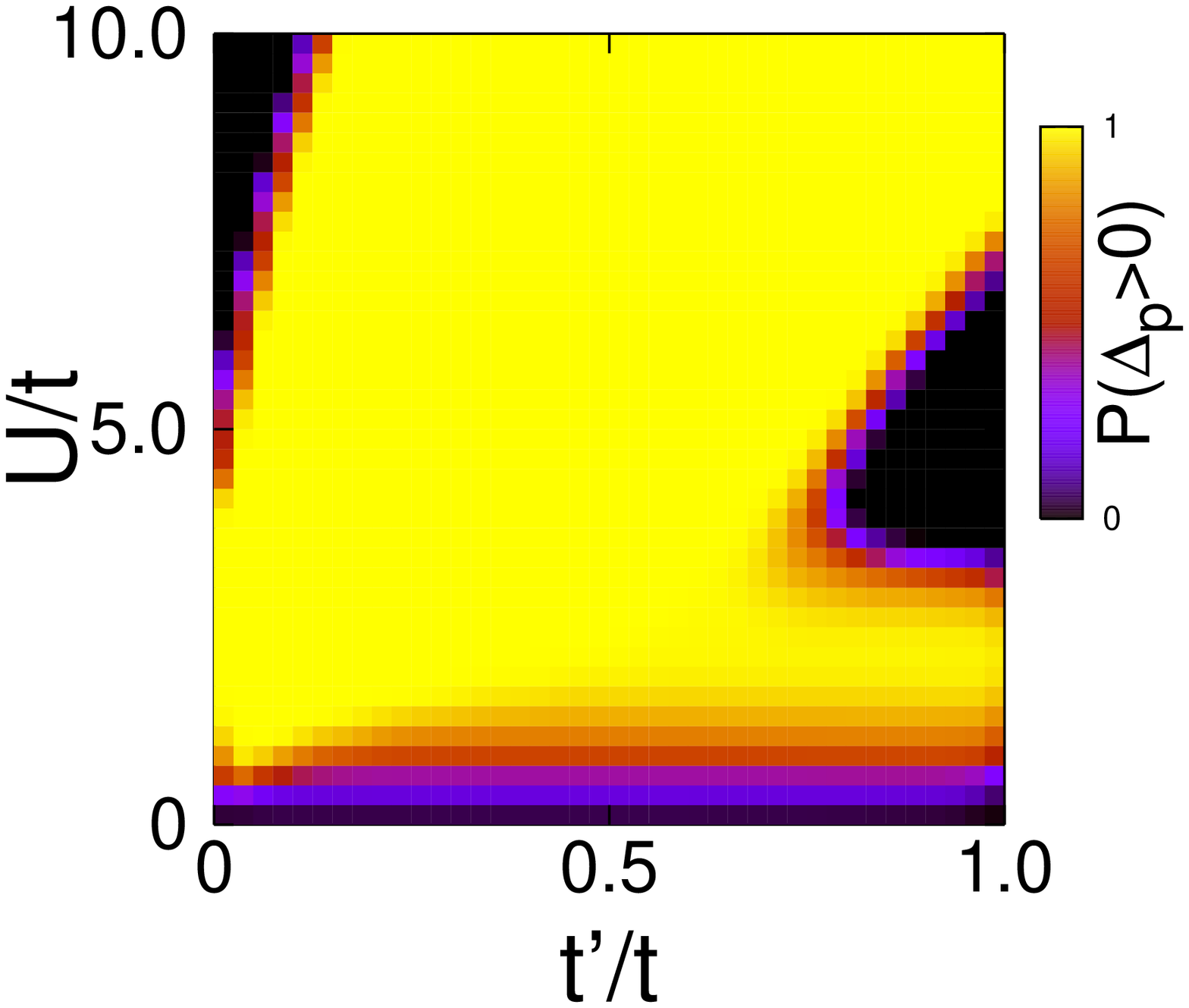}\\

\hspA(c)	& \hspB(d)\vspace{-0.5cm}  	 \\

\hspA	\includegraphics[angle=-90,width=5cm]{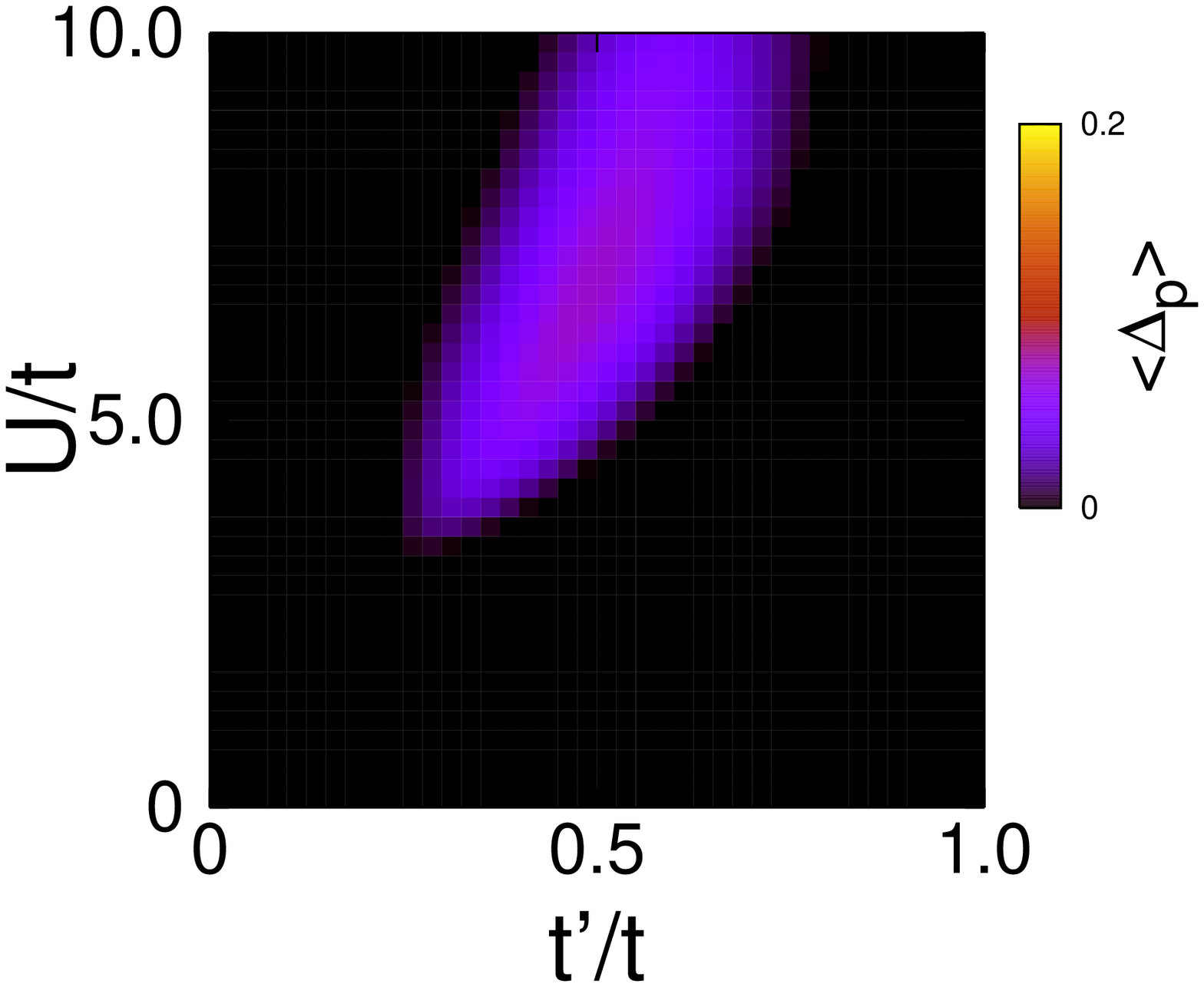} &
\hspB	\includegraphics[angle=-90,width=5cm]{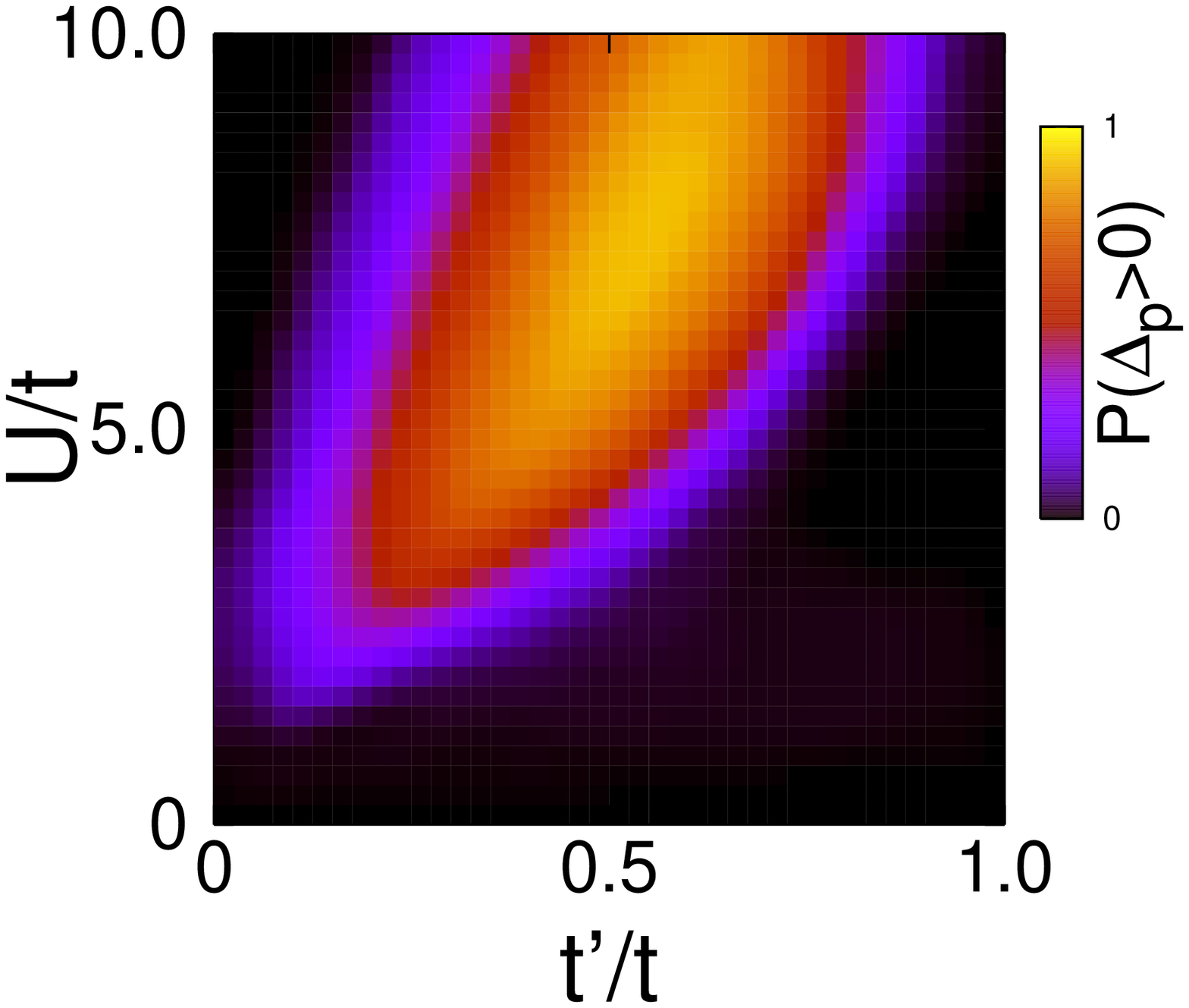}
\end{array}
$
\caption{ 
\label{fig:PBEOnsite08}
Disorder-averaged pair-binding energy, $\left<\Delta_{p}\right>$, and the probability of observing a positive pair-binding energy, $P\left(\Delta_{p}^{k}>0\right)$, for the eight-site ladder at doping $x=1/8$ with potential disorder: 
(a) $\left<\Delta_{p}\right>, W/t = 0.05$;
(b) $P\left(\Delta_{p}^{k}>0\right), W/t = 0.05$;
(c) $\left<\Delta_{p}\right>, W/t = 1.00$;
(d) $P\left(\Delta_{p}^{k}>0\right), W/t = 1.00$.
}
\end{figure}

\begin{figure}
$
\begin{array}{cc}
\hspA(a)	& 
\hspB(b)\vspace{-0.5cm}  	 \\

\hspA\includegraphics[angle=-90,width=5cm]{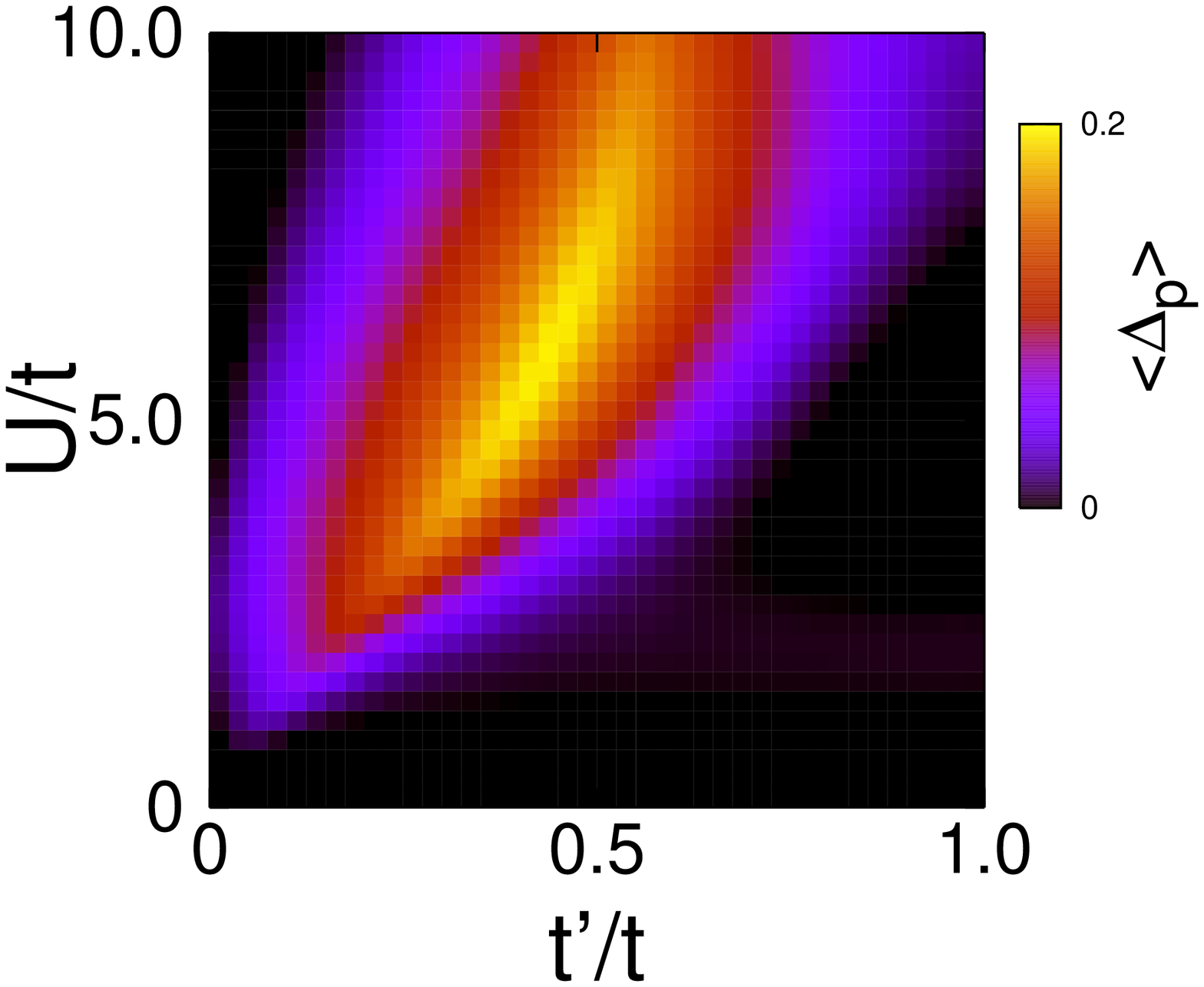} 	& 
\hspB\includegraphics[angle=-90,width=5cm]{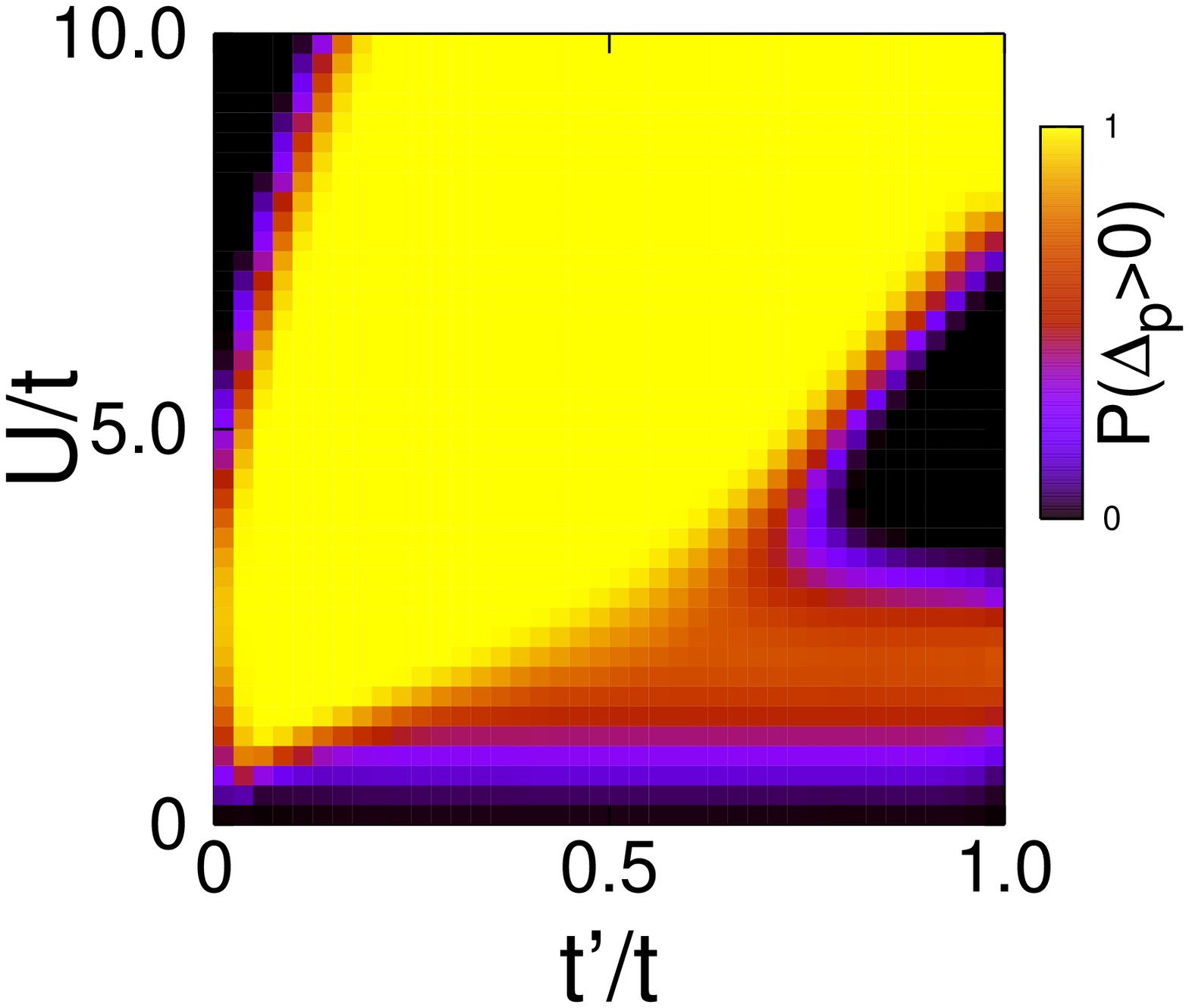}\\

\hspA(c)	& 
\hspB(d)\vspace{-0.5cm}  	 \\

\hspA\includegraphics[angle=-90,width=5cm]{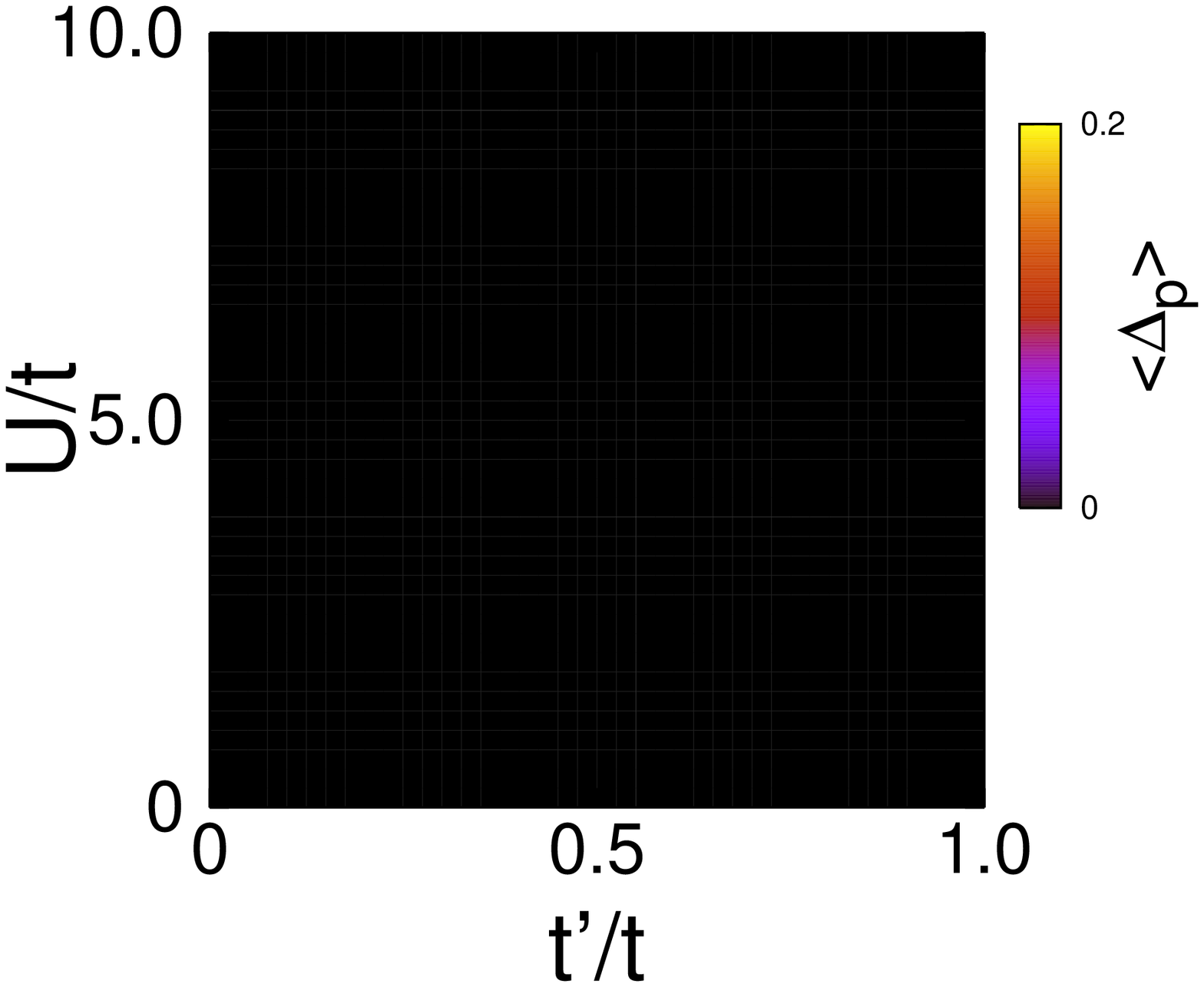} &
\hspB\includegraphics[angle=-90,width=5cm]{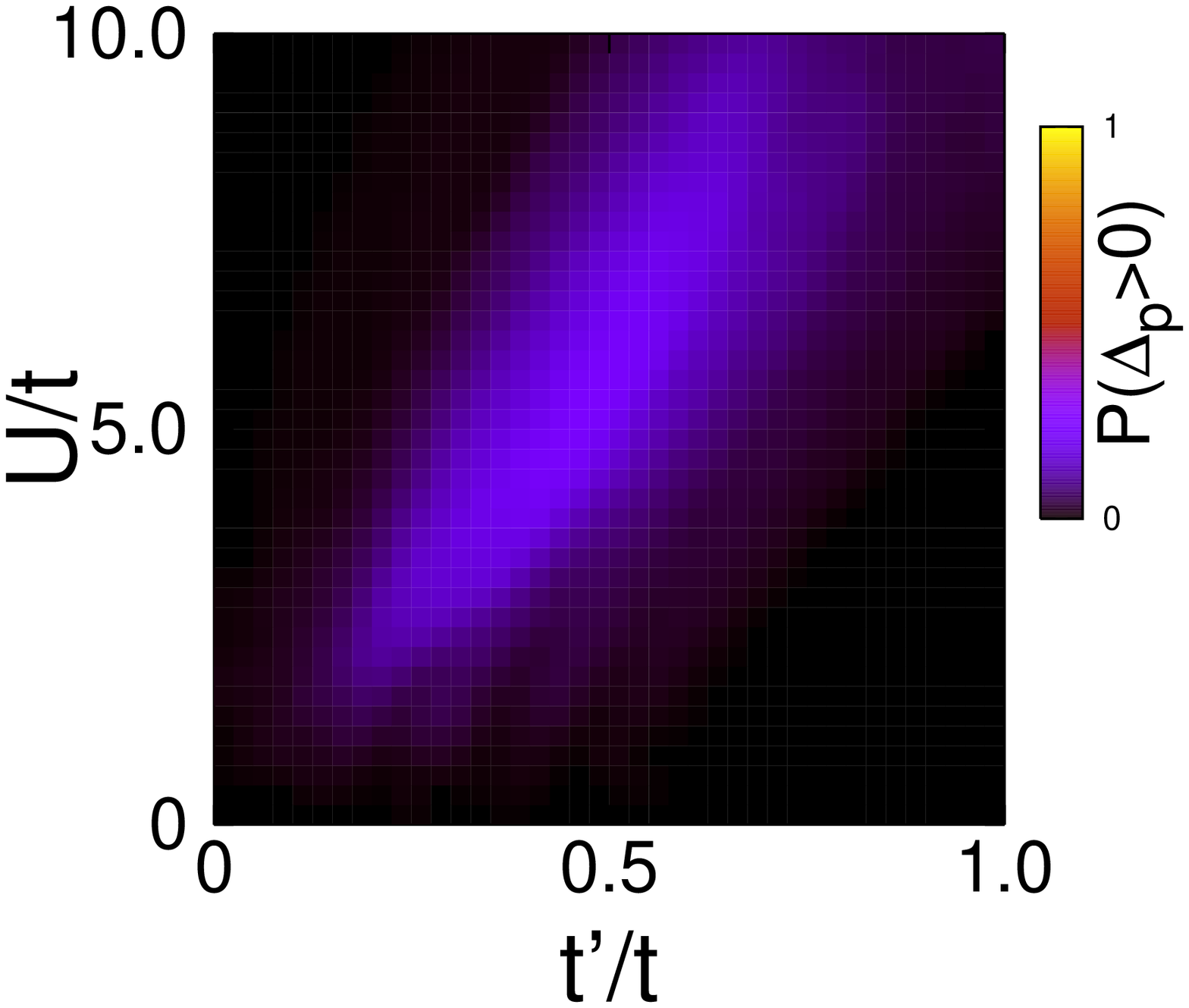}
\end{array}
$
\caption{
\label{fig:PBEBond08} 
Disorder-averaged pair-binding energy, $\left<\Delta_{p}\right>$, and the probability of observing a positive pair-binding energy, $P\left(\Delta_{p}^{k}>0\right)$, for the eight-site ladder at doping $x=1/8$ with hopping disorder: 
(a) $\left<\Delta_{p}\right>, W/t = 0.05$; 
(b) $P\left(\Delta_{p}^{k}>0\right), W/t = 0.05$; 
(c) $\left<\Delta_{p}\right>, W/t = 1.00$;  
(d) $P\left(\Delta_{p}^{k}>0\right), W/t = 1.00$.
}
\end{figure}

\begin{figure}
$
\begin{array}{cc}
\hspA(a)	& 
\hspB(b)\vspace{-0.5cm}  	 \\

\hspA\includegraphics[angle=-90,width=5cm]{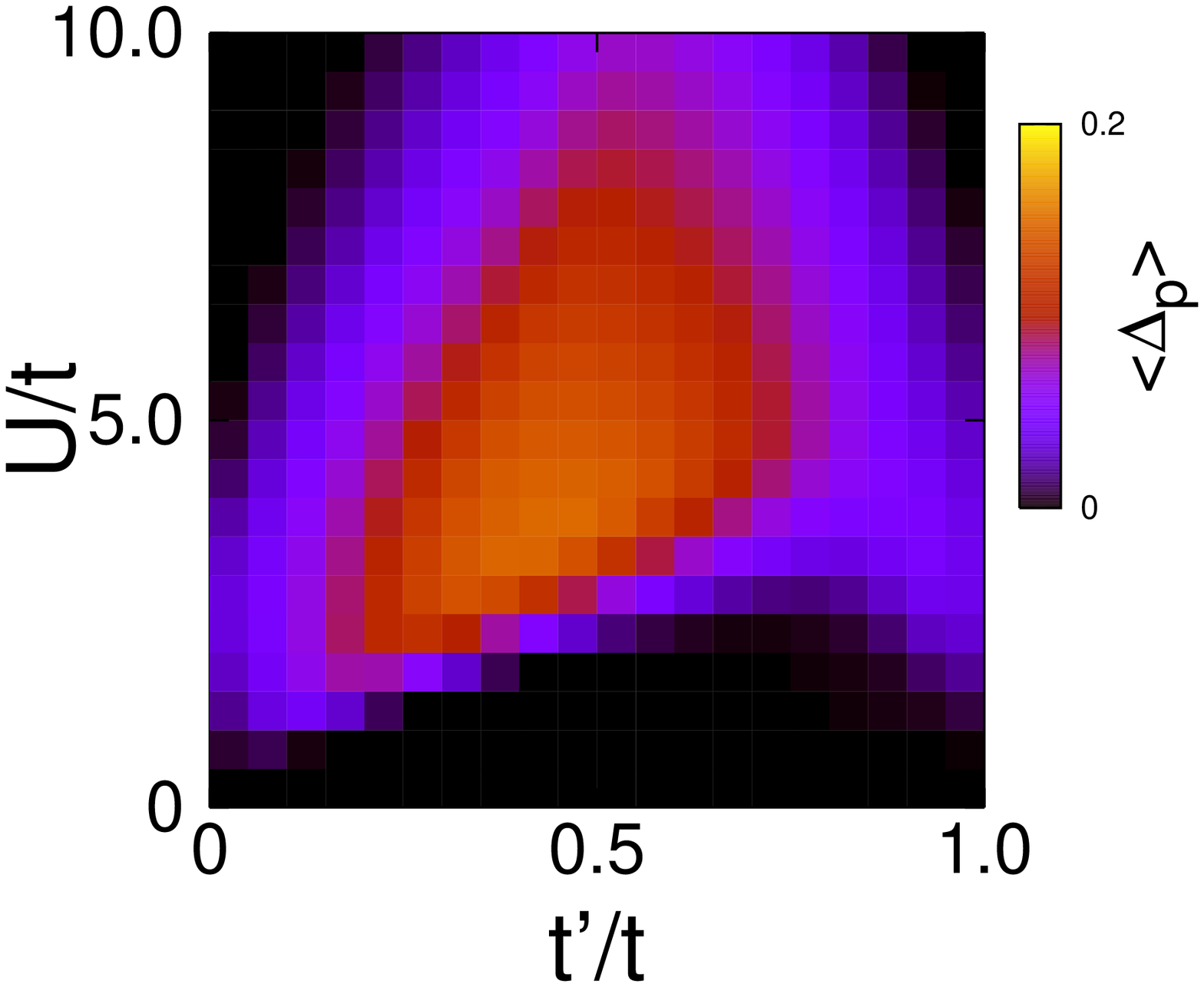} 	& 
\hspB\includegraphics[angle=-90,width=5cm]{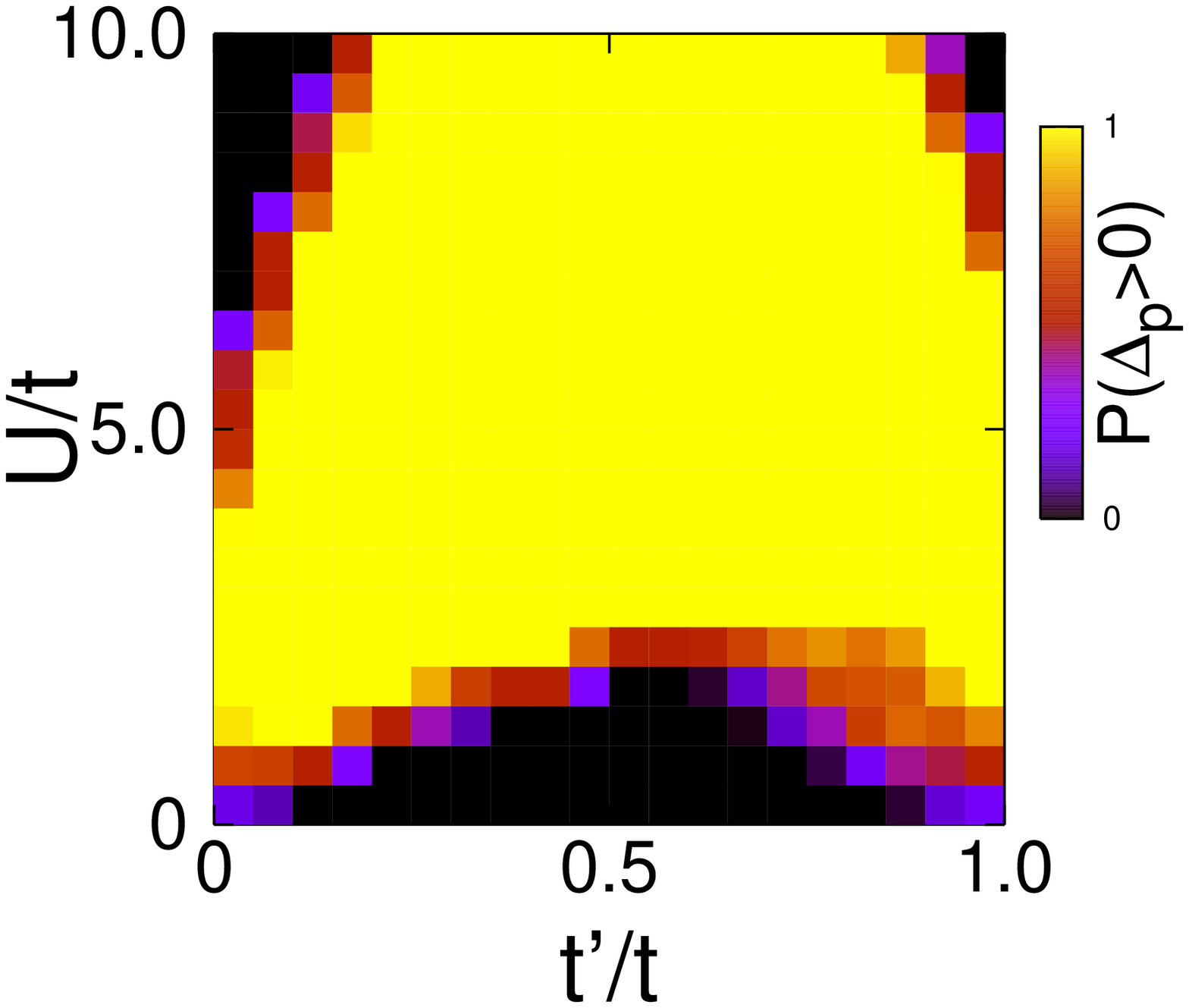}\\

\hspA(c)	& 
\hspB(d)\vspace{-0.5cm}  	 \\

\hspA\includegraphics[angle=-90,width=5cm]{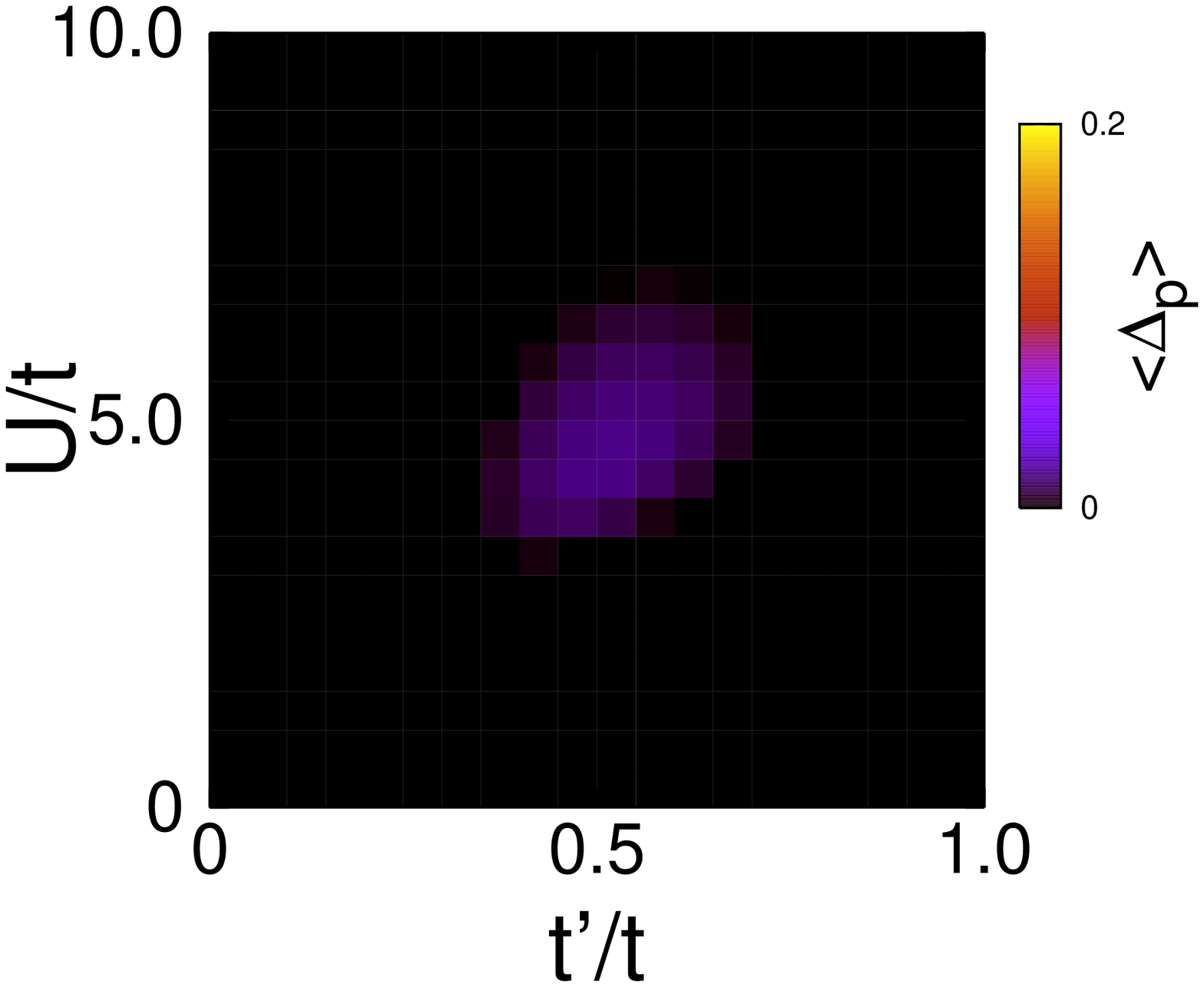} &
\hspB\includegraphics[angle=-90,width=5cm]{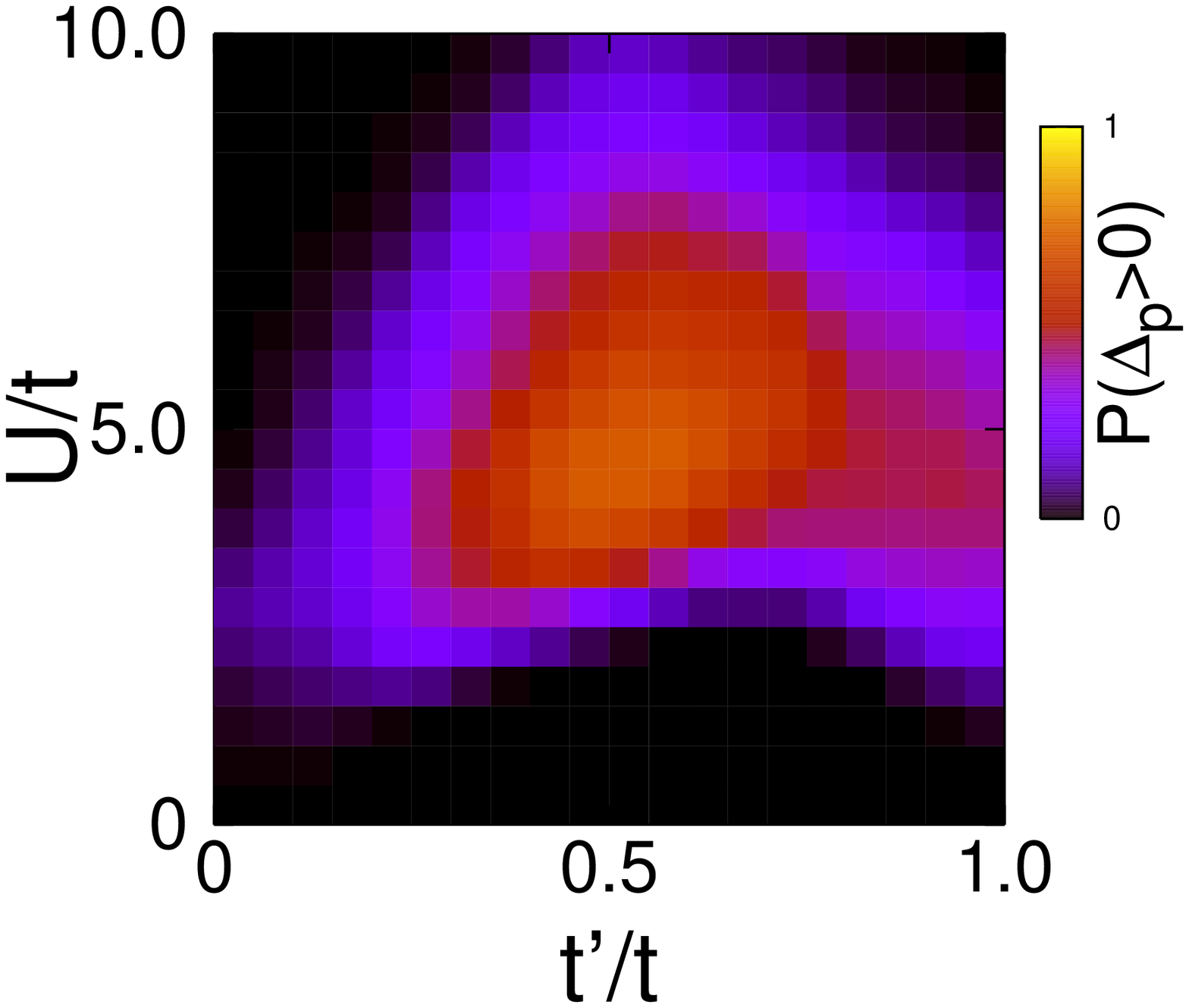}
\end{array}
$
\caption{
\label{fig:PBEOnsite12L} 
Disorder-averaged pair-binding energy, $\left<\Delta_{p}\right>$, and the probability of observing a positive pair-binding energy, $P\left(\Delta_{p}^{k}>0\right)$, for the twelve-site ladder at doping $x=1/12$ with potential disorder: 
(a) $\left<\Delta_{p}\right>, W/t = 0.05$; 
(b) $P\left(\Delta_{p}^{k}>0\right), W/t = 0.05$; 
(c) $\left<\Delta_{p}\right>, W/t = 1.00$; 
(d) $P\left(\Delta_{p}^{k}>0\right), W/t = 1.00$.
}
\end{figure}

\begin{figure}
$
\begin{array}{cc}
\hspA(a)	& 
\hspB(b)\vspace{-0.5cm}  	 \\

\hspA\includegraphics[angle=-90,width=5cm]{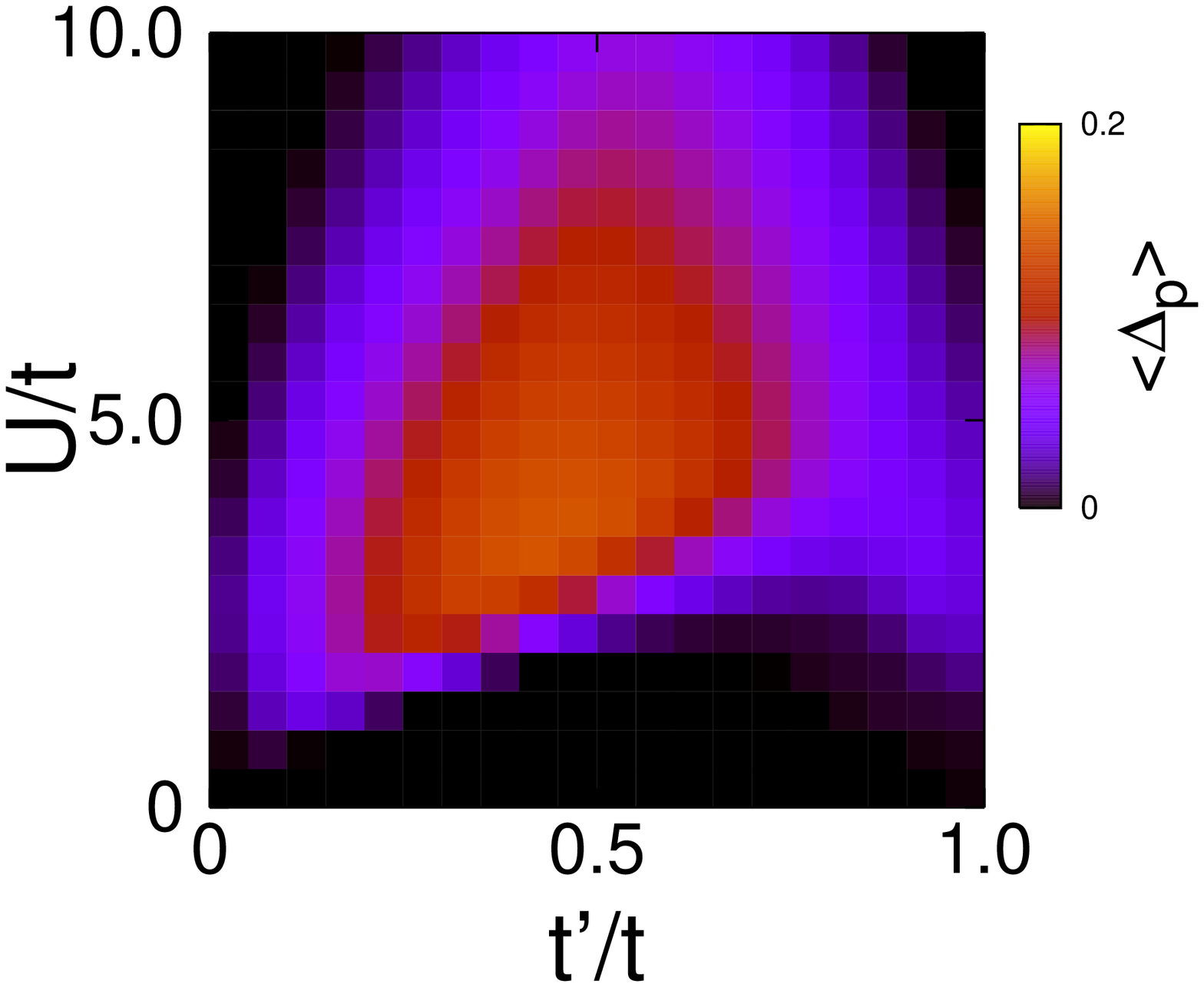} 	& 
\hspB\includegraphics[angle=-90,width=5cm]{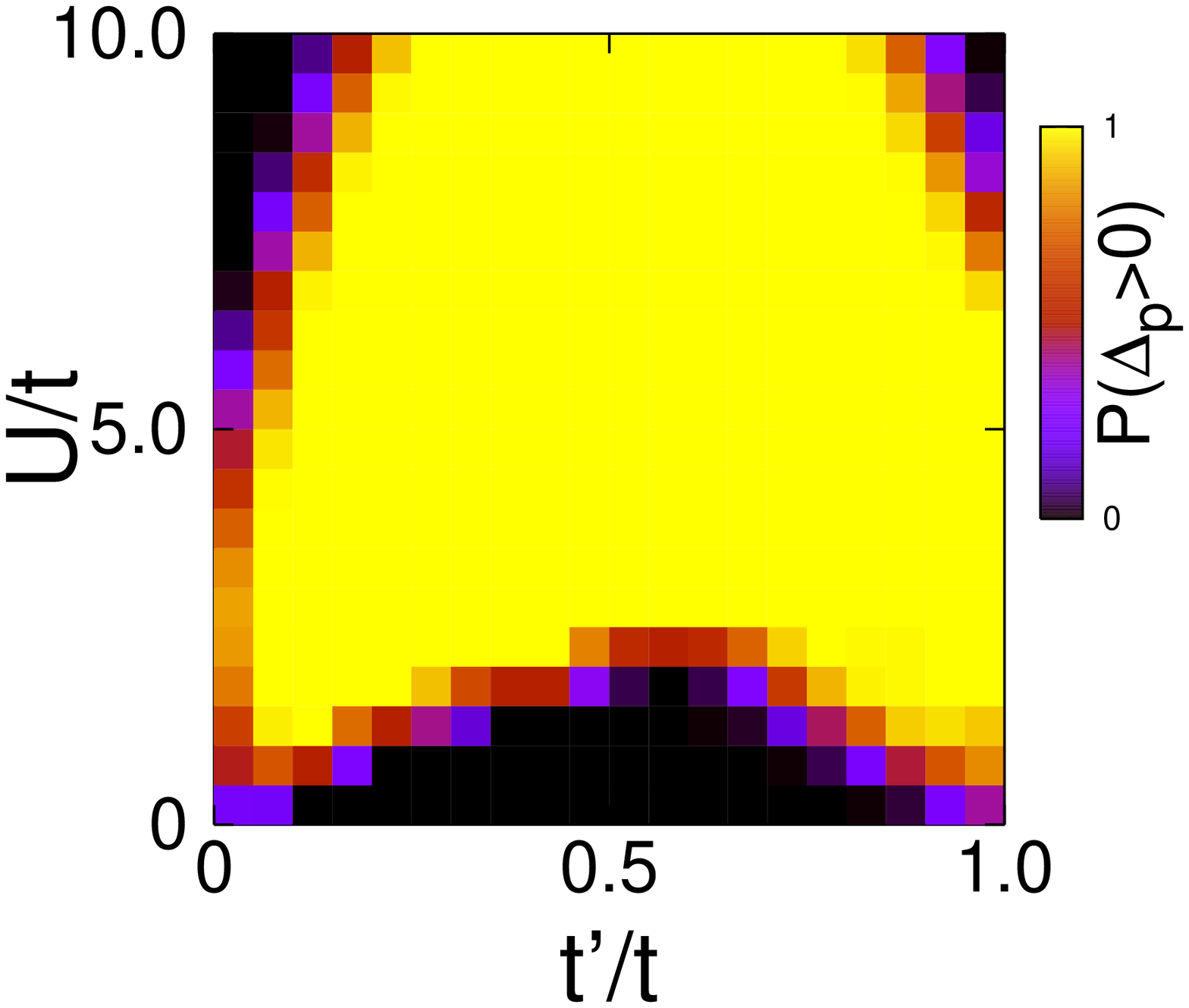}\\

\hspA(c)	& 
\hspB(d)\vspace{-0.5cm}  	 \\

\hspA\includegraphics[angle=-90,width=5cm]{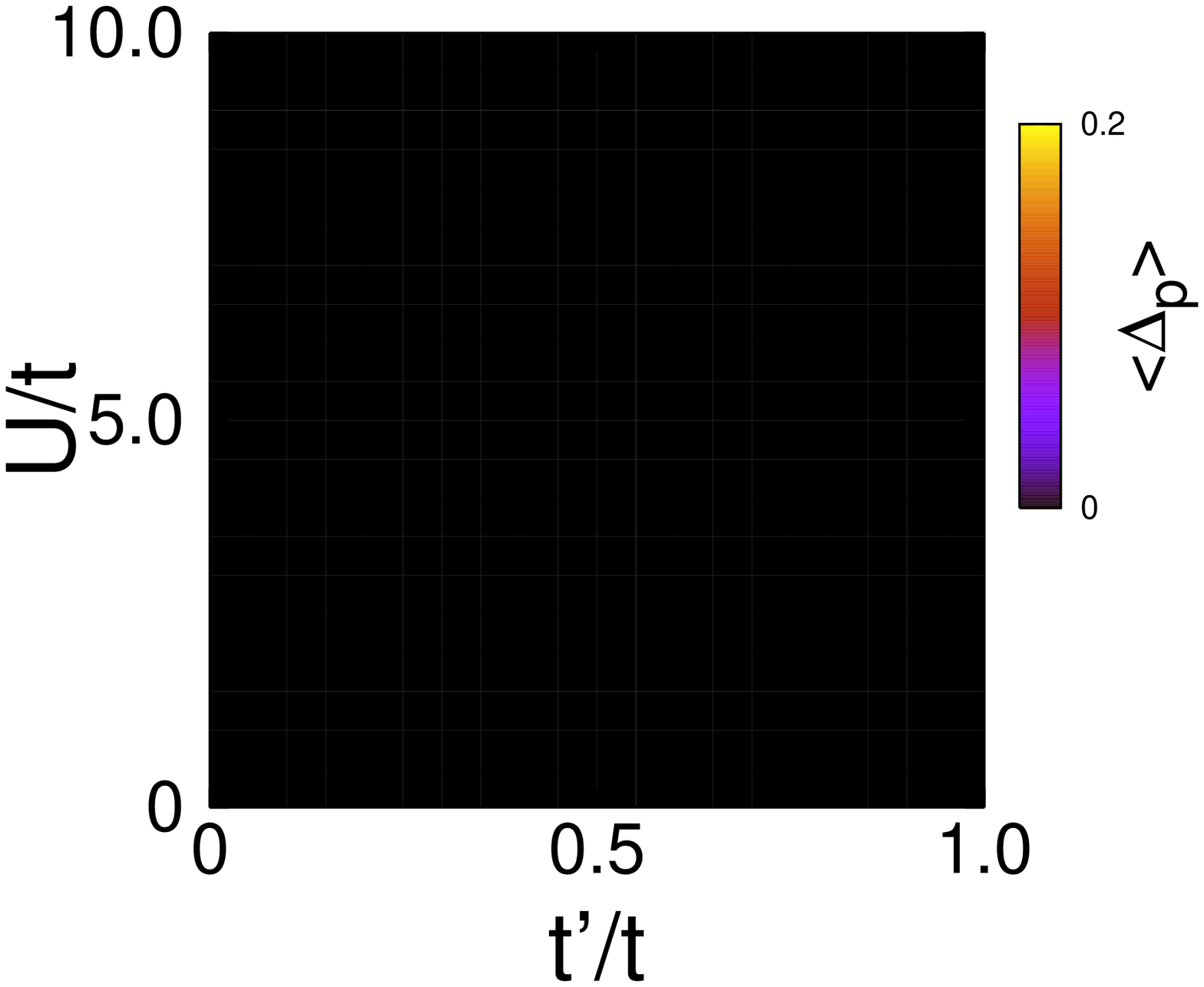} &
\hspB\includegraphics[angle=-90,width=5cm]{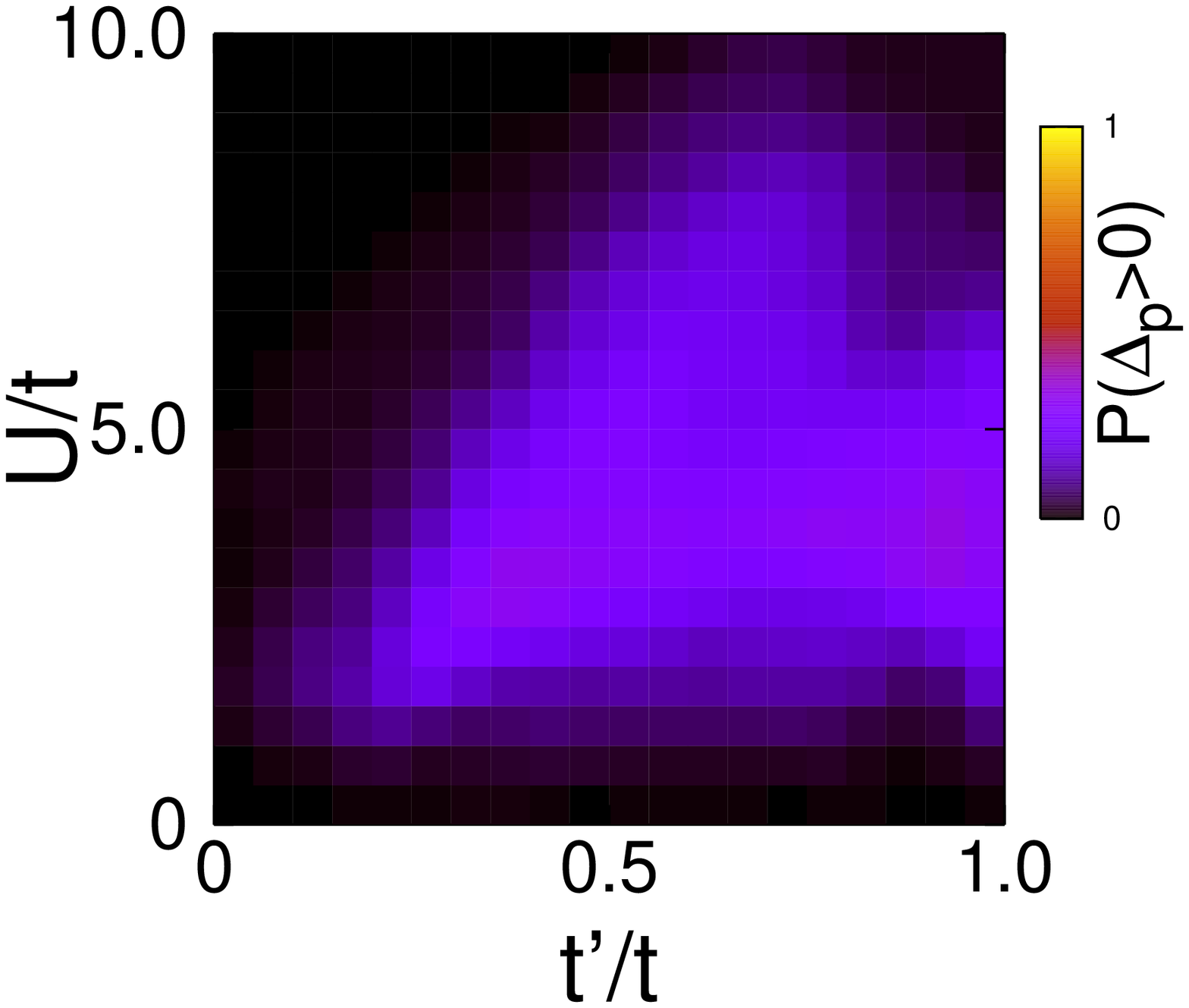}
\end{array}
$
\caption{
\label{fig:PBEBond12L} 
Disorder-averaged pair-binding energy, $\left<\Delta_{p}\right>$, and the probability of observing a positive pair-binding energy, $P\left(\Delta_{p}^{k}>0\right)$, for the twelve-site ladder at doping $x=1/12$ with hopping disorder: 
(a) $\left<\Delta_{p}\right>, W/t = 0.05$; 
(b) $P\left(\Delta_{p}^{k}>0\right), W/t = 0.05$; 
(c) $\left<\Delta_{p}\right>, W/t = 1.00$; 
(d) $P\left(\Delta_{p}^{k}>0\right), W/t = 1.00$.
}
\end{figure}

Previous work on the superconducting 
properties of the CHM was in the clean limit.\cite{TsaiKivelson1, TsaiKivelson2, Doluweera, Tremblay, Karakonstantakis}  
Here, we ask how disorder affects these properties.  We determine the degree to which disorder enhances or 
suppresses $d$-wave superconductivity in such systems. In Figs.~\ref{fig:PBEOnsite08}-\ref{fig:PBEBond12L}, 
we plot  $\left<\Delta_{p}\right>$ and $P\left(\Delta_{p}^{k}>0\right)$ as functions of $t'/t$ and 
$U/t$ at $W/t=0.05$ (Fig. ($a$)) and $W/t=$1.00 (Fig. ($c$)) for the eight-site ladder and twelve-site ladder
geometries for potential and hopping disorder. In order to see the regions where pair binding is favoured as 
disorder is increased, we plot only the values $\left<\Delta_{p}\right> > 0$. For both types of disorder,
 there exists an intermediate range of parameters $t'/t$ and $U/t$ where pair binding is enhanced 
relative to the standard Hubbard model ($t'/t=1$).


For the eight-site ladder, at weak disorder, we find that $\Dp$ appears to be maximum near $t'/t\approx 0.42$ 
at $U/t\approx5.6$ for $x=1/8$. For the twelve site ladder, at weak disorder, the optimal parameters are qualitatively 
similar, as $\Dp$ is maximal near $t'/t\approx 0.4$ and $U/t\approx 5.0$ for $x=1/12$. This is similar to the results of 
Tsai \etal.,\cite{TsaiKivelson1,TsaiKivelson2} who found that the pair binding energy to be maximal at 
$U/t\approx 8$ and $t'/t\approx0.5$ for doping $x=1/16$ on a sixteen-site cluster.  

We find the maximum value of $\Dp$ is always less than the maximum value of the PBE in the absence of disorder. 
As disorder strength is increased, the maximum value of $\Dp$ decreases. Furthermore, the region in $t'-U$ 
parameter space where $\Dp > 0$ shrinks as $W/t$ is increased, with the greatest persistence in the region of 
maximum PBE in the absence of disorder. For the eight-site ladder, we find that $\left<\Delta_{p}\right>$ becomes negative 
at $W/t\approx1.50$ for all $t'/t$ and $U/t$ studied for on-site disorder and  $W/t\approx1.00$ for hopping disorder, 
with similar cutoff values for the twelve-site ladders. For weak disorder, the pair binding properties of 
the system in the region of maximum $\Dp$ are qualitatively unchanged in the case of either on-site disorder or hopping disorder. 
The range in $t'-U$ space where there is a nonzero probability that pair binding is favoured for some disorder 
configurations is much wider than the region where pair binding is favoured on average and persists to much higher values of $W/t$. 

If one interprets the PBE as a measure of the tendency of the system to become superconducting, then our results 
can be interpreted in the following way: as disorder is increased, the region in parameter space where pair binding 
is favoured \emph{on average} decreases, consistent with finite disorder suppressing superconductivity. However, 
the fact that $P\left(\Delta_{p}^{k}>0\right)\ne0$ even when $\left<\Delta_{p}\right> < 0 $ suggests that there 
are local regions in real space where superconductivity can persist even when it is suppressed on average. 
Our results also suggest that pair binding in the CHM is more robust to disorder than in the uniform Hubbard model
as $W/t$ increases.  As disorder is increased, $\Dp$ is suppressed rapidly near $t'/t=1$, while pair binding 
persists at intermediate values of inhomogeneity for large values of $W/t$.

\begin{figure}
$
\begin{array}{cc}
\hspA(a)	& 
\hspB(b)\vspace{-0.5cm}  	 \\

\hspA	\includegraphics[angle=-90,width=5cm]{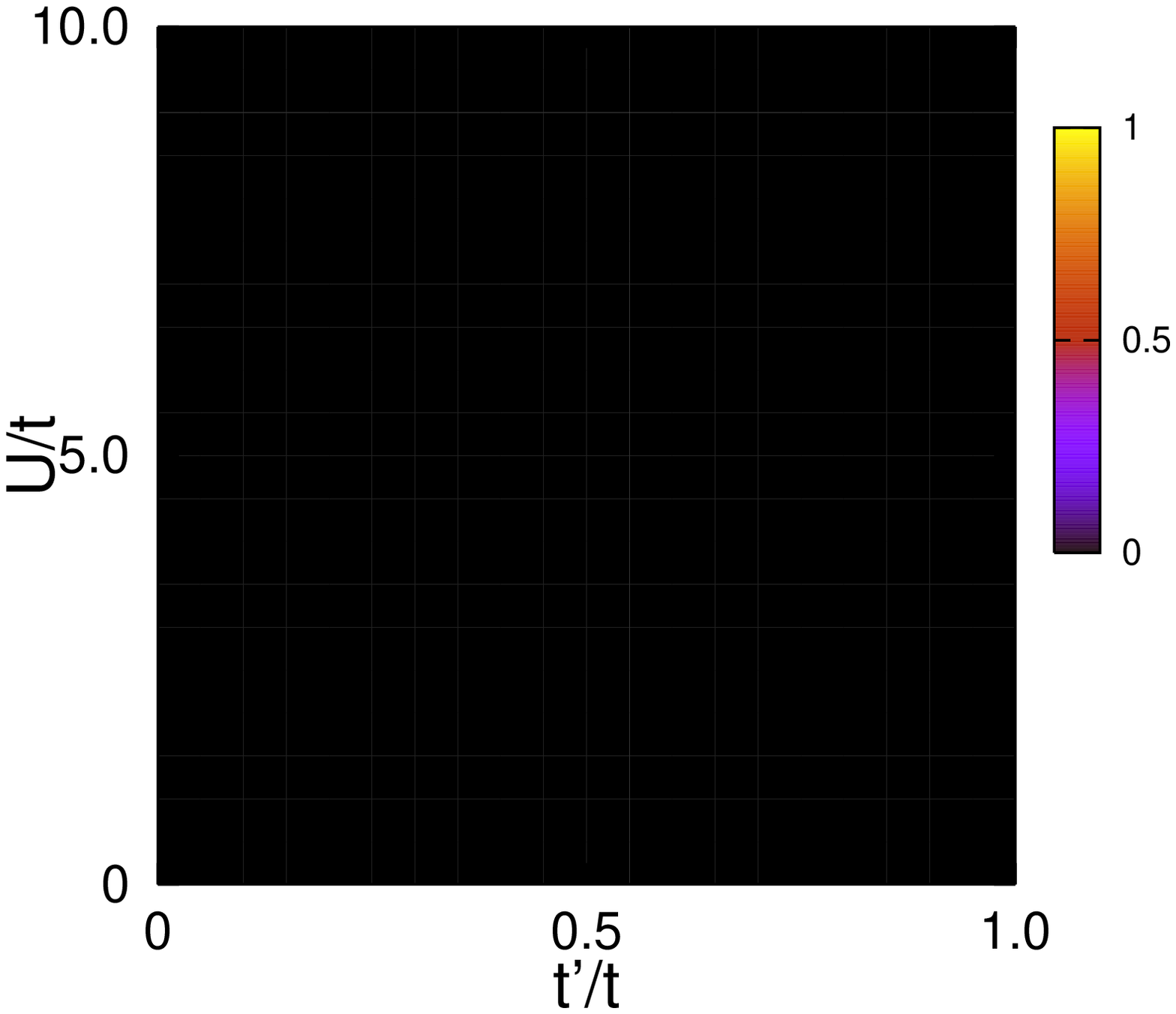} 	& 
\hspB  	\includegraphics[angle=-90,width=5cm]{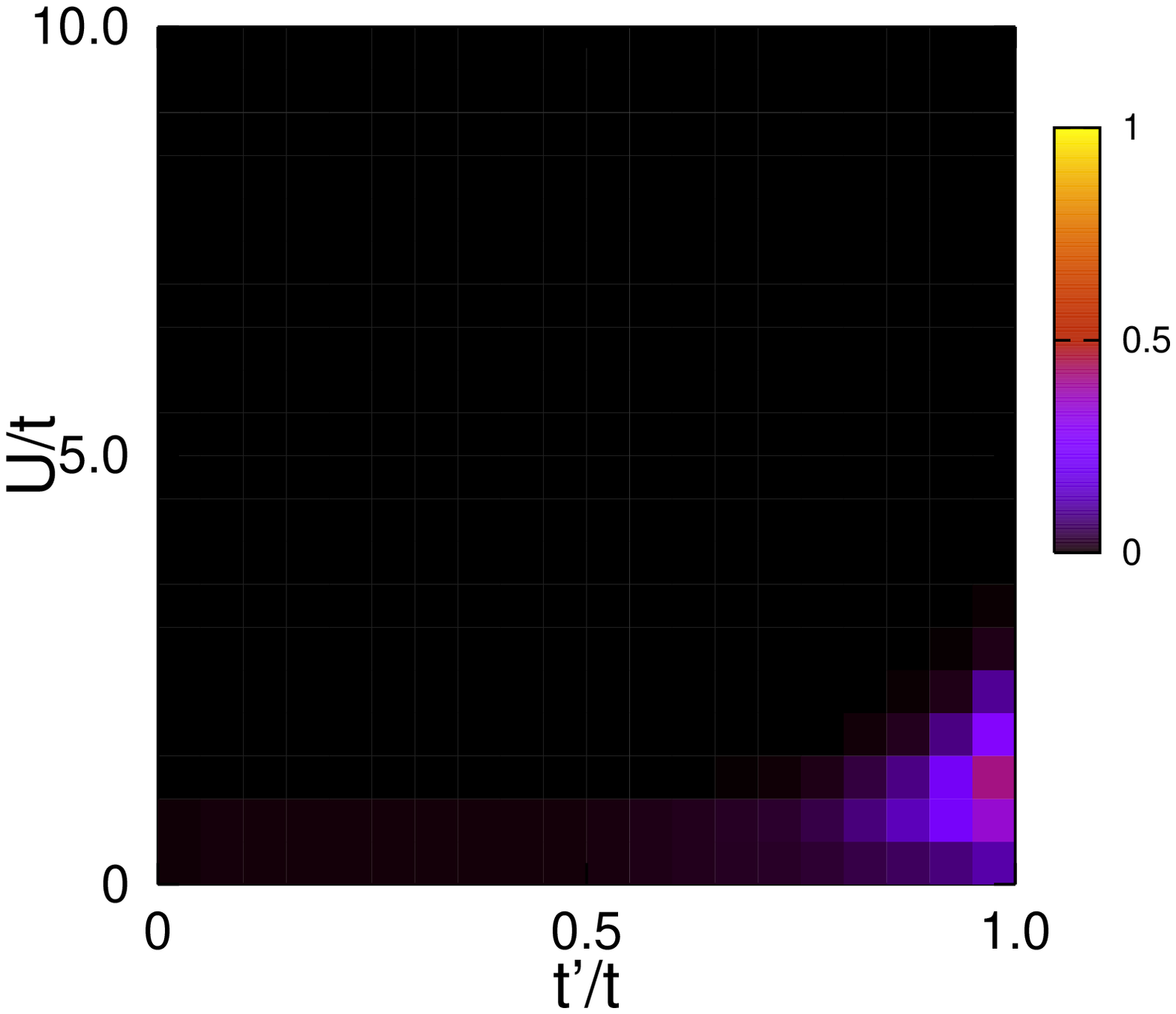}\\

\hspA(c)	& \hspB(d)\vspace{-0.5cm}  	 \\

\hspA	\includegraphics[angle=-90,width=5cm]{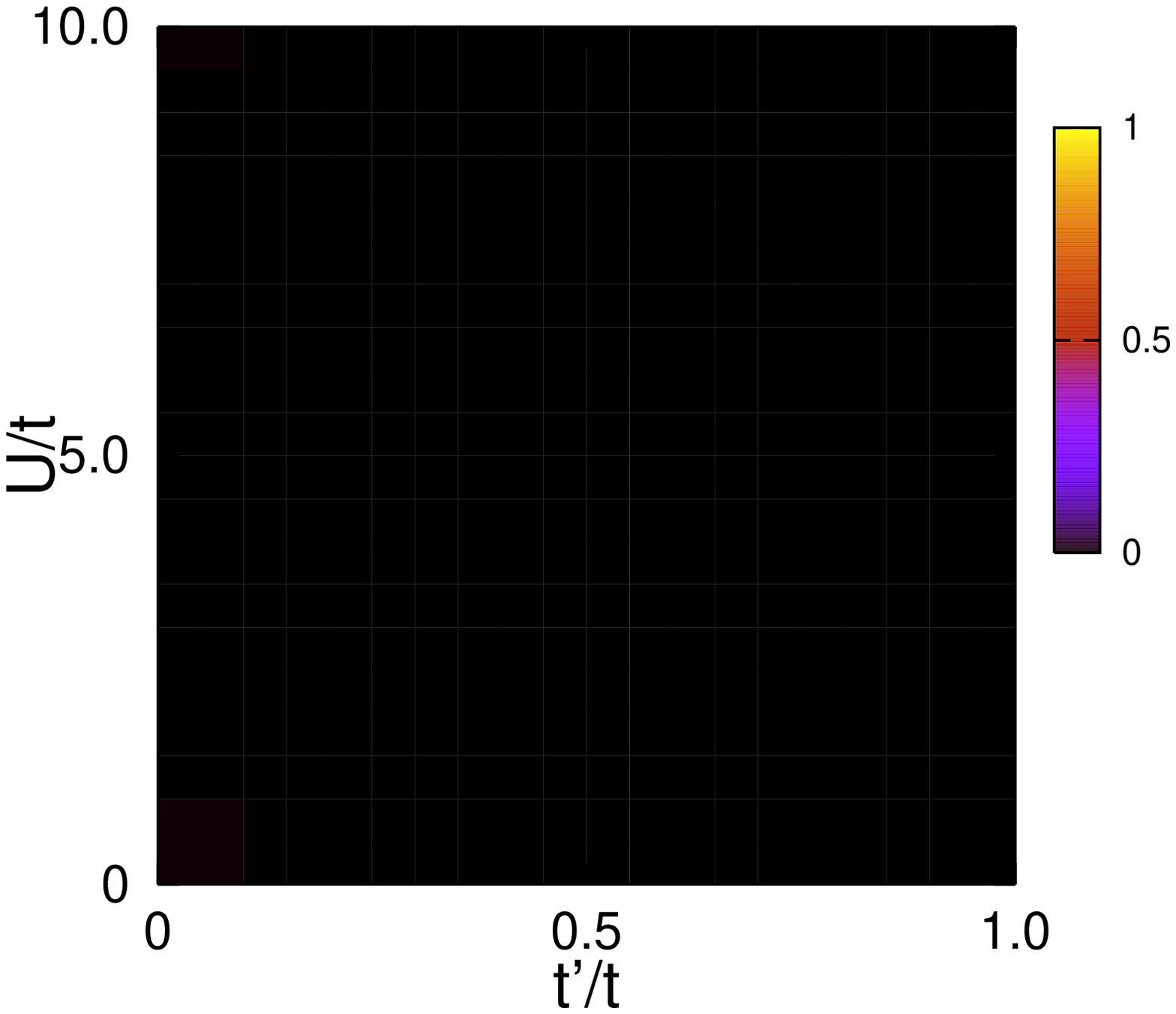} &
\hspB	\includegraphics[angle=-90,width=5cm]{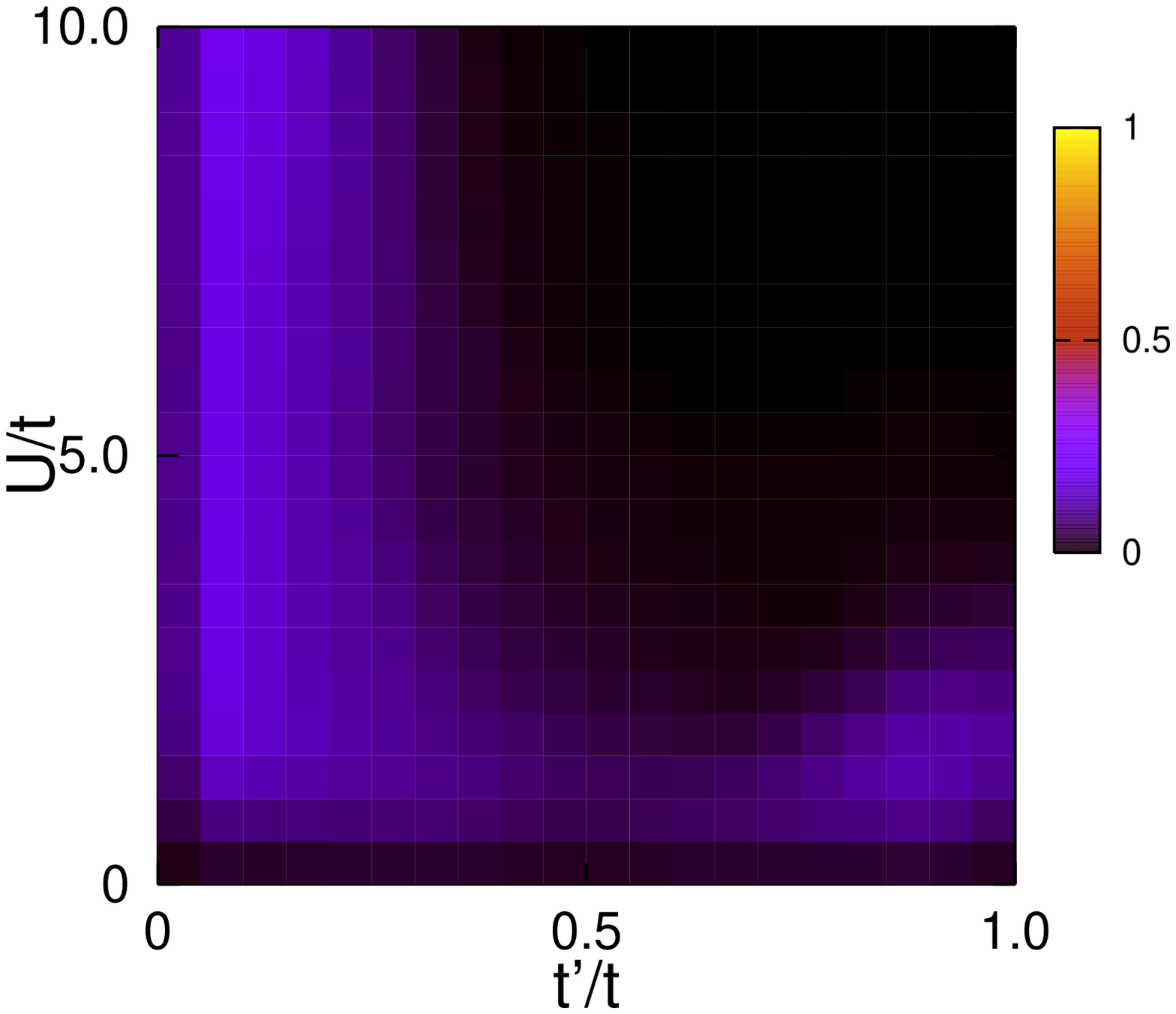}
\end{array}
$
\caption{
\label{fig:PS1OnsiteN08} 
Probability of observing a $S=1$ ground state in the $m=2$-hole doped system, $\PSS$, for the eight-site ladder cluster at doping $x=1/8$: 
(a) potential disorder, $W/t = 0.05$; 
(b) potential disorder, $W/t = 1.00$; 
(c) hopping disorder, $W/t = 0.05$; 
(d) hopping disorder, $W/t = 1.00$
}
\end{figure}
\begin{figure}
$
\begin{array}{cc}
\hspA(a)	& 
\hspB(b)\vspace{-0.5cm}  	 \\

\hspA	\includegraphics[angle=-90,width=5cm]{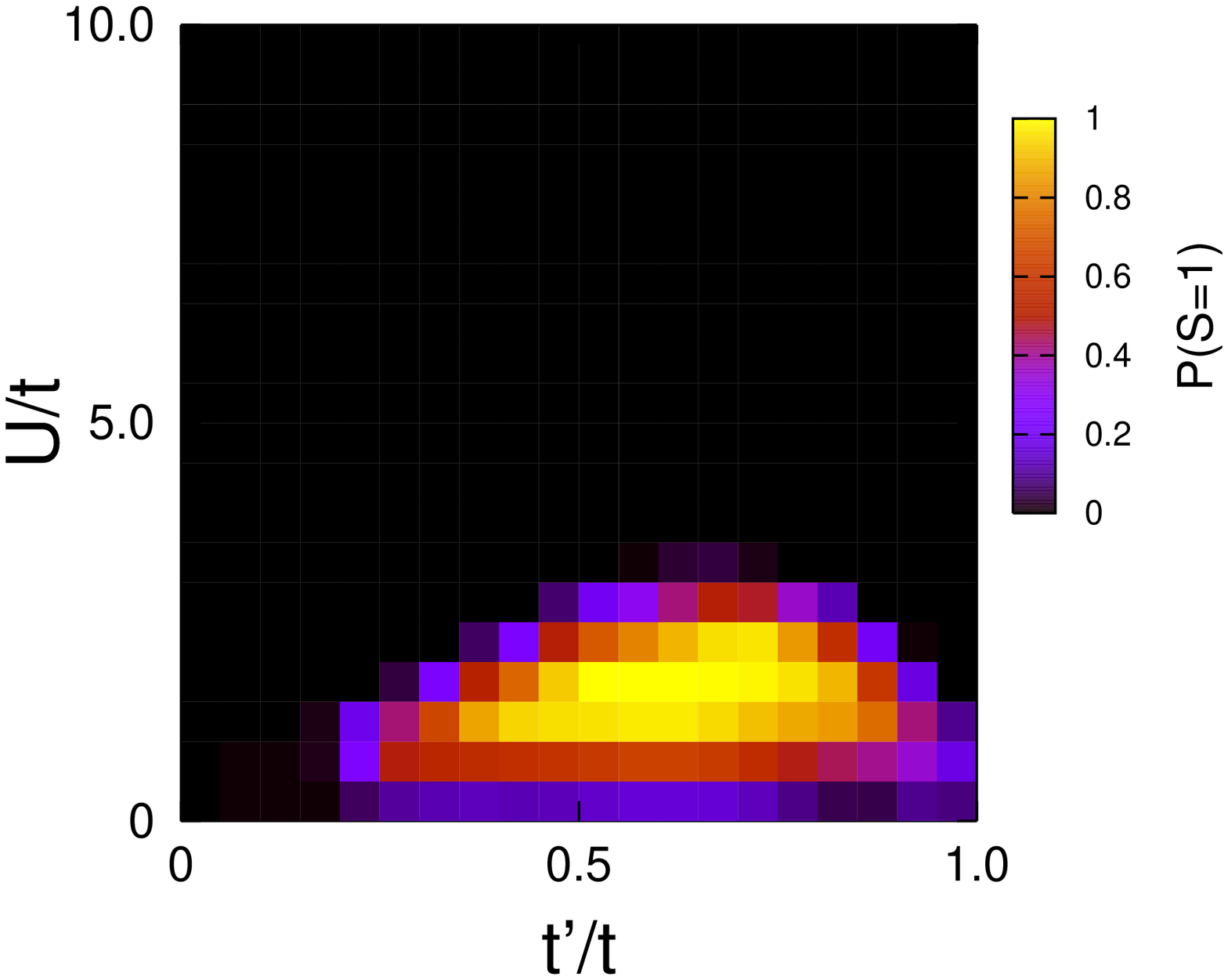} 	& 
\hspB  	\includegraphics[angle=-90,width=5cm]{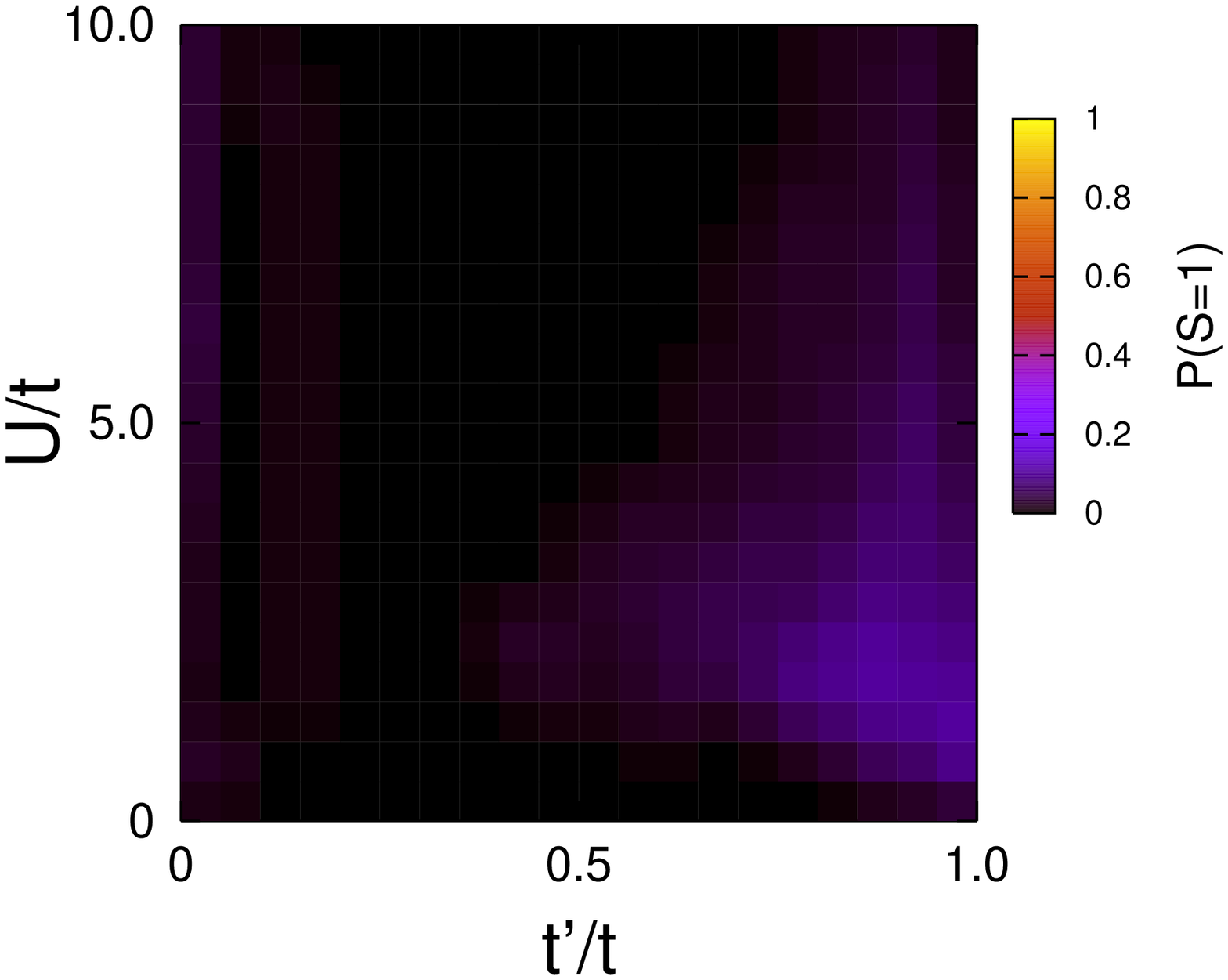}\\

\hspA(c)	& \hspB(d)\vspace{-0.5cm}  	 \\

\hspA	\includegraphics[angle=-90,width=5cm]{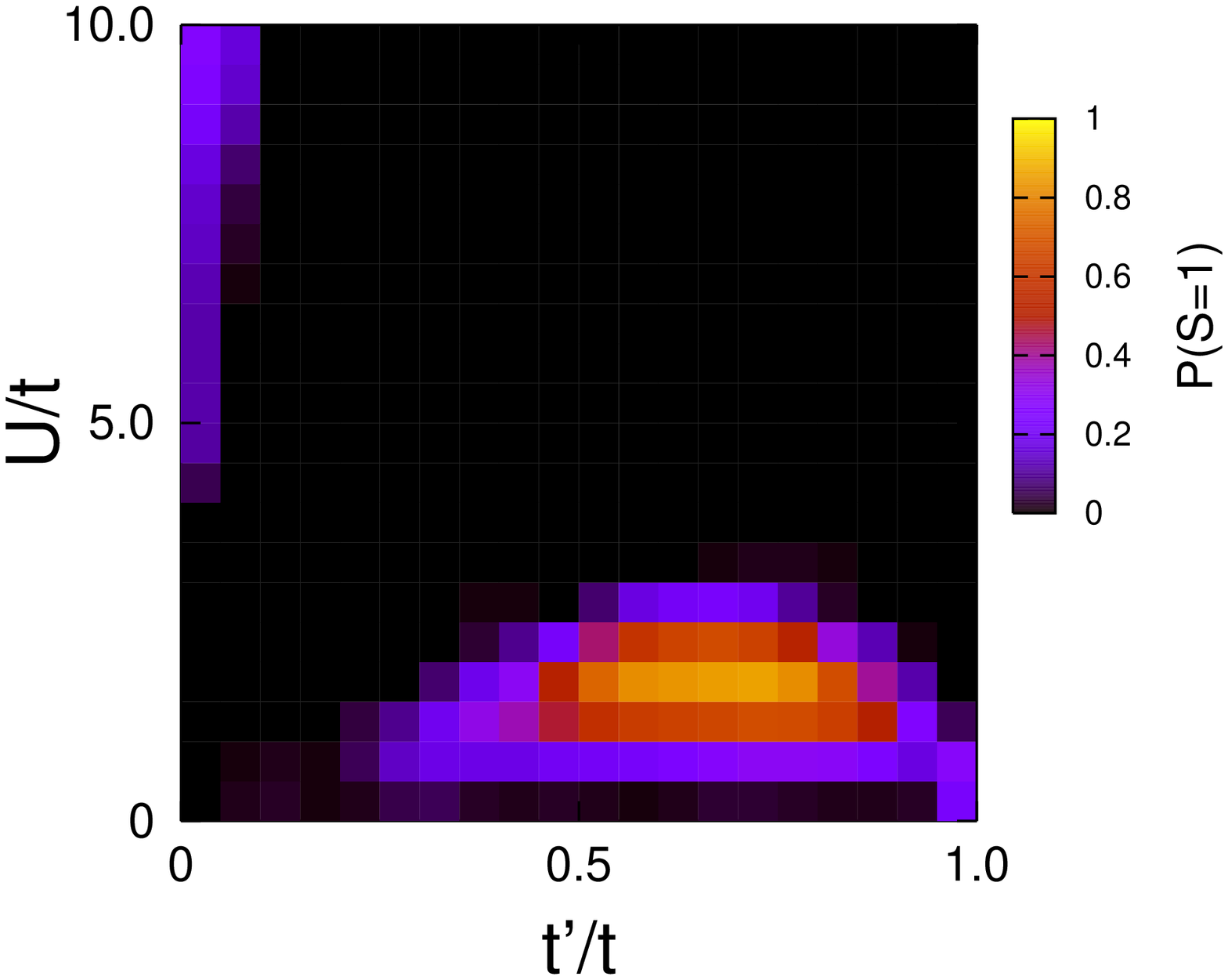} &
\hspB	\includegraphics[angle=-90,width=5cm]{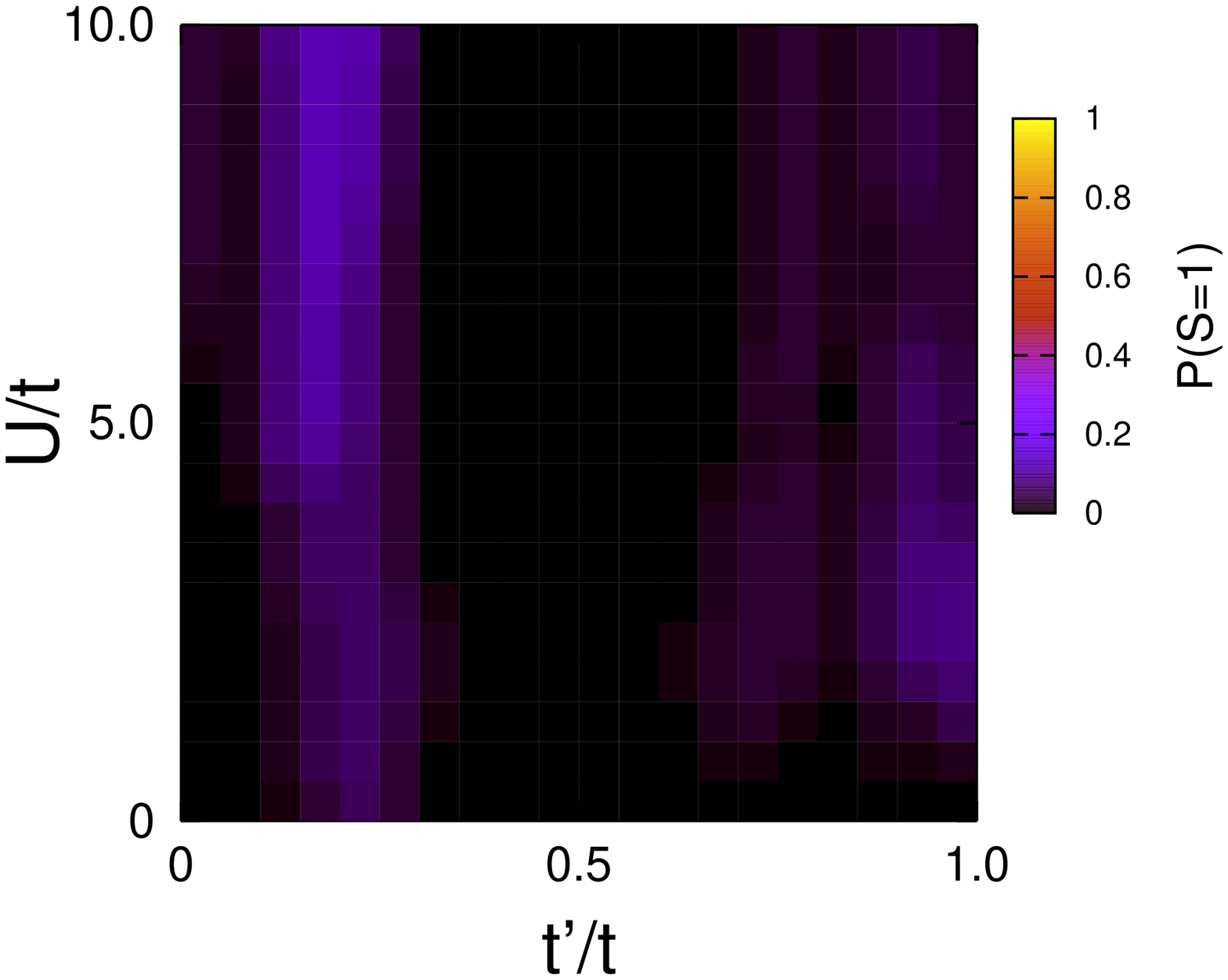}
\end{array}
$
\caption{
\label{fig:PS1OnsiteN12} 
Probability of observing a $S=1$ ground state in the $m=2$-hole doped system, $\PSS$, for the twelve-site ladder cluster at doping $x=1/12$: 
(a) potential disorder, $W/t = 0.05$; 
(b) potential disorder, $W/t = 1.00$; 
(c) hopping disorder, $W/t = 0.05$; 
(d) hopping disorder, $W/t = 1.00$
}
\end{figure}

We also find that disorder affects the spin eigenvalue of the ground state.  
In Figs.~\ref{fig:PS1OnsiteN08} and \ref{fig:PS1OnsiteN12} we plot the probability of finding 
$S=1$ eigenvalues in the $m=2$-hole ground state, $P(S=1)$ for the eight- and twelve-site ladder clusters. 
We find that the ground state of the eight-site ladder always has $S=0$ for weak disorder. 
However, at $W/t=1$, $P(S=1)$ appears to be maximum at $t'/t=1$ and $U/t\approx2$, which corresponds 
to the maximum of the PBE of the uniform, homogeneous eight-site ladder cluster. In the twelve-site ladder 
cluster, there is a region with $0.4 \lesssim t^\prime/t \lesssim 0.8$ for $U/t<3$ where the ground state 
is mainly $S=1$ even for weak disorder. As disorder strength increases, the value of 
$\PSS$ decreases in this region while simultaneously increasing around $t'/t\approx0$ and $t'/t\approx1$. 
At $W/t=1$, $\PSS=0$ for intermediate values of $t'/t$ and $U/t>5$ for both the eight- and twelve-site ladder clusters. 

For both eight and twelve site clusters, the region in parameter space where $P(S=1)>0$ 
is larger for hopping disorder than for potential disorder as disorder strength increases.  
In both cases, at weak disorder the largest values of $P(S=1)$ are found for intermediate
$t^\prime/t$ and $U/t \lesssim 3$.  From the definition of the $d$-wave order 
parameter, when the total spin of the ground state is $S=1$, $\Psi_{d}=0$. We expect 
$\Dd$ to be most robust against potential disorder in the regions where $\PSS=0$ up to $W/t=1$.

\subsection{Spin gap}
\label{Sec:SpinGap}

Karakonstantakis \etal.\cite{Karakonstantakis} calculated the spin gap and the PBE in the ladder CHM using 
density matrix renormalization group (DMRG) methods and argued that there is an optimal inhomogeneity that 
leads to maximal values of $\Delta_{p}$ and $\Delta_{s}$.  Similarly to Ref.~\onlinecite{Karakonstantakis}, we calculate the disorder averaged spin gap, $\Ds$, as the average gap between the $S=0$ state and the lowest-energy $S=1$ state in the $m=2$-hole doped system. The spin gap, like the pair binding energy, may be interpreted as a measure of the pairing scale of the system.\cite{Maier3}

In order to evaluate the spin gap, we calculate the eigenvalues and associated eigenvectors of the ground state and the first few 
excited states, determine the $S^{2}$ eigenvalue of each state, isolate the lowest-lying $S=1$ eigenstate, and then 
calculate $E_{0}(S=1)-E_{0}(S=0)$. There also exist several low-lying states with $S=0$ near the $S=1$ state, and 
for large enough $t'/t$, we see crossings between the lowest lying $S=1$ state and $S=0$ excited states. We plot $\Ds$ 
as a function of $t'/t$ and $U/t$ at fixed $W/t$ for the eight-site ladder, but focus on  $U/t=8.0$ for the 
twelve-site ladder. The reason for the latter situation is that there are multiple level crossings between 
excited states in the twelve-site model, which makes it difficult to survey numerically the spectrum of excited 
states for many values of $t'/t$ and $U/t$. We also calculate the disorder averaged gap between the ground 
state and the first excited $S=0$ state, $\Dex$, to get a better sense of how the lowest-lying $S=0$ 
states behave as $t'/t$ is varied. The value of $\Dex$ is calculated only for those configurations where the ground state has $S=0$. 

\begin{figure}[htb]
$
\begin{array}{cc}
\hspA(a)	& 
\hspB(b)\vspace{-0.5cm}  	 \\
\hspA\includegraphics[angle=-90,width=5cm]{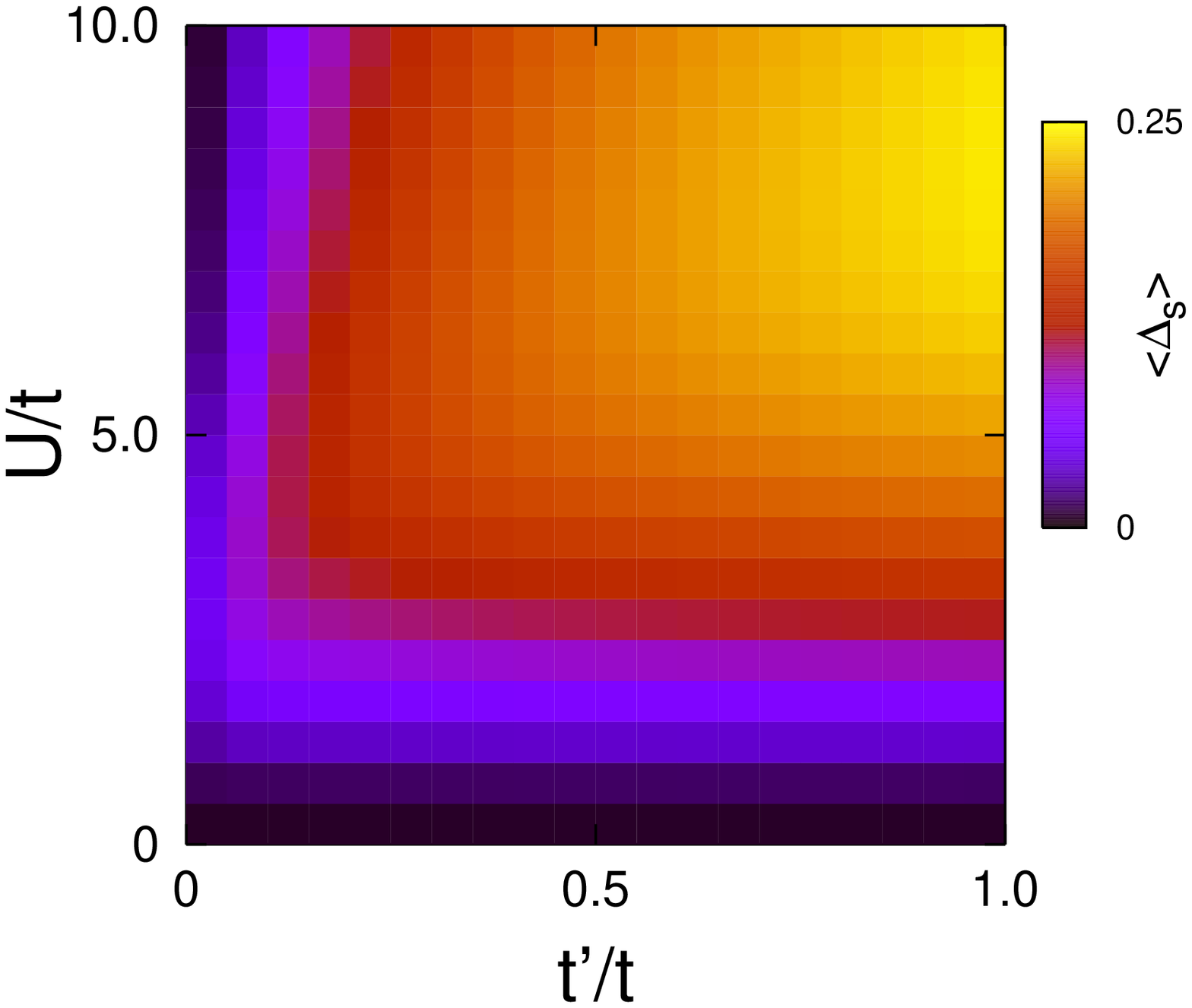}	&
\hspB\includegraphics[angle=-90,width=5cm]{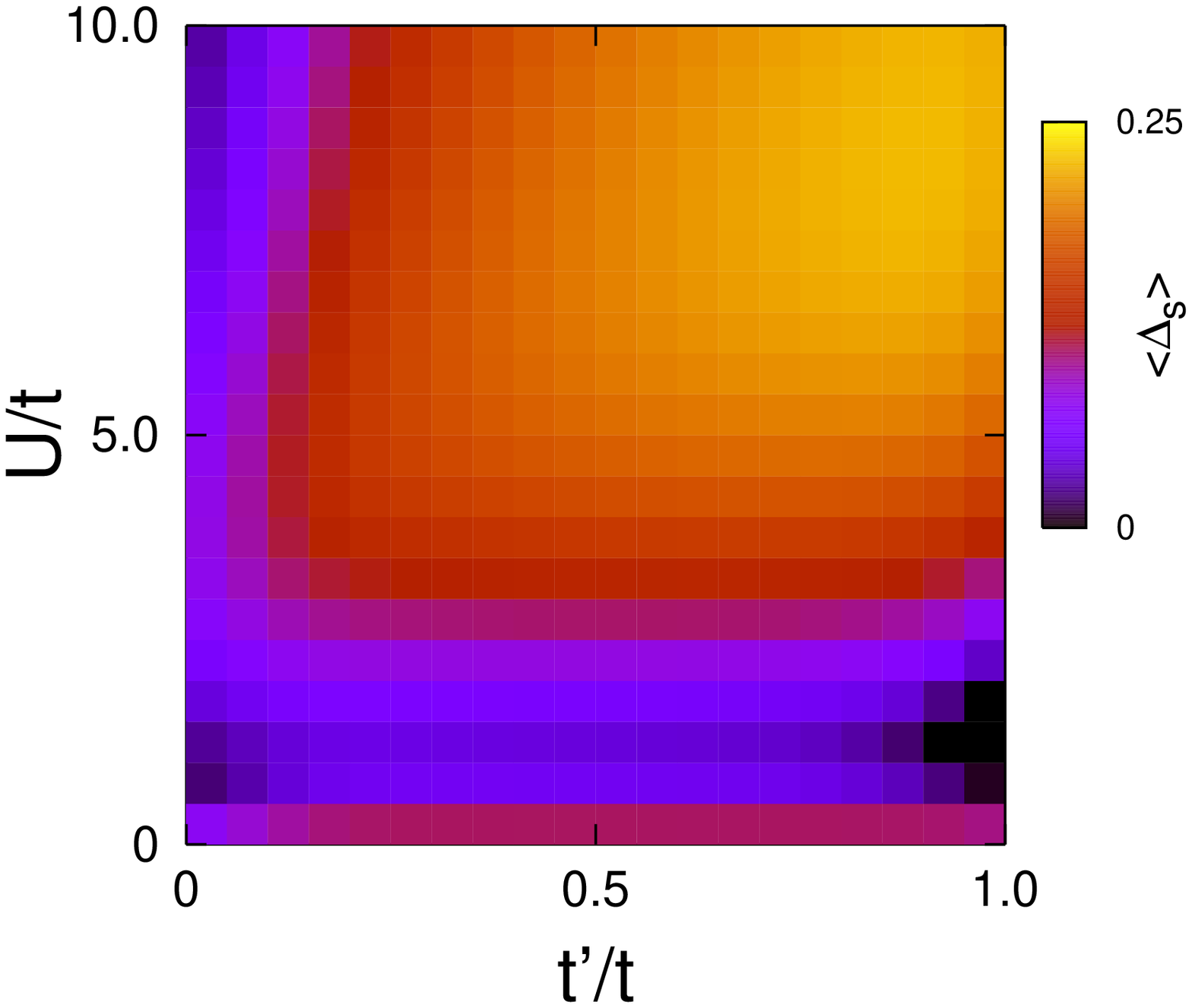}
\end{array}
$
\caption{
\label{fig:SG1N08ORL}
Disorder averaged spin gap, $\Ds$, as a function of $t'/t$ and $U/t$  at doping $x=1/8$ on the eight-site ladder with potential disorder: (a) $W/t = 0.05$ and (b) $W/t = 1.00$.
}
\end{figure}

\begin{figure}[htb]
$
\begin{array}{cc}
\hspA(a)	& 
\hspB(b)\vspace{-0.5cm}  	 \\
\hspA\includegraphics[angle=-90,width=5cm]{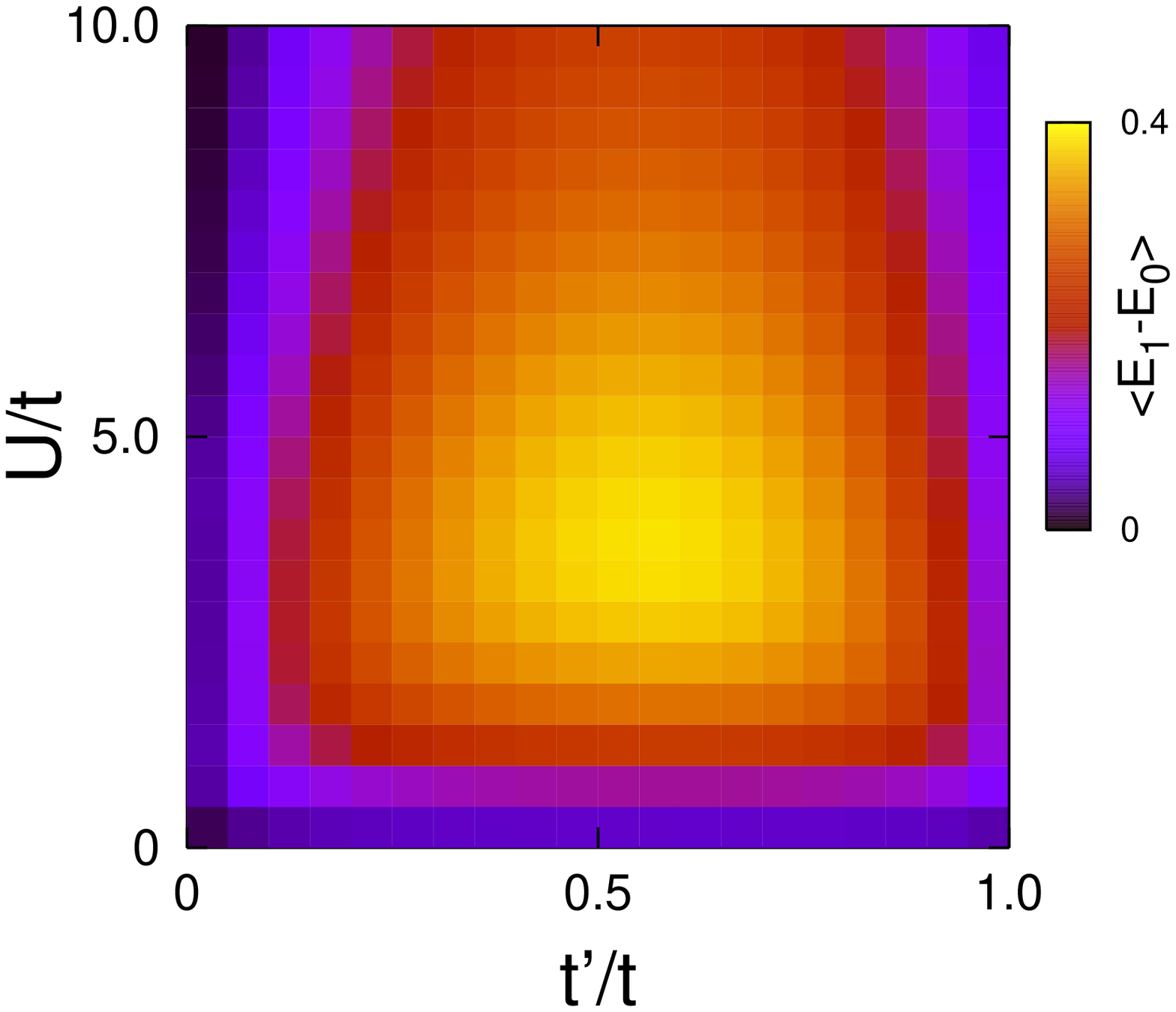}	&
\hspB\includegraphics[angle=-90,width=5cm]{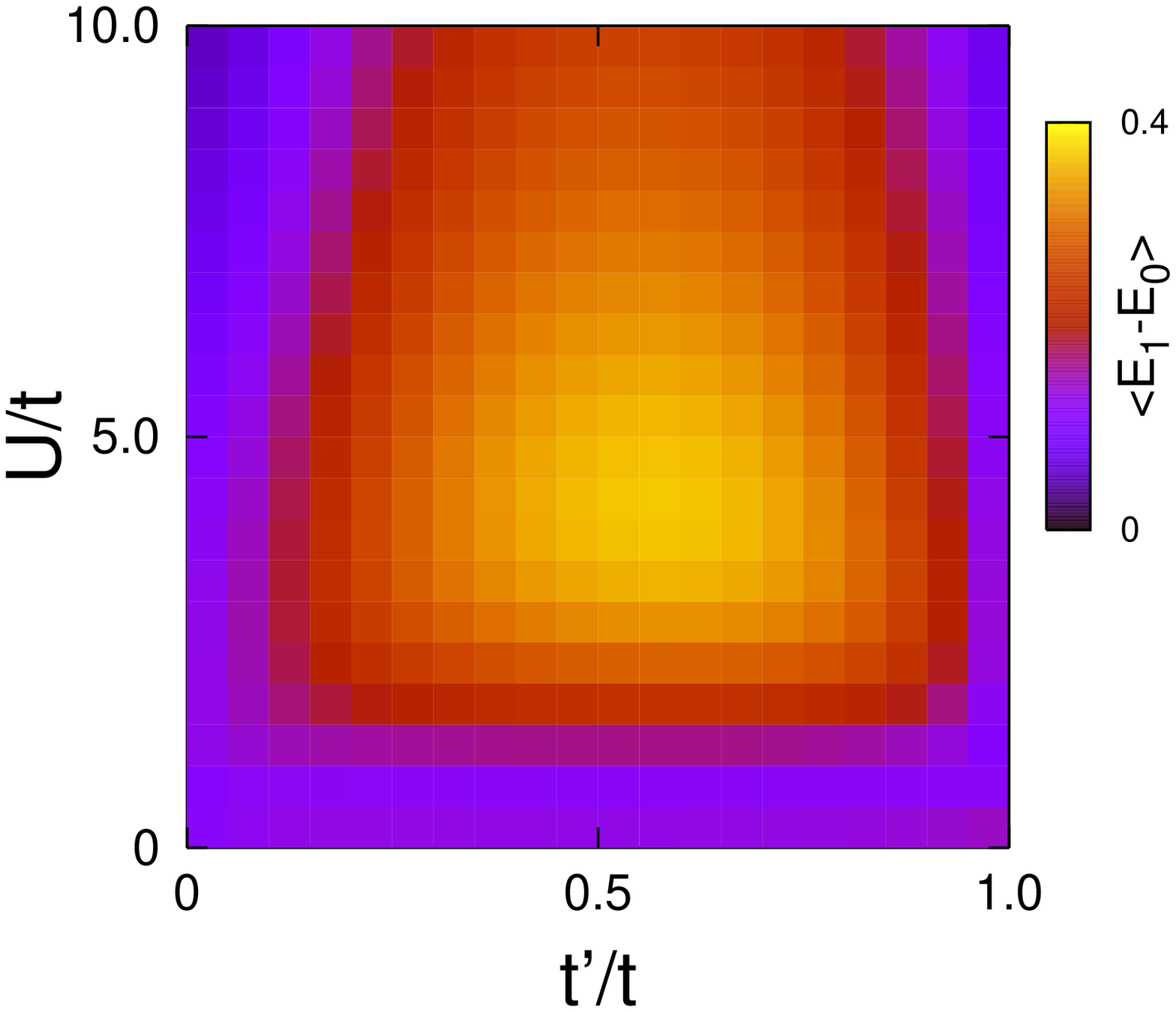}
\end{array}
$
\caption{
\label{fig:SG2N08ORL}
Disorder-averaged gap to the lowest energy $S=0$ excited state, $\Dex$, as a function of $t'/t$ and $U/t$ at doping $x=1/8$  on the eight-site ladder with potential disorder: (a) $W/t = 0.05$ and (b) $W/t = 1.00$.
}
\end{figure}

\begin{figure}[htb]
$
\begin{array}{cc}
\hspA(a)	& 
\hspB(b)\vspace{-0.5cm}  	 \\
\hspA\includegraphics[angle=-90,width=5cm]{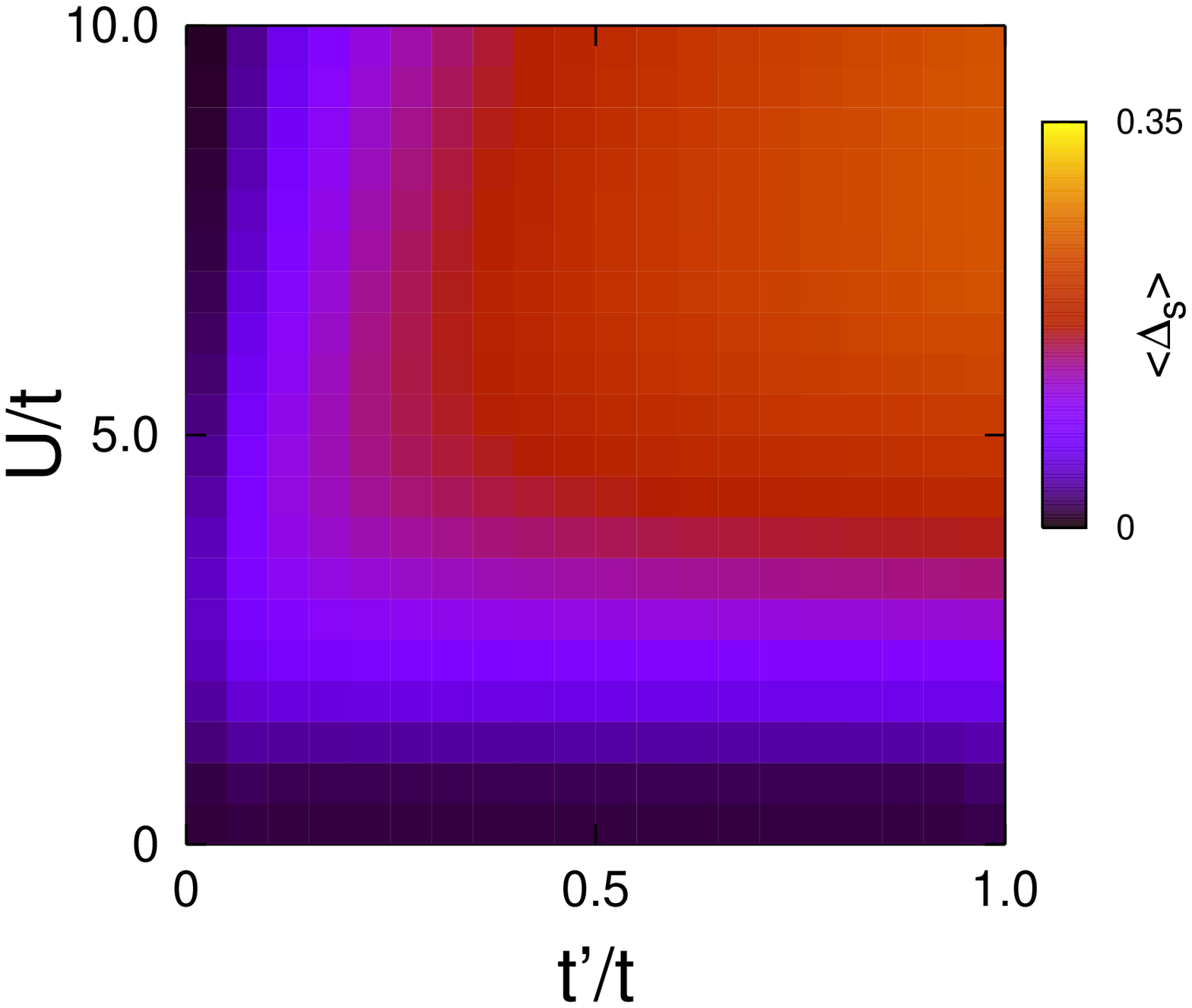}	&
\hspB\includegraphics[angle=-90,width=5cm]{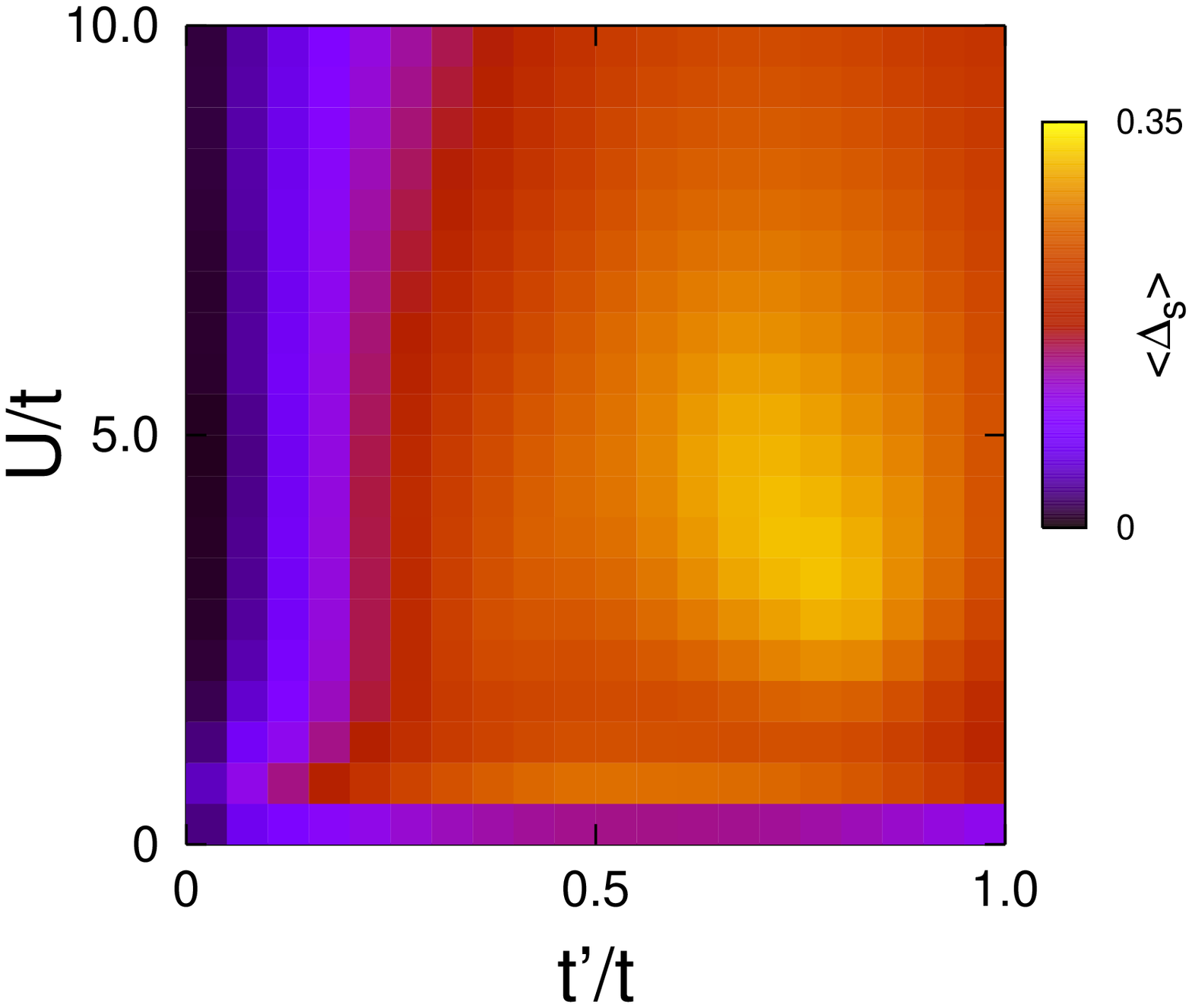}
\end{array}
$
\caption{
\label{fig:SG1N08BRL}
Disorder averaged spin gap, $\Ds$, as a function of $t'/t$ and $U/t$  at doping $x=1/8$ on the eight-site ladder with hopping disorder: (a) $W/t = 0.05$ and (b) $W/t = 1.00$.
}
\end{figure}

\begin{figure}[htb]
$
\begin{array}{cc}
\hspA(a)	& 
\hspB(b)\vspace{-0.5cm}  	 \\
\hspA\includegraphics[angle=-90,width=5cm]{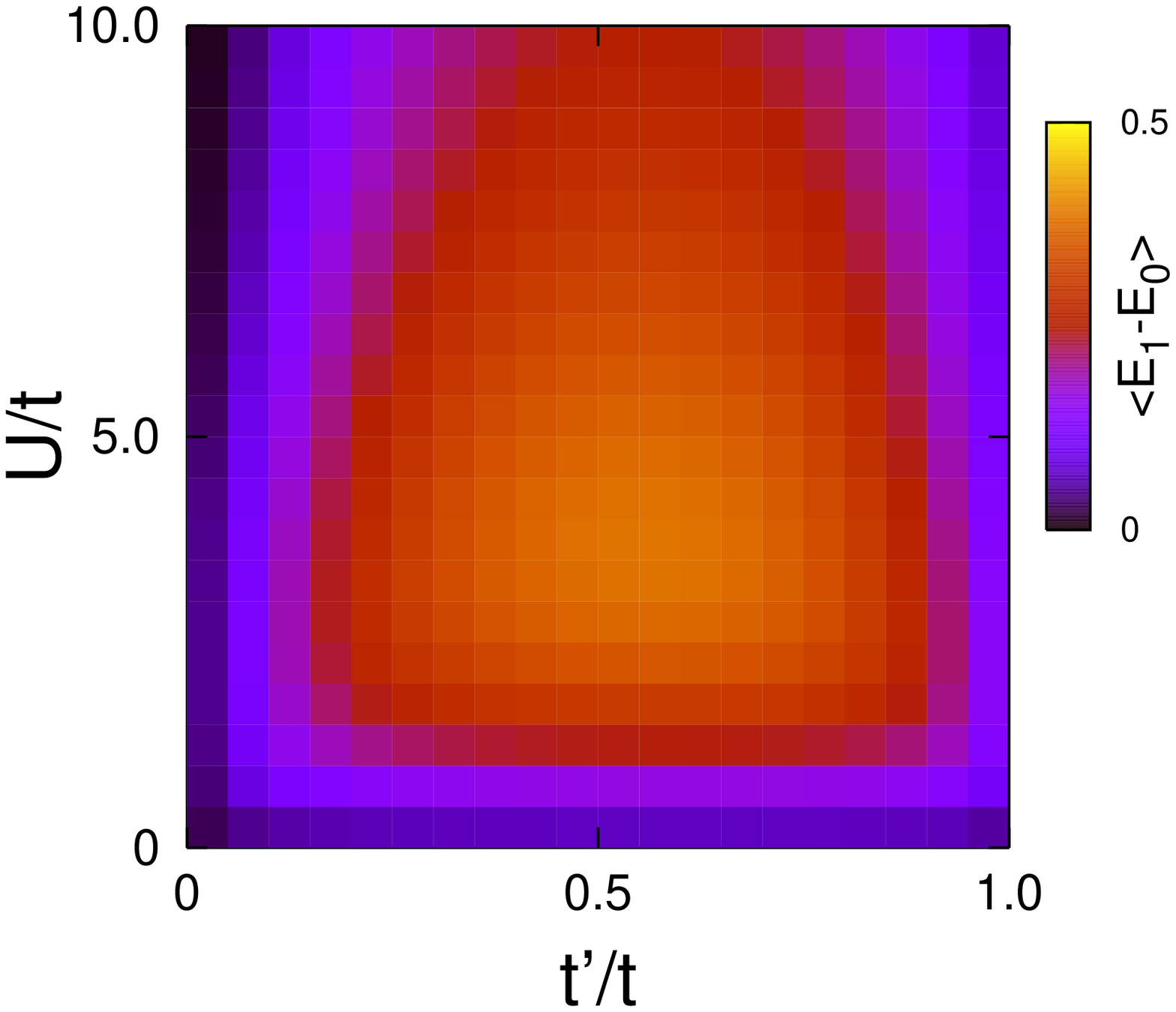}	&
\hspB\includegraphics[angle=-90,width=5cm]{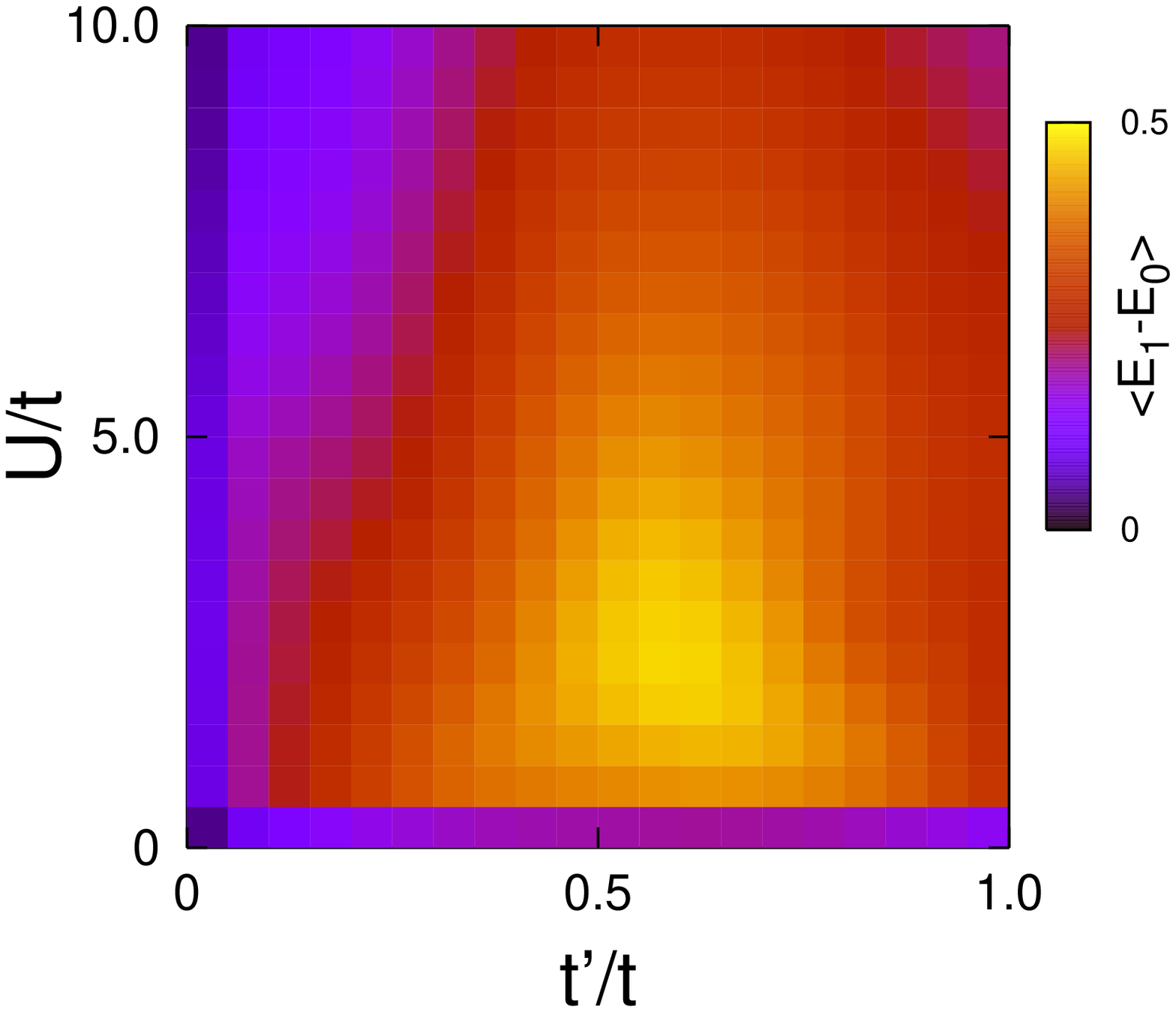}
\end{array}
$
\caption{
\label{fig:SG2N08BRL}
Disorder-averaged gap to the lowest energy $S=0$ excited state, $\Dex$, as a function of $t'/t$ and $U/t$ at doping $x=1/8$ for the eight-site ladder with hopping disorder: (a) $W/t = 0.05$ and (b) $W/t = 1.00$.
}
\end{figure}

\begin{figure}[htb]
$
\begin{array}{cc}
(a)	&  \hspA(b)\vspace{-0.125cm}  	 \\
\includegraphics[angle=-90,width=5cm]{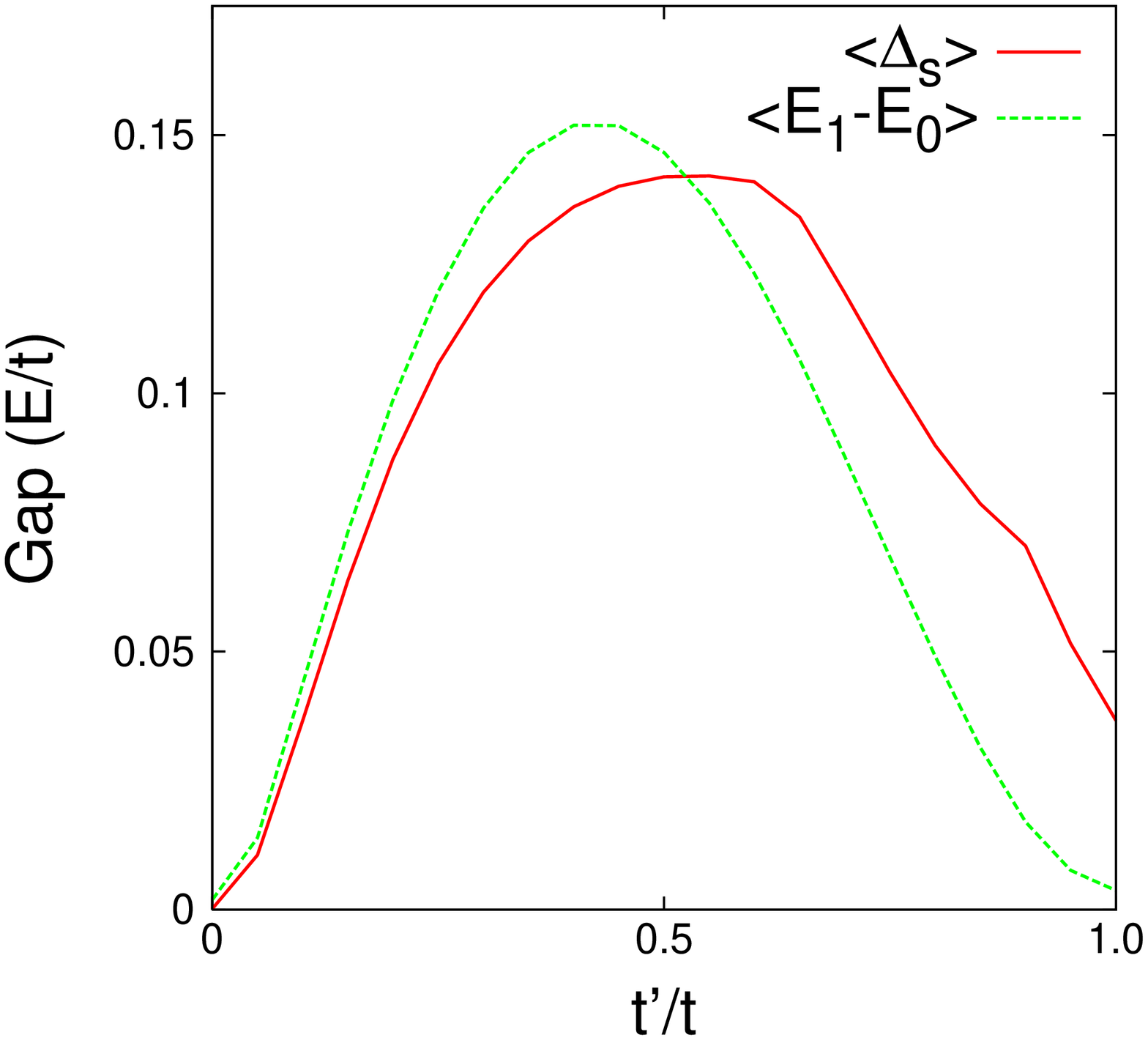}	&
\hspA\includegraphics[angle=-90,width=5cm]{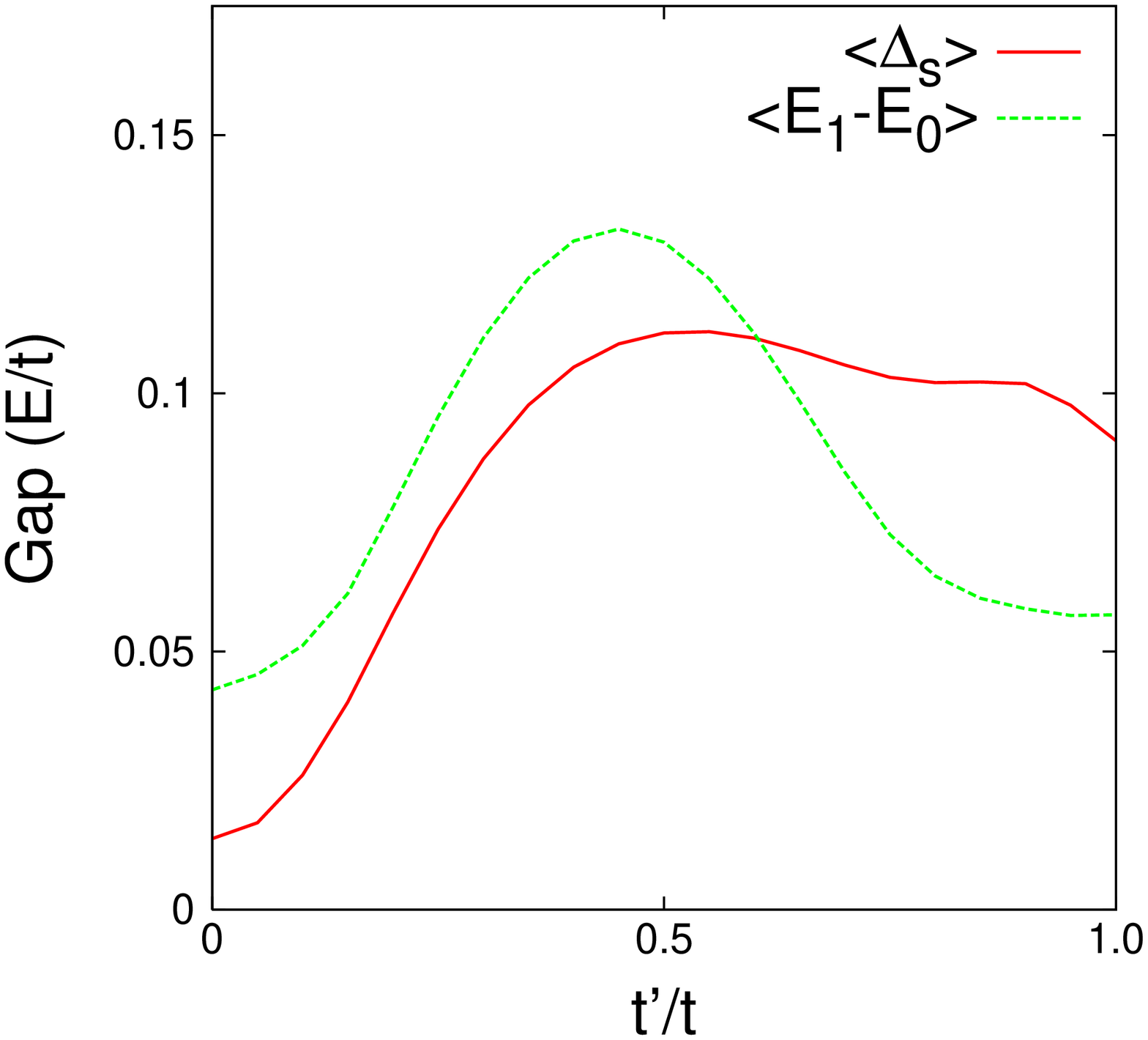}
\end{array}
$
\caption{
\label{fig:SGN12ORL}
Disorder-averaged spin gap, $\Ds$, and disorder-averaged gap to the lowest energy $S=0$ excited state, $\Dex$, as a function of $t'/t$ 
for $U/t=8$ and doping $x=1/12$ for the twelve-site ladder with potential disorder: (a) $W/t = 0.05$ and (b) $W/t = 1.00$.
}
\end{figure}

\begin{figure}[htb]
$
\begin{array}{cc}
(a)	& 
\hspA(b)\vspace{-0.125cm}  	 \\
\includegraphics[angle=-90,width=5cm]{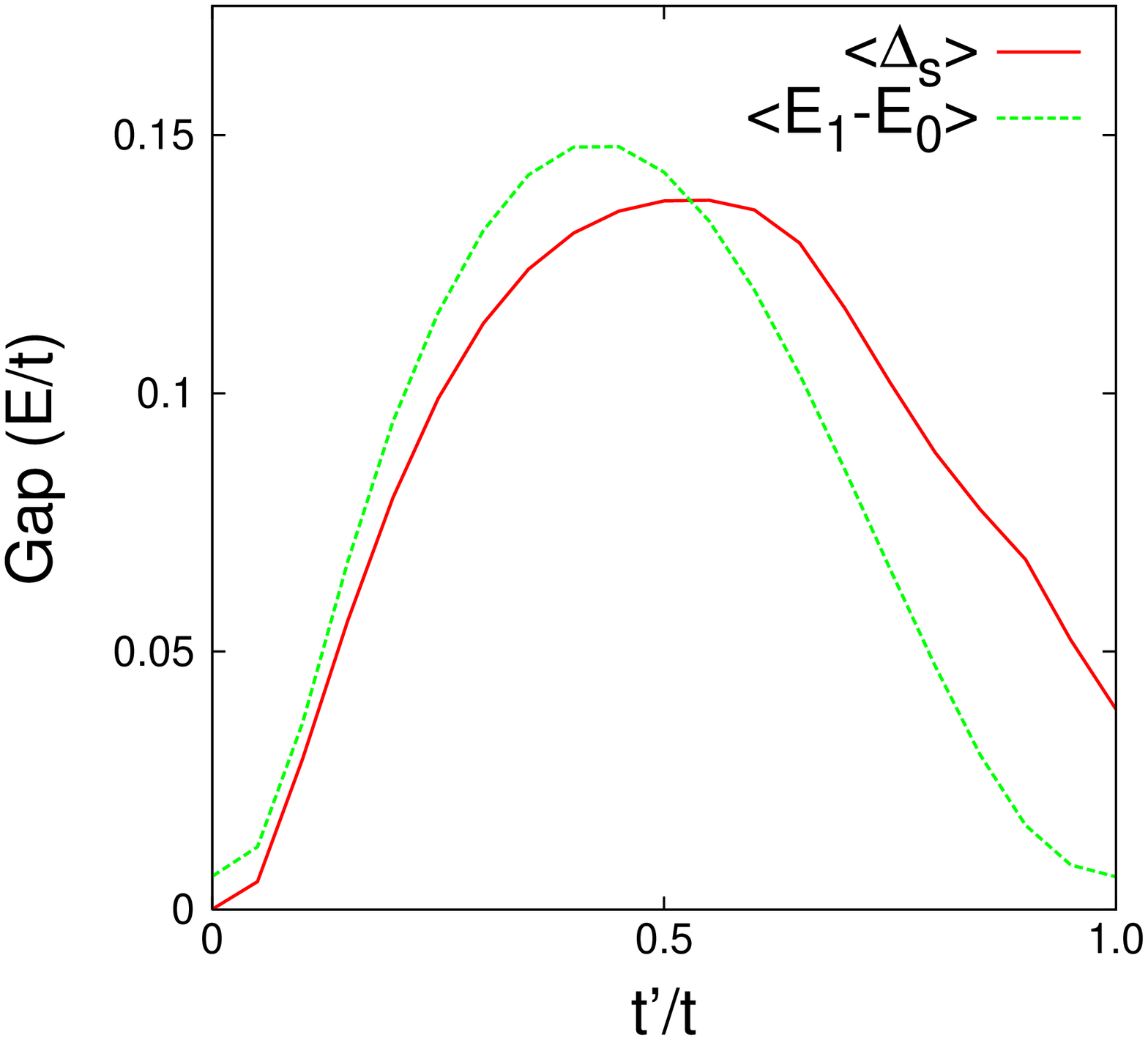}	&
\hspA\includegraphics[angle=-90,width=5cm]{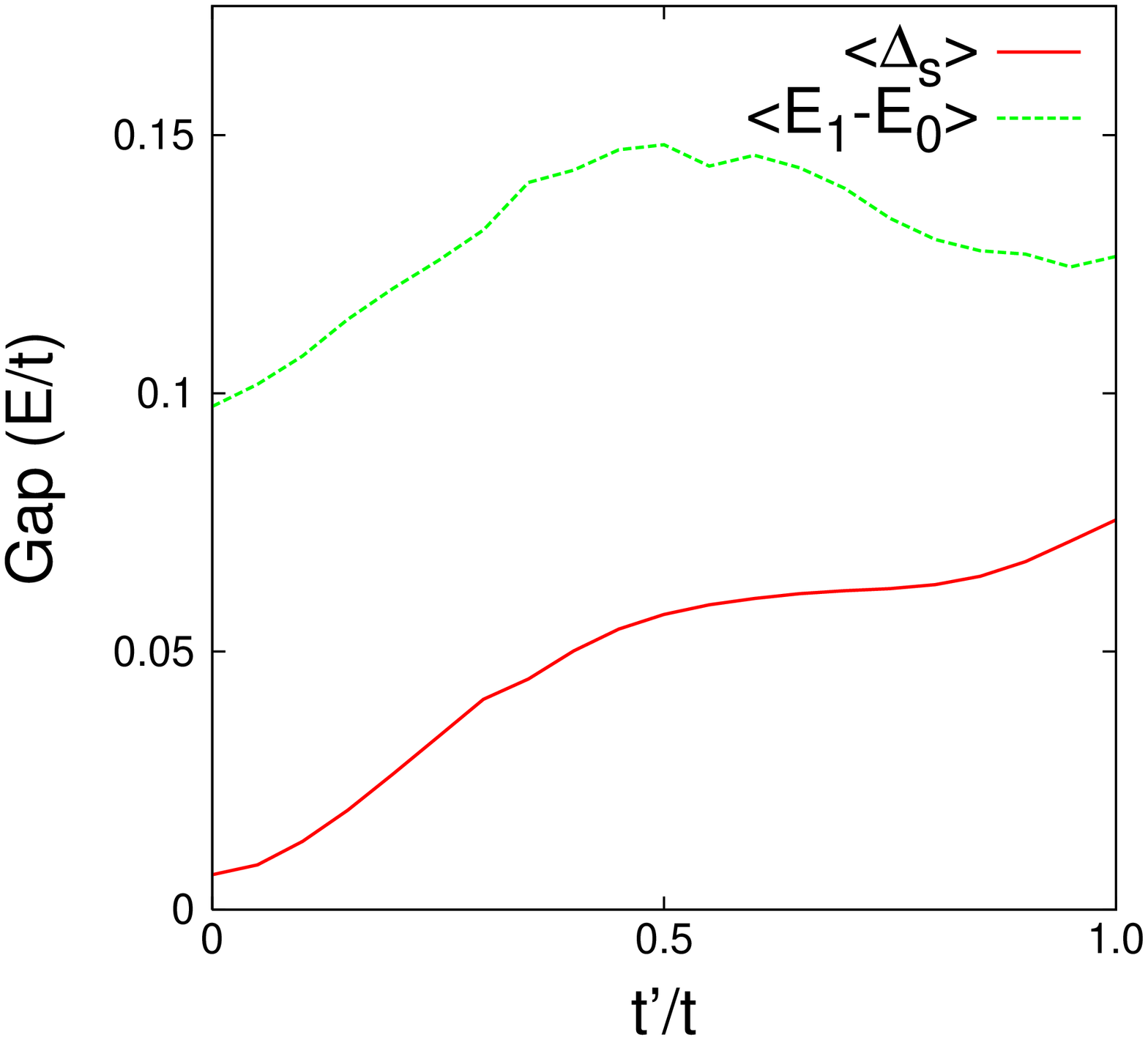}
\end{array}
$
\caption{
\label{fig:SGN12BRL}
Plots of the disorder-averaged spin gap, $\Ds$, and disorder-averaged gap to the lowest energy $S=0$ excited state, $\Dex$, as a function of $t'/t$ 
for $U/t=8$ and doping $x=1/12$ for the twelve-site ladder with hopping disorder: (a) $W/t = 0.05$ and (b) $W/t = 1.00$.
}
\end{figure}

We plot $\Ds$ and $\Dex$ for the eight- and twelve-site ladders in 
Figs.~\ref{fig:SG1N08ORL}-\ref{fig:SGN12BRL}. For eight-site clusters with weak disorder and 
moderate interactions, the spin gap increases monotonically as $t'/t\to 1$. As disorder increases, 
the gap appears to soften slightly as $t'/t\to 1$, which leads to the optimal $\Ds$ occurring for $t'/t<1$, 
albeit with different $U/t$ and $t'/t$ to where pair binding is favoured. On the other hand, the results 
for $\Dex$ for eight-site clusters show that this quantity is maximized for intermediate values of inhomogeneity. 
Also of interest is that for $t'/t\gtrsim0.7$, $\Dex$ $<\Ds$, indicating that the lowest lying excitations  
are $S=0$ states. 
However, there is a maximum in $\Dex$ at $t'/t\approx0.5$ for which $\Dex>\Ds$. As $t'/t\to1$, we find $\Dex<\Ds$.

At weak disorder in eight-site clusters, the results for $\Dp$ and $\Dex$ show little qualitative 
difference between potential and hopping disorder. However, we see a clear difference between both types of disorder at large $W/t$.
As illustrated in Figs.~\ref{fig:SG1N08BRL} and \ref{fig:SG2N08BRL}, for hopping disorder, $\Ds$ and $\Dex$ 
grow with increasing $W/t$, whereas for potential disorder large values of $W/t$ appear to suppress these quantities slightly. 

For twelve-site ladder clusters with $U/t=8$, $\Ds$ appears to be maximized at $t'/t\approx0.5$, while $\Dex$ appears 
to be maximized at $t'/t\approx0.4$. A comparison between this result to the data for the spin gap calculated 
on eight-site clusters, which show no signs of ``optimization'' for intermediate $t'/t$, suggests 
that  doping effects are significant when calculating the spin gap. However, similarly to the eight-site clusters, 
we see a crossover in the gaps as a function of intraplaquette hopping around $t'/t\approx0.6$ for weak disorder. Depending on disorder 
strength and configuration, there may be several $S=0$ states with lower energy than the lowest $S=1$. Although this may be an 
artifact of small system size and/or geometry, our results still suggest that the spin gap is less than the gap to 
other low-lying $S=0$ states for $t'/t<0.5$; above this value, other low-lying states may lie below the spin gap. 
Similarly to the eight-site model, it appears that $\Ds$ is affected more by hopping disorder tahn by potential disorder, 
as can be seen in Figs.~\ref{fig:SG1N08BRL} and \ref{fig:SGN12BRL}.

\subsection{d-wave Pairing}
\label{Sec:Dwave}
\begin{figure}[htb]
$
\begin{array}{cc}
\hspA(a)	& 
\hspB(b)\vspace{-0.5cm}  	 \\
\hspA\includegraphics[angle=-90,width=5cm]{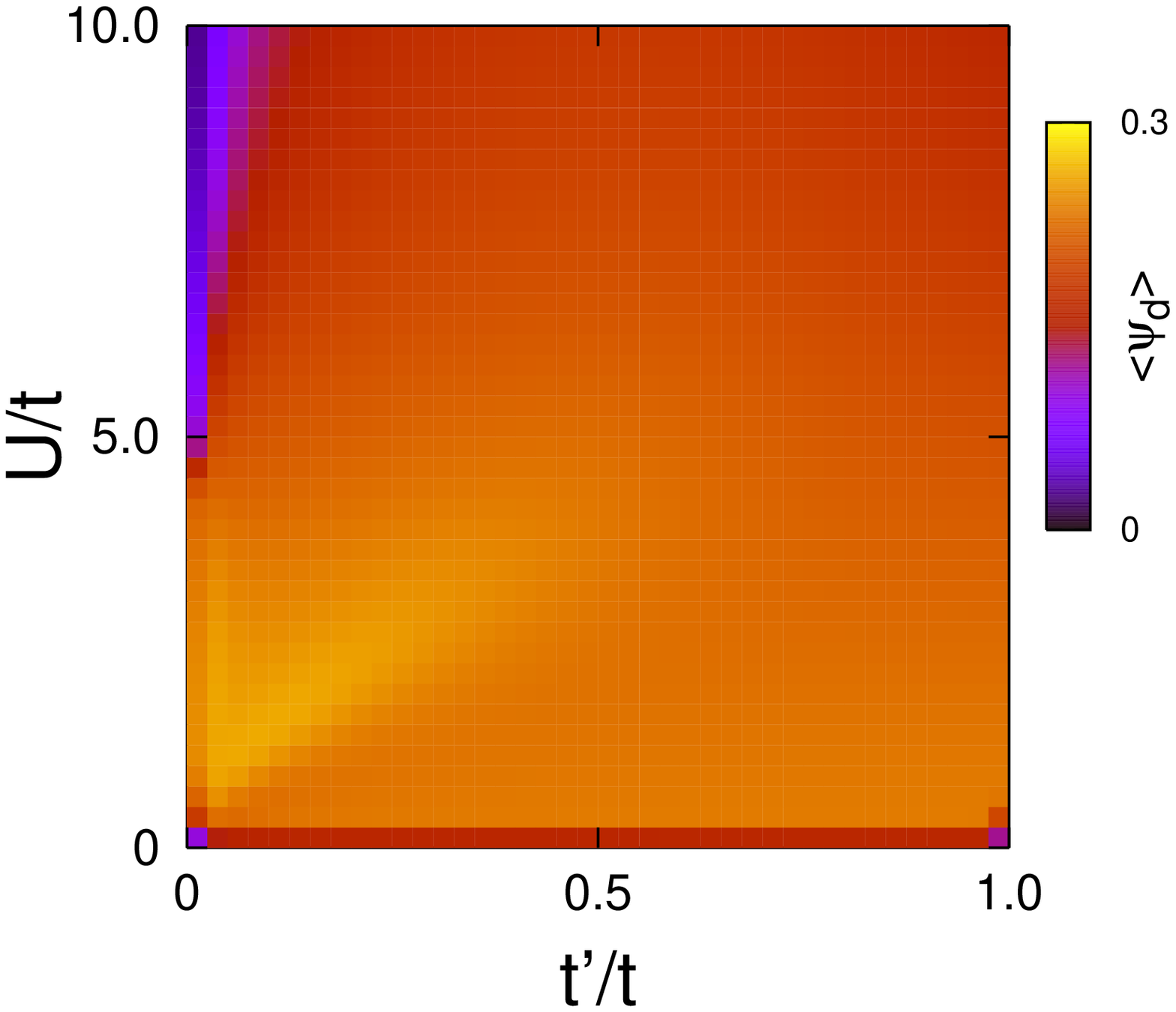}	&
\hspB\includegraphics[angle=-90,width=5cm]{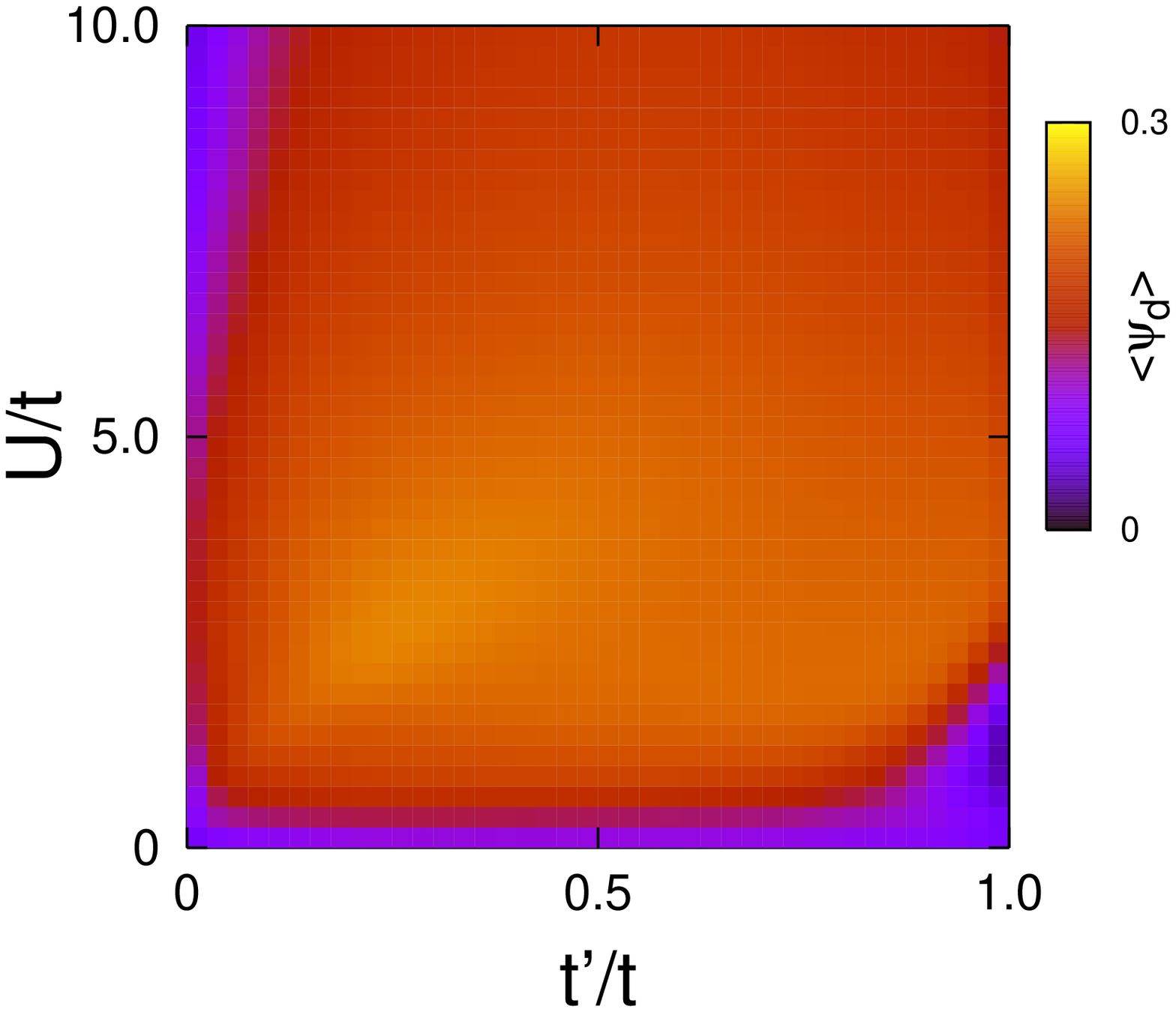}
\end{array}
$
\caption{
\label{fig:DWN08ORL}
Disorder-averaged $d$-wave pairing order parameter, $\Dd$, on the eight-site ladder with potential disorder: (a) $W/t = 0.05$ and (b) $W/t = 1.00$.
}
\end{figure}

\begin{figure}[htb]
$
\begin{array}{cc}
\hspA(a)	& 
\hspB(b)\vspace{-0.5cm}  	 \\
\hspA\includegraphics[angle=-90,width=5cm]{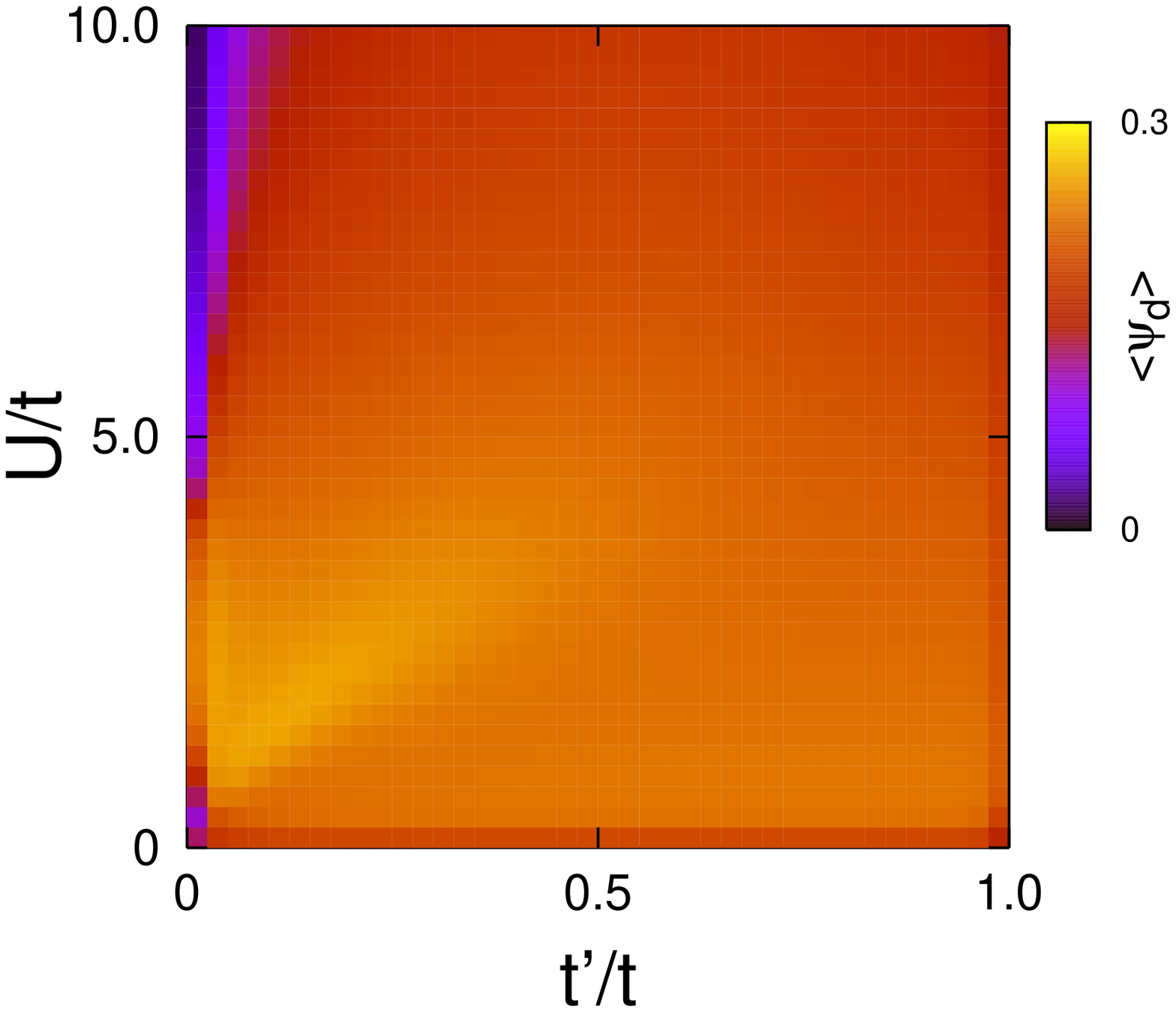}	&
\hspB\includegraphics[angle=-90,width=5cm]{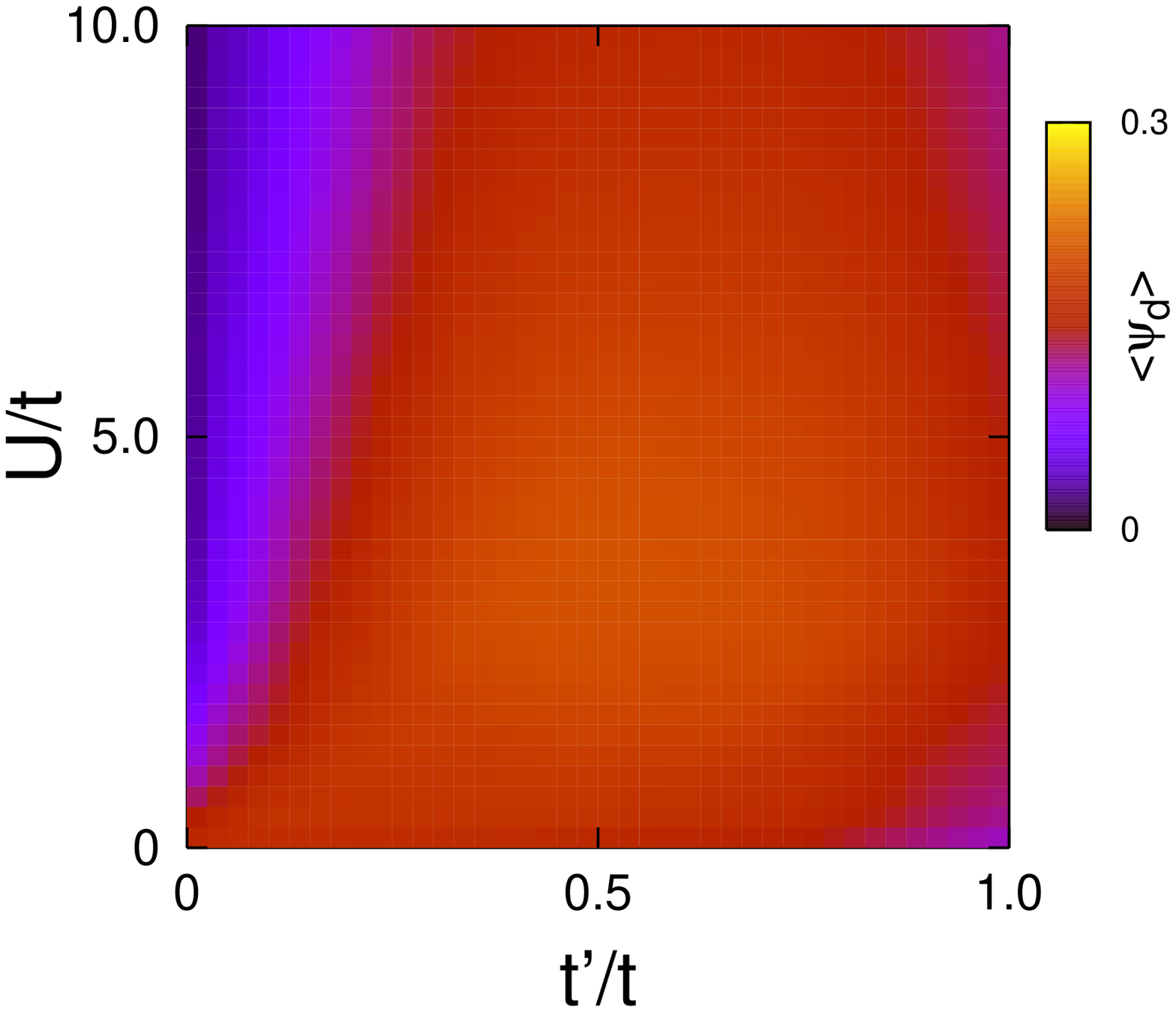}
\end{array}
$
\caption{
\label{fig:DWN08BRL}
Disorder-averaged $d$-wave pairing order parameter, $\Dd$, on the eight-site ladder with hopping disorder: (a) $W/t = 0.05$ and (b) $W/t = 1.00$.
}
\end{figure}

\begin{figure}[htb]
$
\begin{array}{cc}
\hspA(a)	& 
\hspB(b)\vspace{-0.5cm}  	 \\
\hspA\includegraphics[angle=-90,width=5cm]{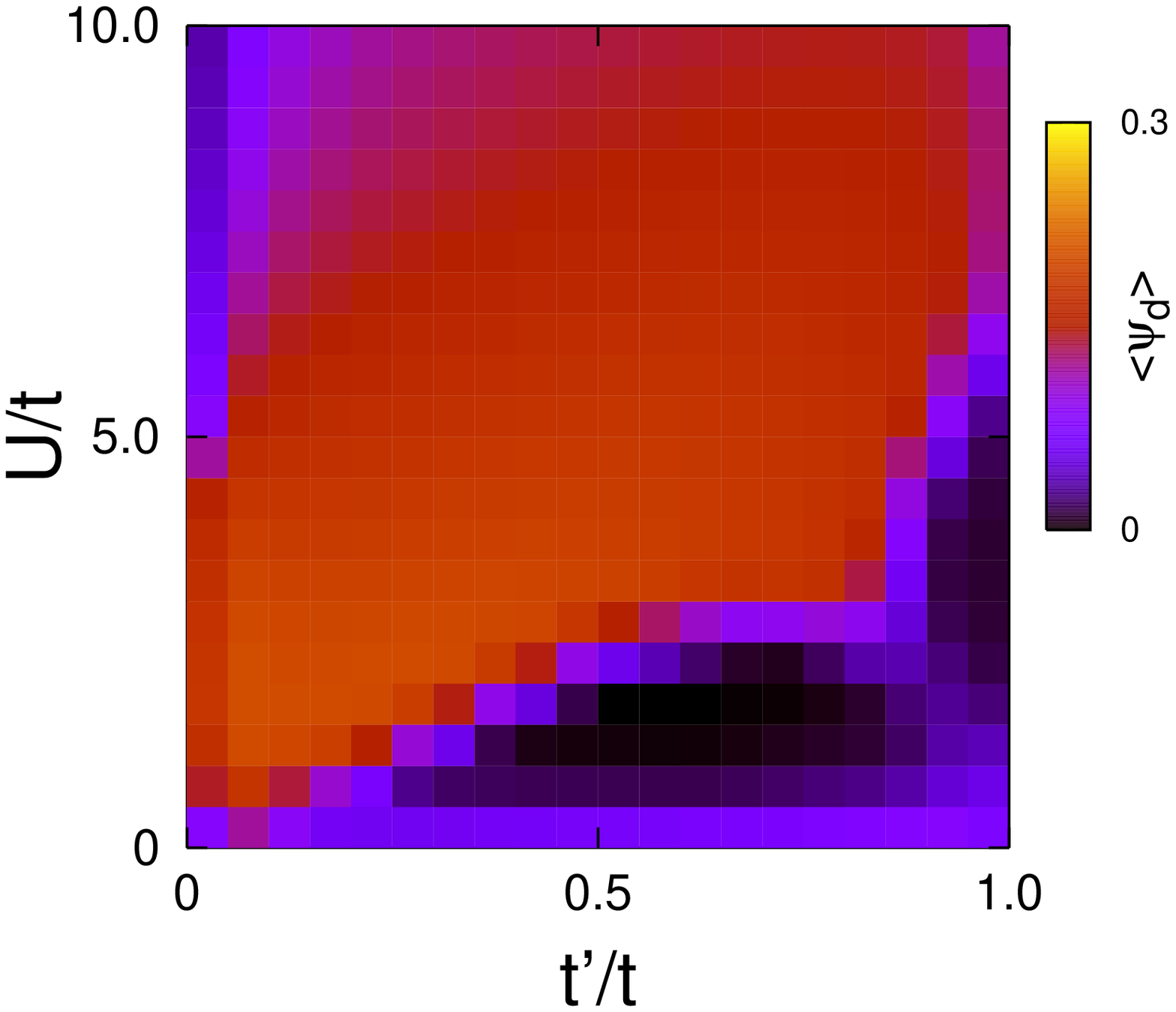}	&
\hspB\includegraphics[angle=-90,width=5cm]{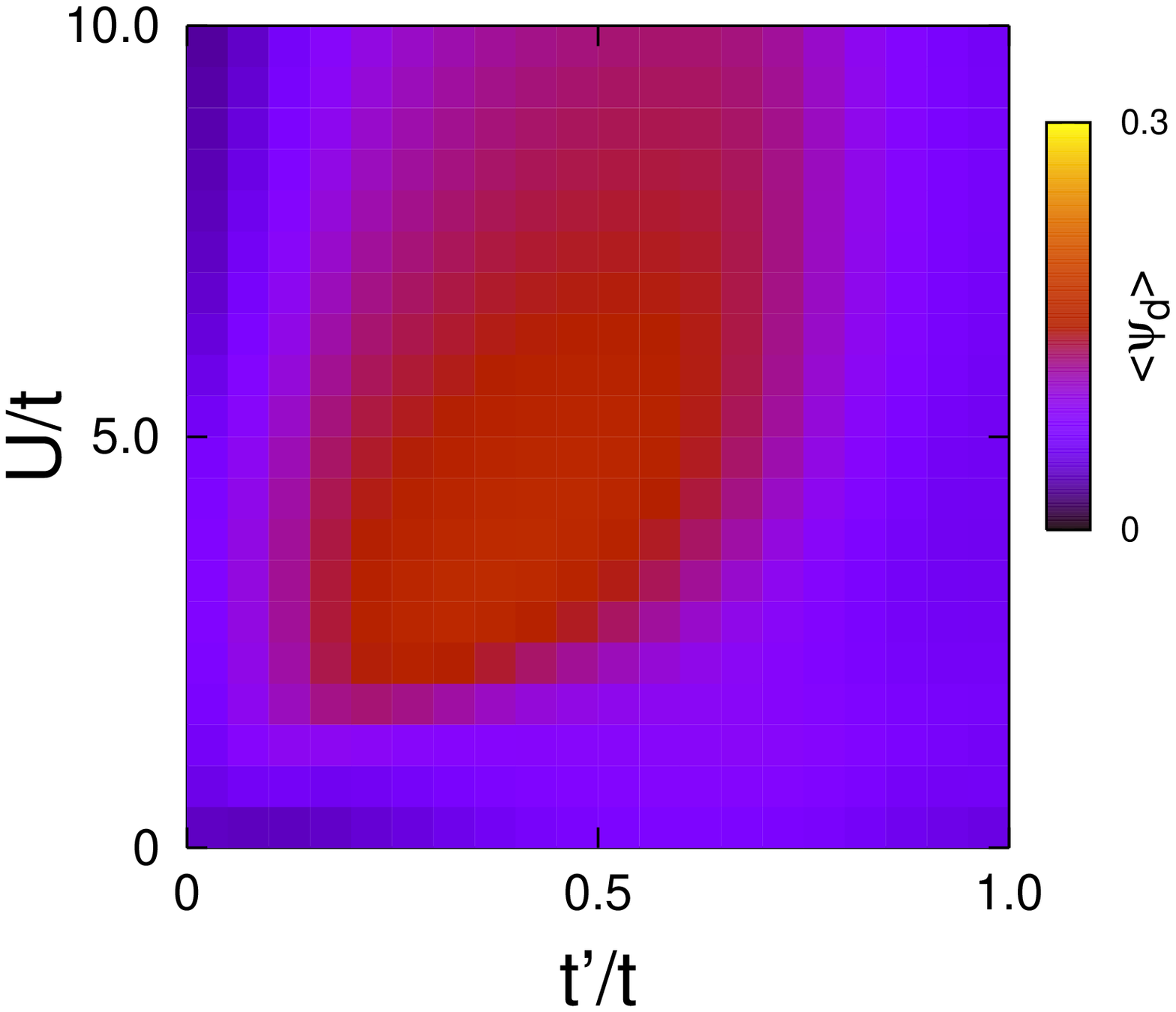}	
\end{array}
$
\caption{
\label{fig:DWN12ORL}
Disorder-averaged $d$-wave pairing order parameter, $\Dd$, on the twelve-site ladder with potential disorder: (a) $W/t = 0.05$ and (b) $W/t = 1.00$. The black regions in (a) correspond to $\PSS\approx1$.
}
\end{figure}

\begin{figure}[htb]
$
\begin{array}{cc}
\hspA(a)	& 
\hspB(b)\vspace{-0.5cm}  	 \\
\hspA\includegraphics[angle=-90,width=5cm]{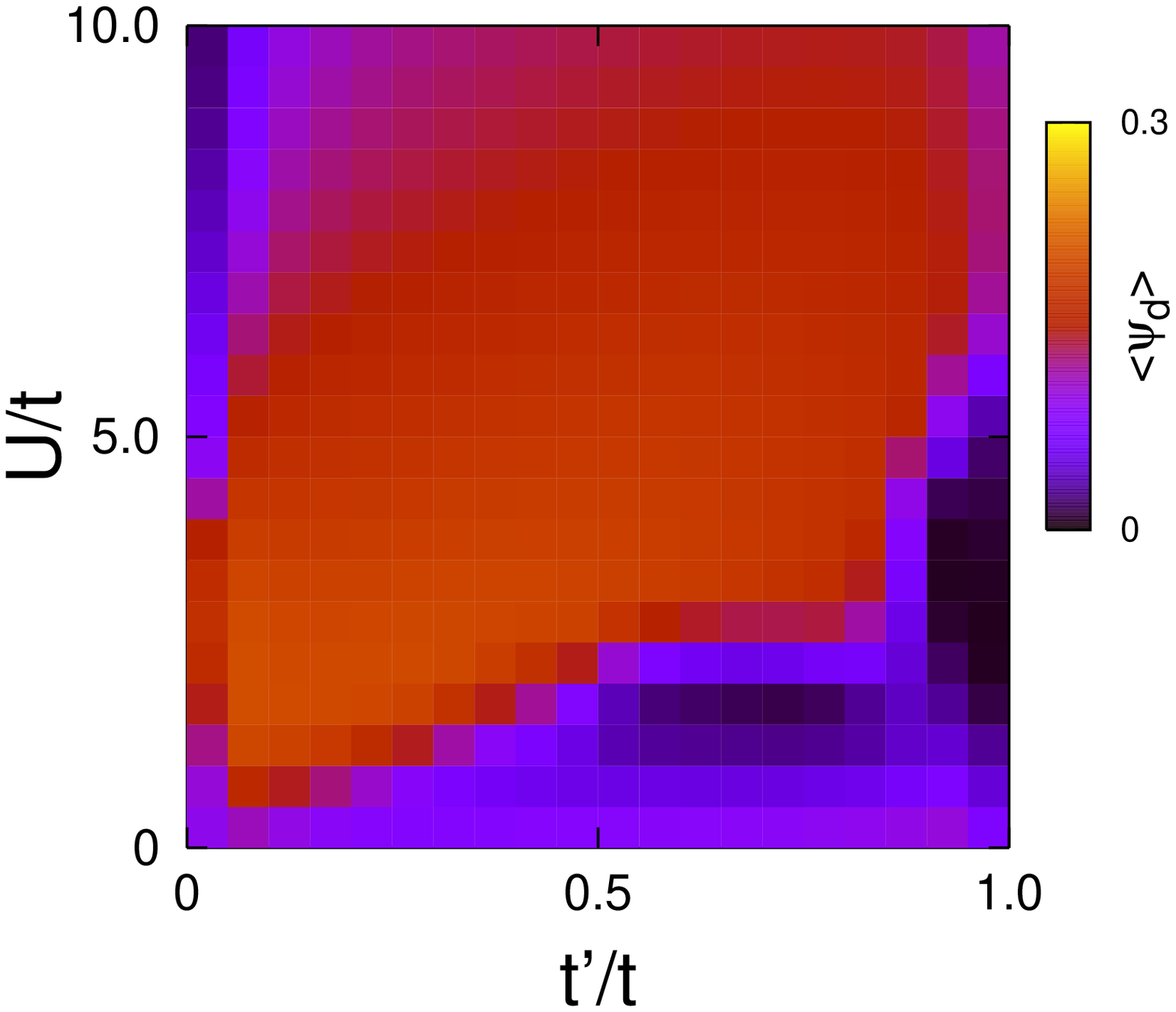}	&
\hspB\includegraphics[angle=-90,width=5cm]{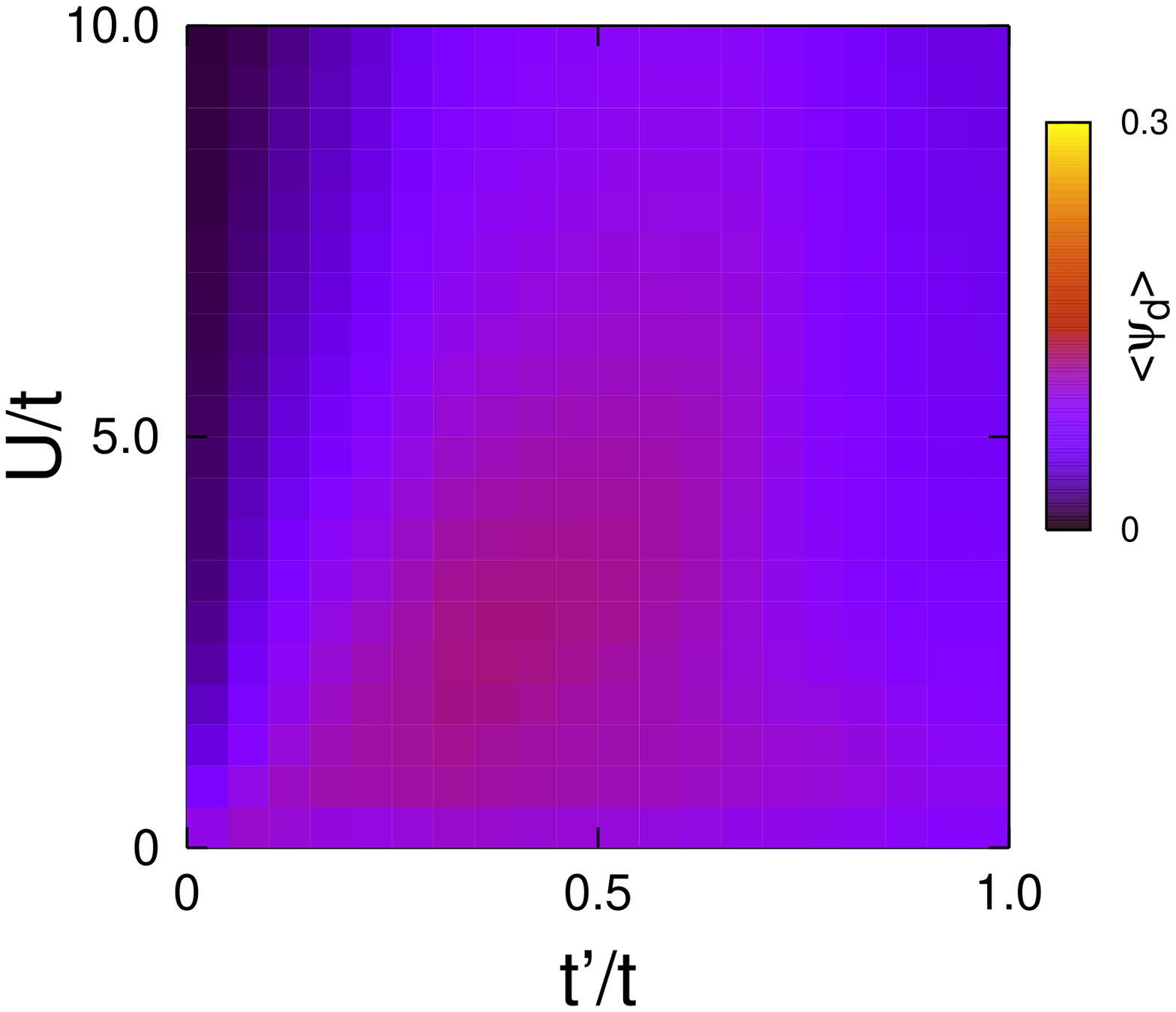}	
\end{array}
$
\caption{
\label{fig:DWN12BRL}
Disorder-averaged $d$-wave pairing order parameter, $\Dd$, on the twelve-site ladder with hopping disorder: (a) $W/t = 0.05$ and (b) $W/t = 1.00$. The black regions in (a) correspond to $\PSS\approx1$.
}
\end{figure}

We investigate $d$-wave symmetry of the ground state by studying the disorder-averaged $d$-wave order parameter, 
$\Dd$. The results for $\Dd$ for the eight- and twelve-site ladder clusters are plotted in Figs.~\ref{fig:DWN08ORL}-\ref{fig:DWN12BRL}. 

The data for the eight-site and twelve-site ladder clusters show that $\Dd$ is maximized for intermediate values of inhomogeneity,
which is consistent with the results for $\Dp$. However, there are some important distinctions between the regions where $\Dd$ and
$\Dp$ are strong.  First, there is a strong maximum in $\Dd$ near $t'/t\approx0$ for small $U/t$ as illustrated
in Figs.~\ref{fig:DWN08ORL} and \ref{fig:DWN08BRL}. Furthermore, the data shown in these graphs suggest that $\Dd$ is strongly dependent on the strength and type of disorder in this region of parameter space.   Second, $\Dd$ appears to be much more robust against increasing disorder for intermediate values of inhomogeneity than $\Dp$. As such, it appears that disorder does not greatly affect the $d$-wave symmetry of the ground state as $t'/t$ is increased from zero to unity. We will discuss this second point in more detail in Sec.~\ref{Sec:Variance}.

Recall from Eqs.~(\ref{eq:DWOP}) that $\Dd$ has  contributions from the strong ($t$) and weak bonds ($t'$).
Tsai \etal.\cite{TsaiKivelson2} showed that for a clean system the contribution to the $d$-wave order parameter comes
strictly from the strong bonds. We find however that the weak bonds contribute to the value of the order parameter for non-zero disorder in the limits of strong and weak inhomogeneity, and this effect is more pronounced for hopping disorder than potential disorder. For intermediate inhomogeneity, the strong bonds make the dominant contributions to the value of the order parameter. These results suggest that the properties of $\Dd$ are more robust against disorder for intermediate to large values of $t'/t$ in comparion to the case of $t'/t\approx0$ or $t^\prime/t \simeq 1$.

For $x=1/12$, we see that for regions in parameter space where $P(S=1) \simeq 1$, we have $\Dd=0$.
As $W/t$ increases, $P(S=1)$ decreases on account of level crossings between the $S=0$ and $S=1$ states,
which in turn leads to an increase in $\Dd$ in this region.

\subsection{Disorder-induced fluctuations}
\label{Sec:Variance}
\begin{figure}[h]
$
\begin{array}{cc}
\hspA(a)        &
\hspB(b)\vspace{-0.5cm}          \\

\hspA   \includegraphics[angle=-90,width=5cm]{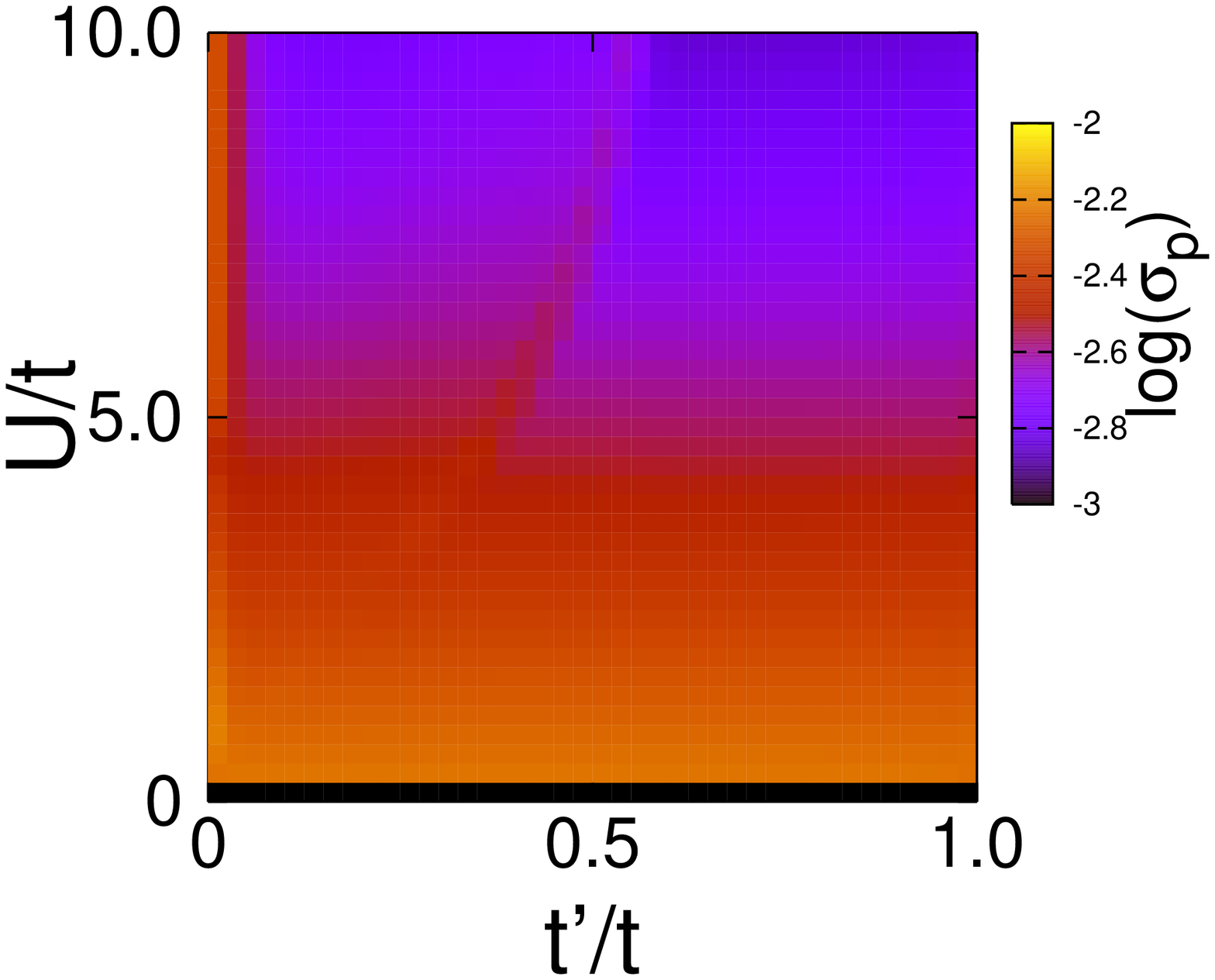}    &
\hspB   \includegraphics[angle=-90,width=5cm]{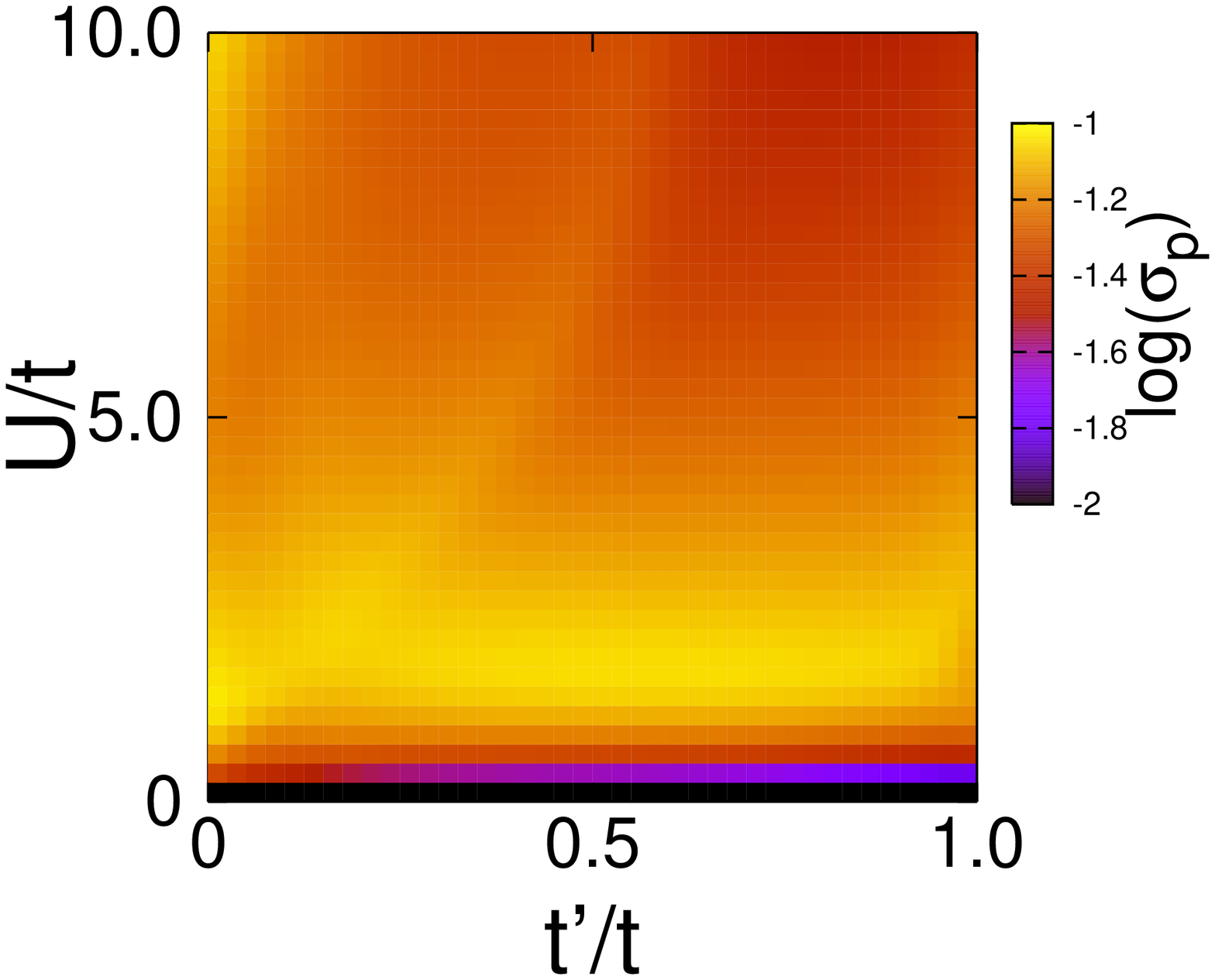}\\

\hspA(c)        & \hspB(d)\vspace{-0.5cm}        \\

\hspA   \includegraphics[angle=-90,width=5cm]{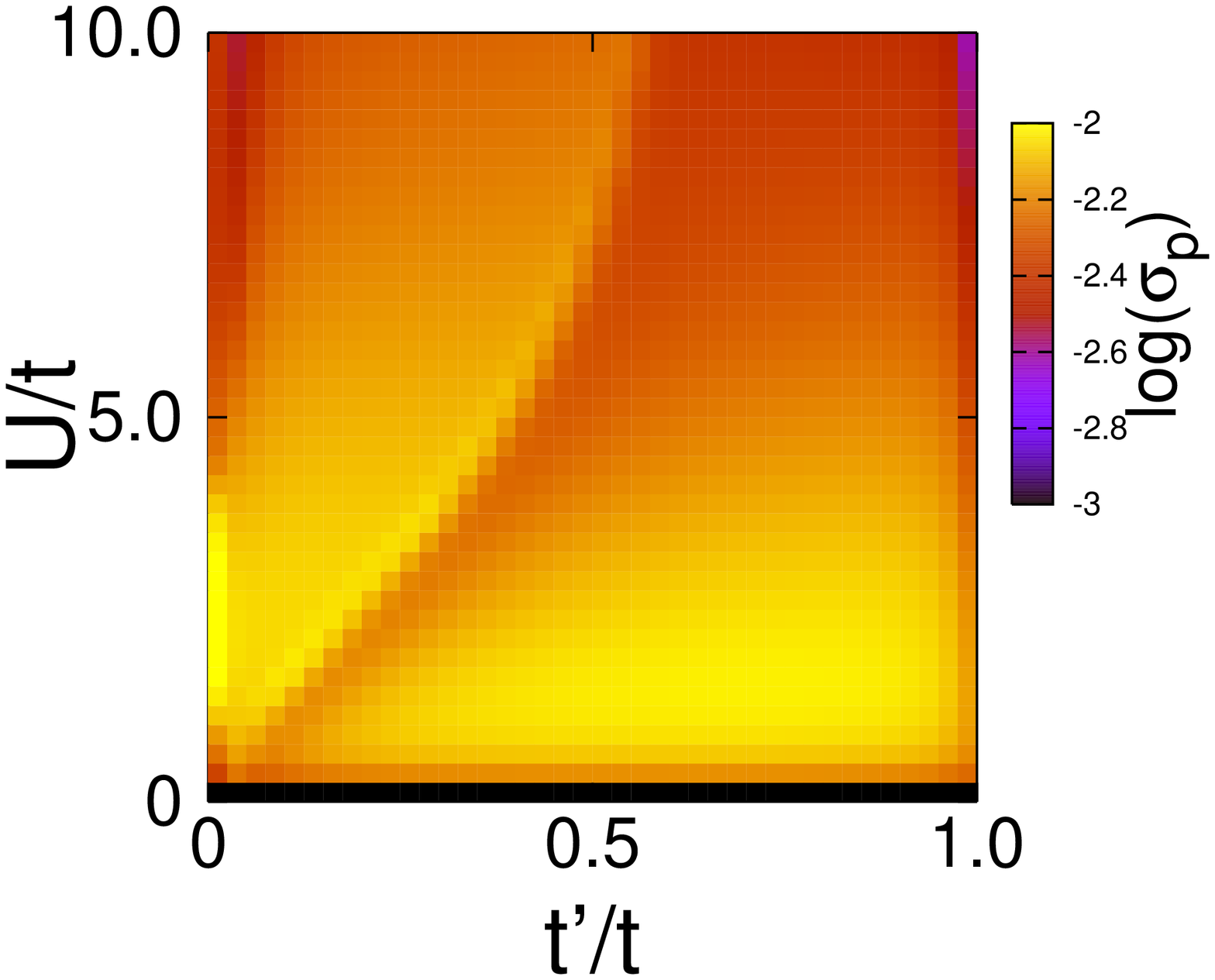} &
\hspB   \includegraphics[angle=-90,width=5cm]{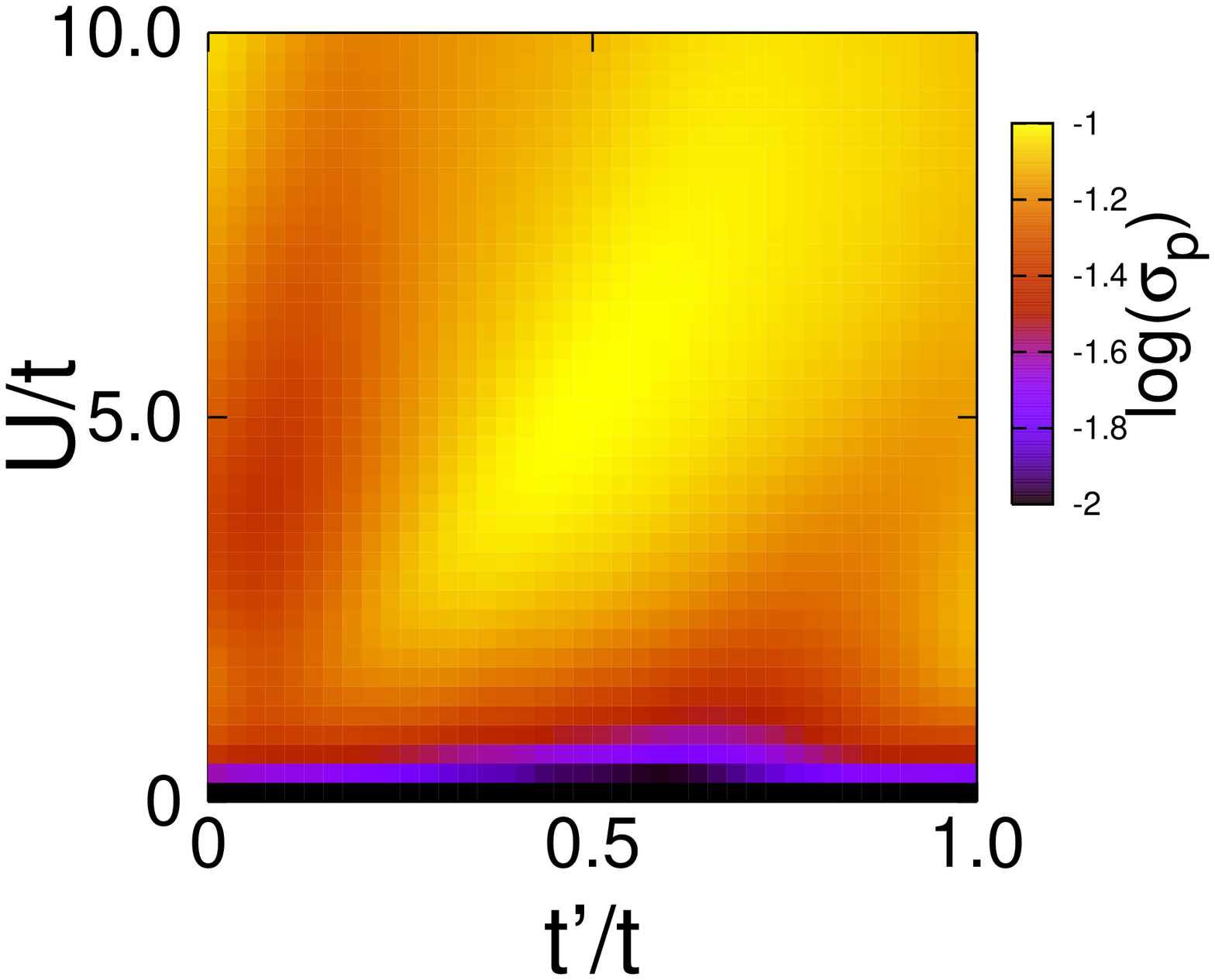}
\end{array}
$
\caption{
\label{fig:SigmaPN08}
Plots of $\log(\ssp)$ for the eight-site ladder cluster at doping $x=1/8$: 
(a) potential disorder, $W/t = 0.05$;
(b) potential disorder, $W/t = 1.00$;
(c) hopping disorder, $W/t = 0.05$;
(d) hopping disorder, $W/t = 1.00$.
}
\end{figure}

\begin{figure}[h]
$
\begin{array}{cc}
\hspA(a)	& 
\hspB(b)\vspace{-0.5cm}  	 \\

\hspA	\includegraphics[angle=-90,width=5cm]{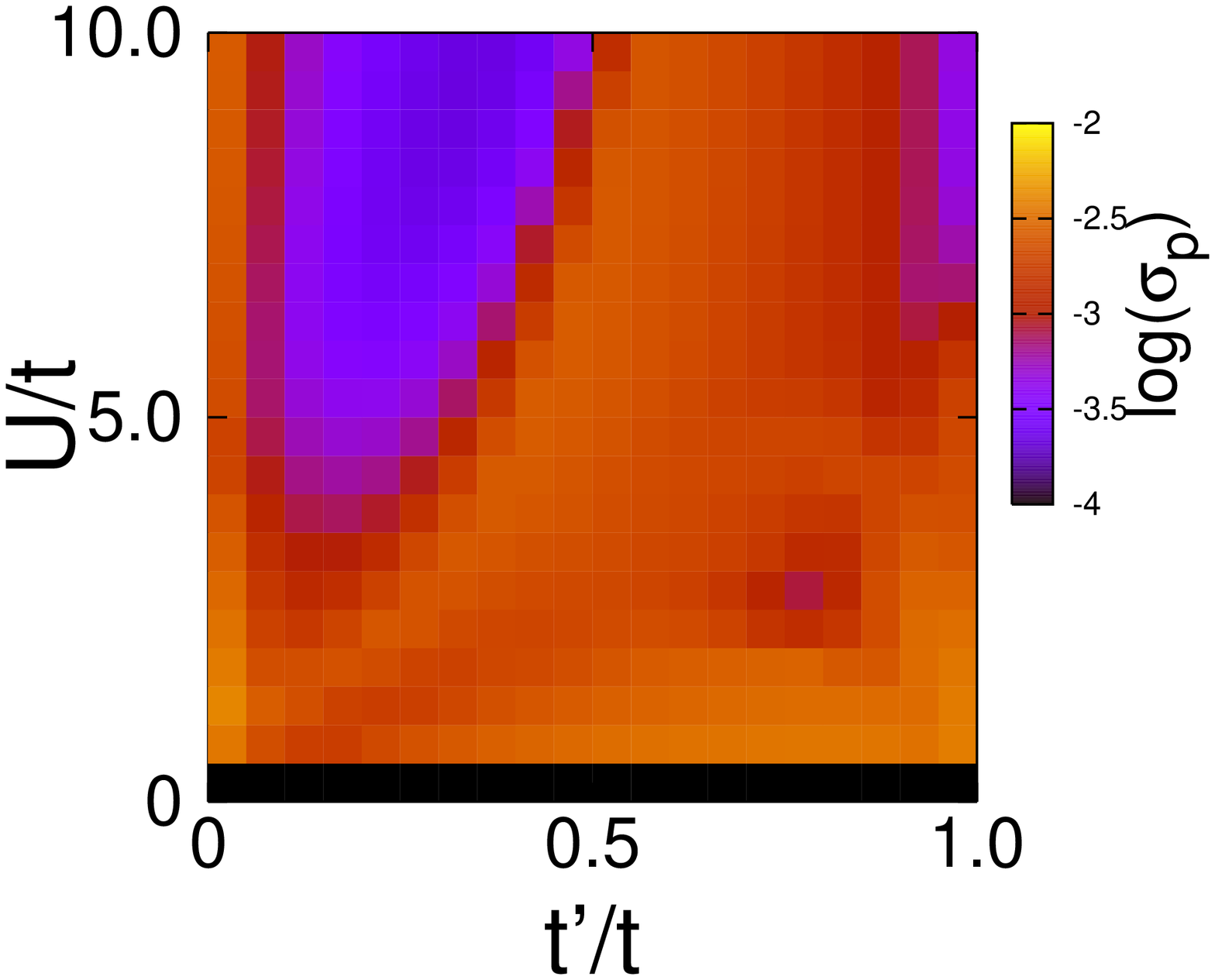} 	& 
\hspB  	\includegraphics[angle=-90,width=5cm]{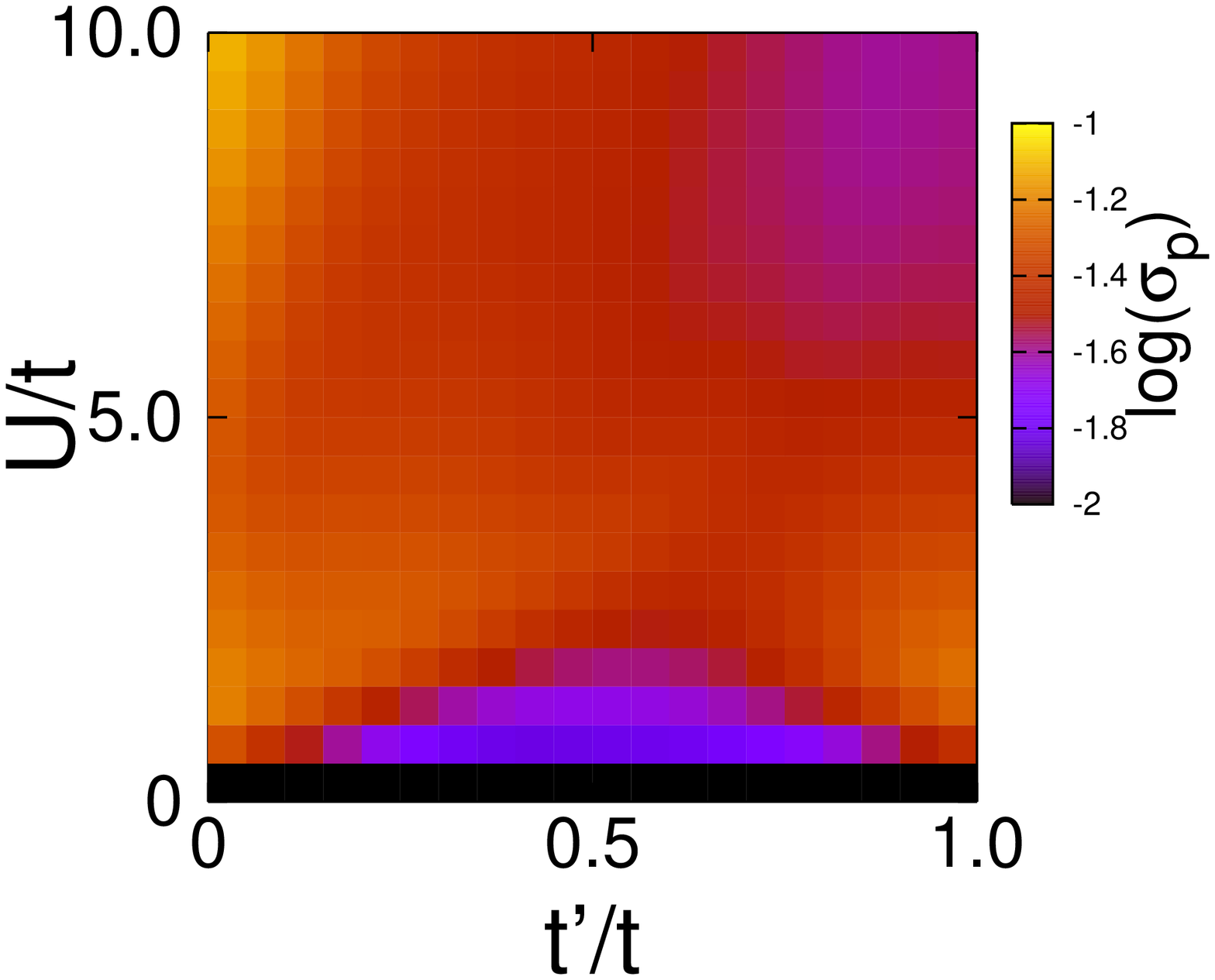}\\

\hspA(c)	& \hspB(d)\vspace{-0.5cm}  	 \\

\hspA	\includegraphics[angle=-90,width=5cm]{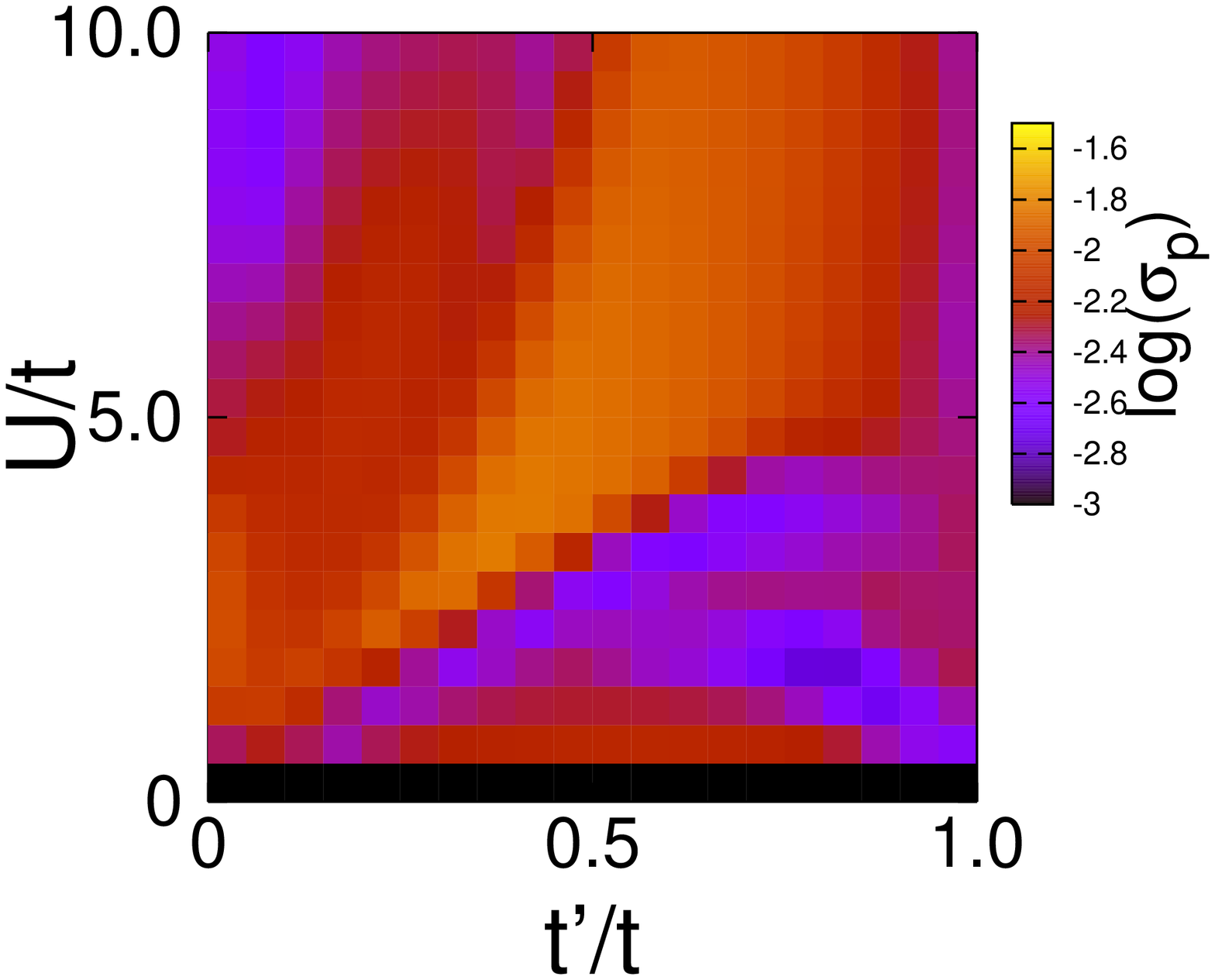} &
\hspB	\includegraphics[angle=-90,width=5cm]{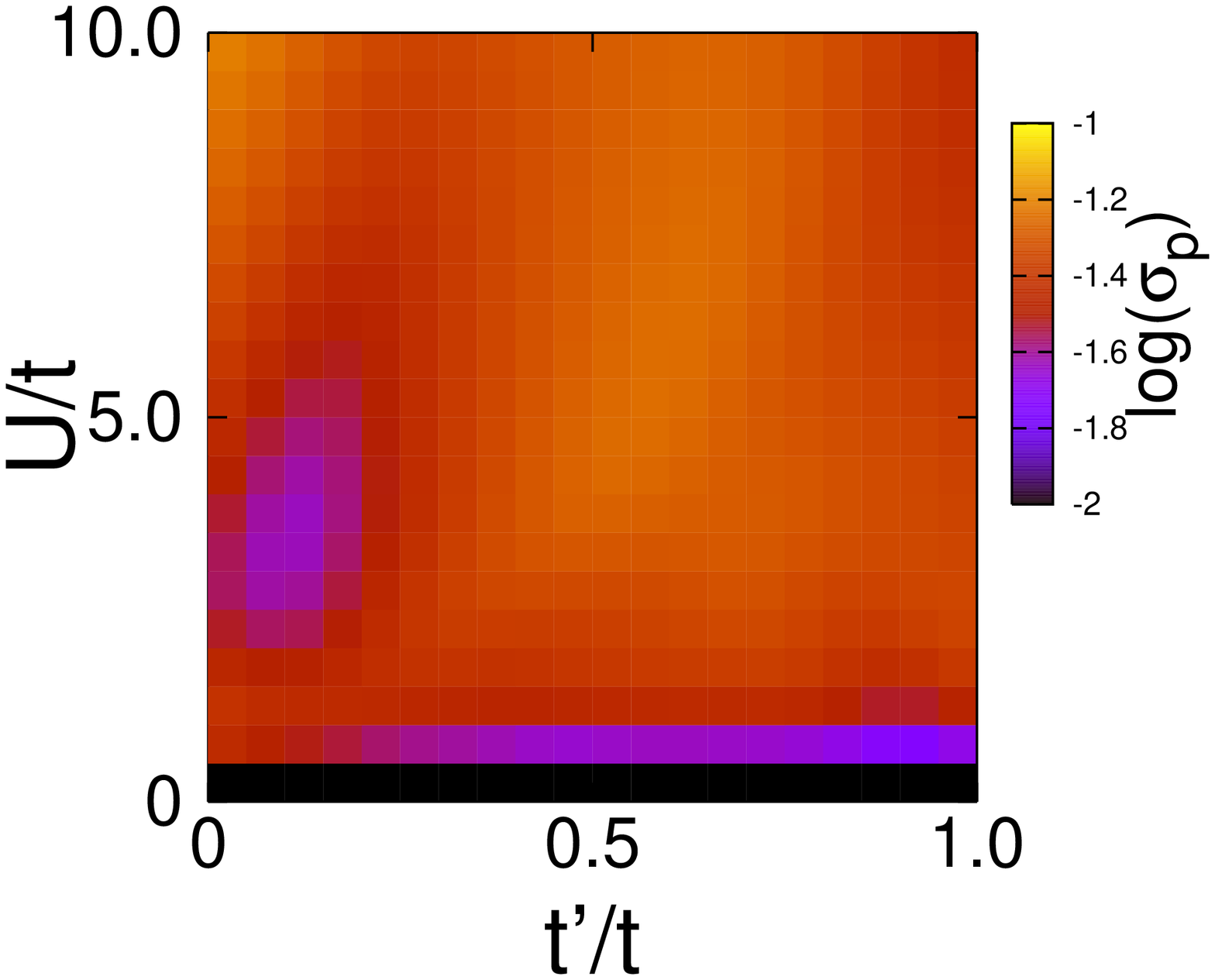}
\end{array}
$
\caption{
\label{fig:SigmaPN12}
Plots of $\log(\ssp)$ for the twelve-site ladder cluster at doping $x=1/12$: 
(a) potential disorder, $W/t = 0.05$;
(b) potential disorder, $W/t = 1.00$;
(c) hopping disorder, $W/t = 0.05$;
(d) hopping disorder, $W/t = 1.00$.
}
\end{figure}

The results for $\Dp$ show that as disorder is increased, pair binding is suppressed on average. However, 
the results for $\PDP$ suggest that pair binding can still be favoured at large $W/t$ for certain configurations 
even when $\Dp<0$.  We investigate disorder-induced fluctuations of the pair binding energy by calculating 
the variance of $\Delta_{p}$:
\begin{eqnarray}
\sigma_{p}&=&\sqrt{\frac{1}{K}\sum_{k=1}^{K}\left(\Delta_{p}^{k}-\Dp\right)^2}.
\end{eqnarray}
In Figs.~\ref{fig:SigmaPN08} and \ref{fig:SigmaPN12} we plot $\log\left(\sigma_p\right)$ to illustrate the magnitude of the disorder-induced fluctuations. 
At weak disorder, $\log\left(\ssp\right)$ shows a clear minimum in the region of maximum $\Dp$ for both potential and hopping disorder. 

This minimum has the form of a cusp, although it is smoothed out with increasing disorder strength.  The cusp is more pronounced for twelve
site clusters: for $t'/t<(t'/t)_{optimal}$, the value of $\ssp$ decreases by almost an order of magnitude from its value at $t'/t=0.5$ at
weak disorder, illustrating the change in the distribution of pair binding energies from $t'/t<(t'/t)_{optimal}$ to $t'/t>(t'/t)_{optimal}$. 
One possible explanation is that the location of the cusp signals a crossover from isolated plaquette physics to inhomogeneous lattice physics.
This is similar to the transition observed  by Peterson \etal.\cite{Peterson} in the case of the $d$-Mott insulator state.

We have also investigated fluctuations in the spin gap and the $d$-wave order parameters. In each case, 
the fluctuations are weakest in the region where the disorder averaged quantities are maximal. 
Unlike $\sigma_{p}$, disorder induced fluctuations in the spin gap and the 
$d$-wave order parameter do not exhibit any cusp-like features.

\section{Staggered Potentials}
\label{Sec:Staggered}

One way of viewing a particular disorder configuration in either potential or hopping strength is as a 
linear combination of staggered on-site potential or hopping  configurations, where the deviation from a 
uniform system on each site or bond can only take values of $\pm \frac{W}{2}$. On an eight-site ladder cluster, with periodic 
boundary conditions along the ladder, there are 32 possible non-trivial inequivalent staggered configurations 
for either potentials or hoppings. Even though for an interacting problem we cannot take a linear combination of 
solutions for different potentials to determine the full behaviour, studying individual staggered potentials can 
lead to insights into their contributions in random potentials. Hence, we study $\Delta_{p}$, $\Delta_{s}$, 
and $\Psi_{d}$ at doping $x=1/8$ for all staggered on-site potentials and hopping patterns.

\begin{figure}
$
\begin{array}{c}
(a) \\ 
\includegraphics[angle=-90,width=8cm]{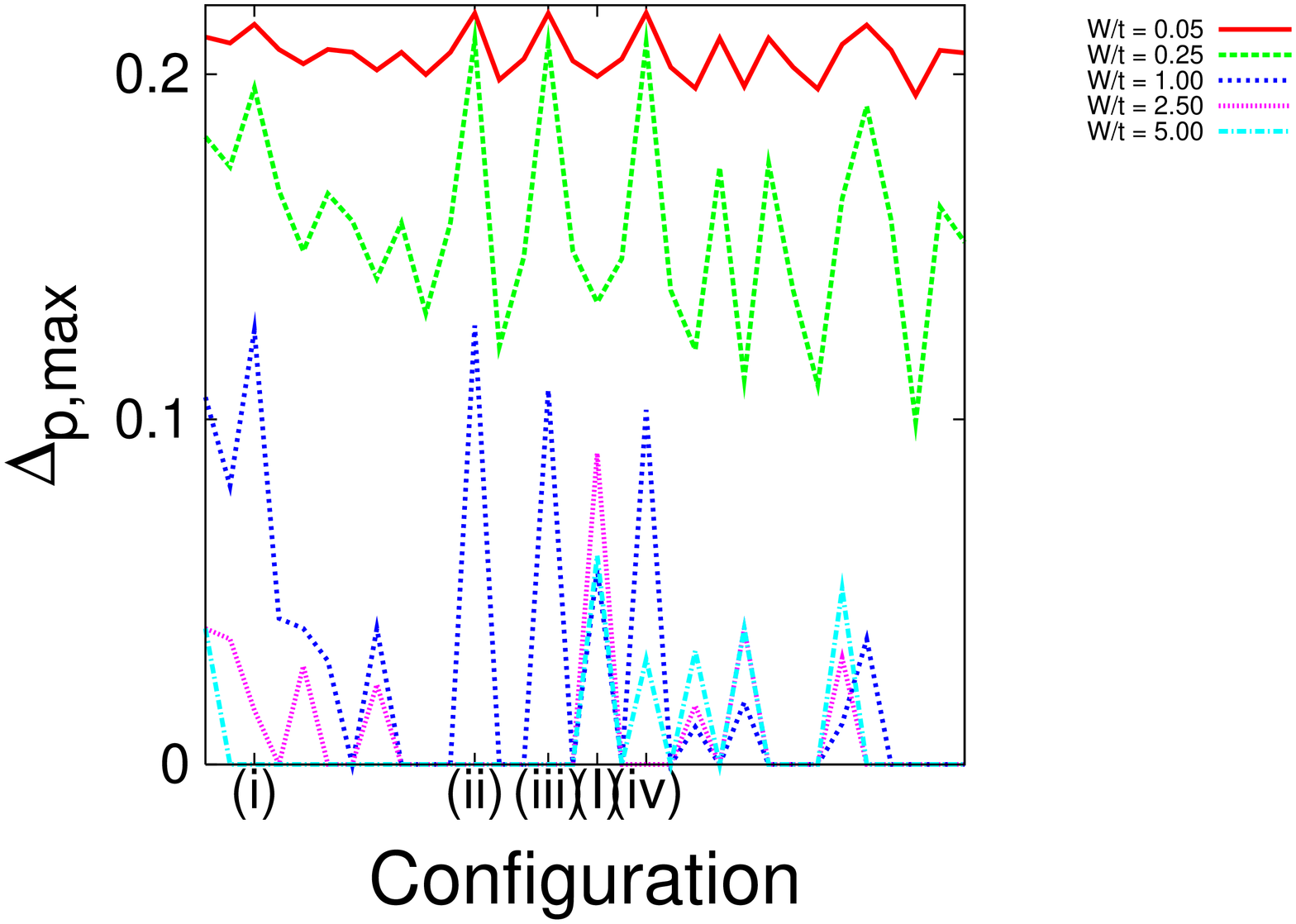}         \vspace{0.125cm}\\
(b)  \\
\includegraphics[angle=-90,width=8cm]{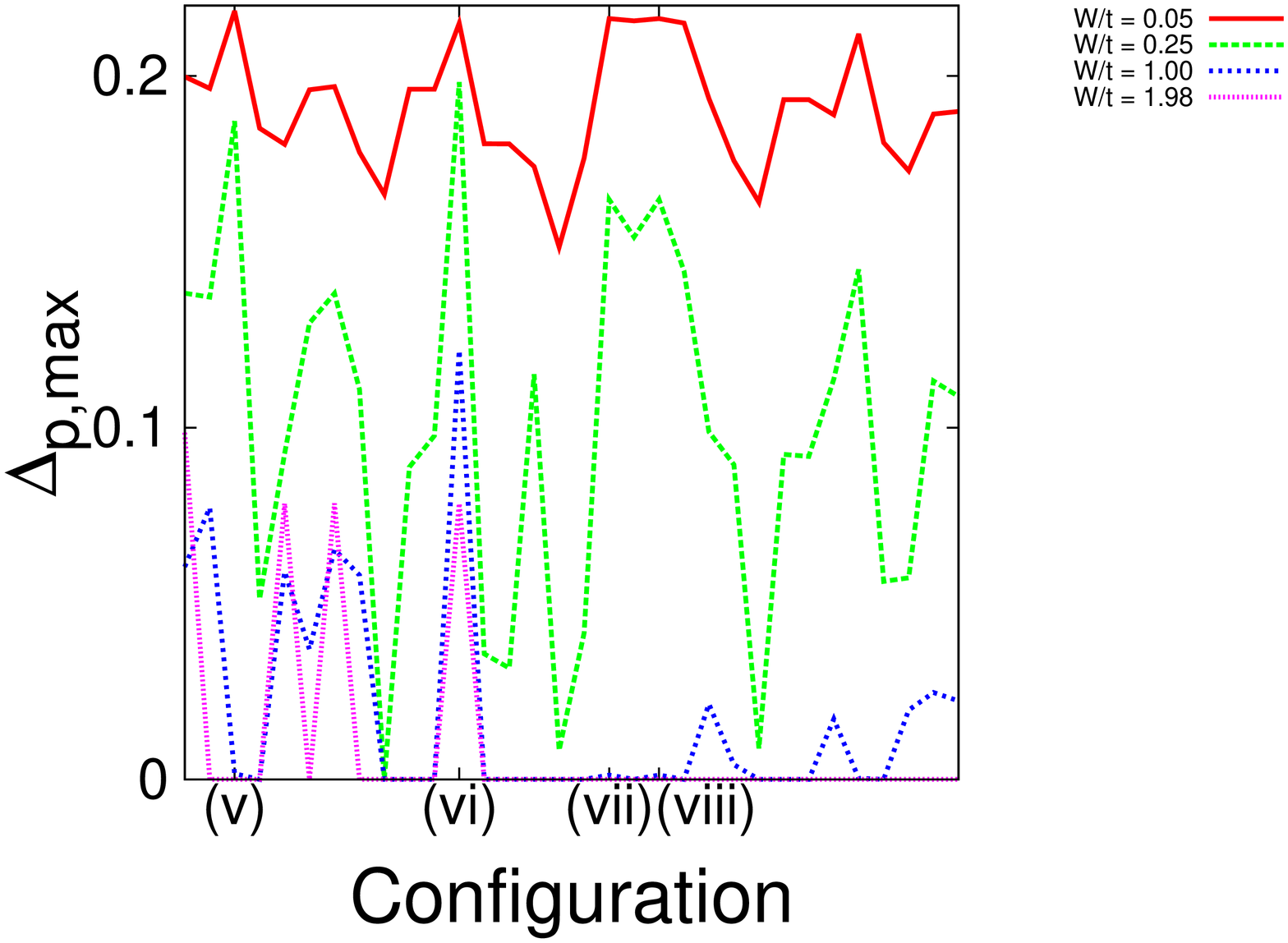}
\end{array}
$
\caption{
\label{fig:MaxPBEStaggered}
Plots of $\Delta_{p,max}$ for each staggered configuration as a function of $W/t$:
(a) staggered potentials;
(b) staggered hopping.}
\end{figure}

In Figs.~\ref{fig:MaxPBEStaggered} $(a)$ and $(b)$ we plot the maximum $\Delta_{p}$ in the $U-t^\prime$ plane for
all 32 staggered potential and hopping configurations for several different values of disorder strength.
For most configurations the maximum value of $\Delta_{p}$ is suppressed as $W/t\to1$. 
However, as disorder strength is increased beyond this limit, there are some configurations where 
pair binding persists to large $W/t$. This is in contrast to the behaviour of the 
disorder-averaged PBE, which is suppressed beyond some maximum $W/t$. 

In the interest of brevity, we do not present data for all 32 configurations for the two types 
of disorder in this paper. Instead, we have identified  configurations of staggered on-site potentials 
and hopping patterns, shown in Fig.~\ref{fig:StaggeredConfigurationsSmallW}, for which the pair 
binding properties appear to be most robust against increasing $W/t$. 

We observe no individual disorder configuration that enhances the PBE above its value in the clean limit for any of the disordered ladder models at any value of $W/t$. However, there appears to be no consistent response to increasing $W/t$ between all configurations considered -- the PBE of some configurations is suppressed rapidly as disorder increases, while the pair binding properties of other configurations appear to be much more resistant to disorder. The maximum of the PBE for each configuration as a function of $U/t$ and $t'/t$ is plotted in Fig.~\ref{fig:MaxPBEStaggered}.

\subsection{Small $W/t$}
\label{Sec:SmallW}

\begin{figure}[htb]
$
\begin{array}{cccc}
(i)	& 
(ii)   	&
(iii)	&
(iv)	\\
\includegraphics[angle=90,scale=0.3]{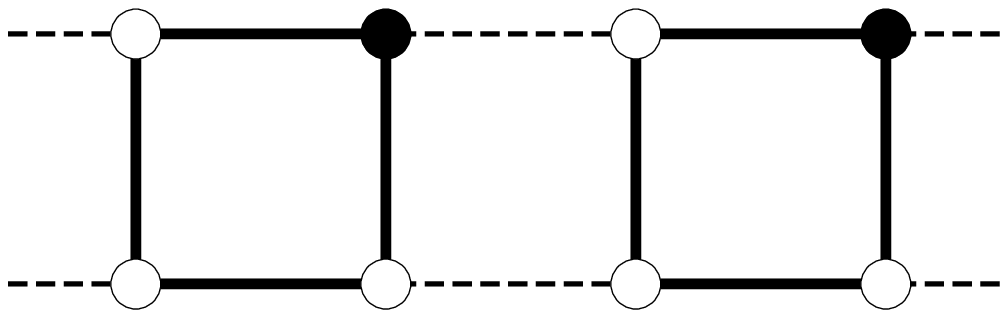}	&
\includegraphics[angle=90,scale=0.3]{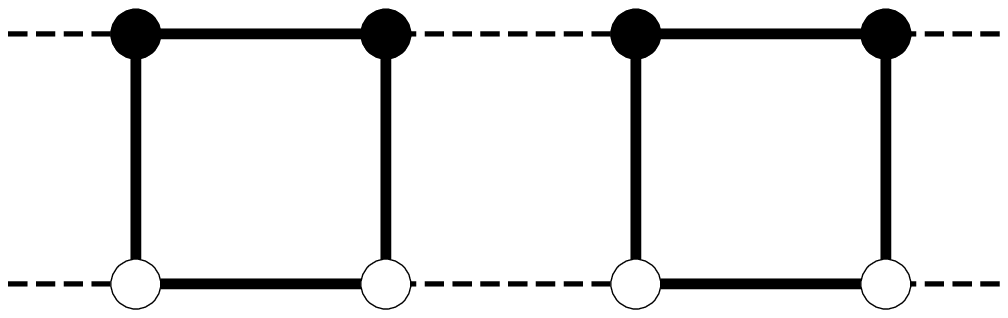}	&
\includegraphics[angle=90,scale=0.3]{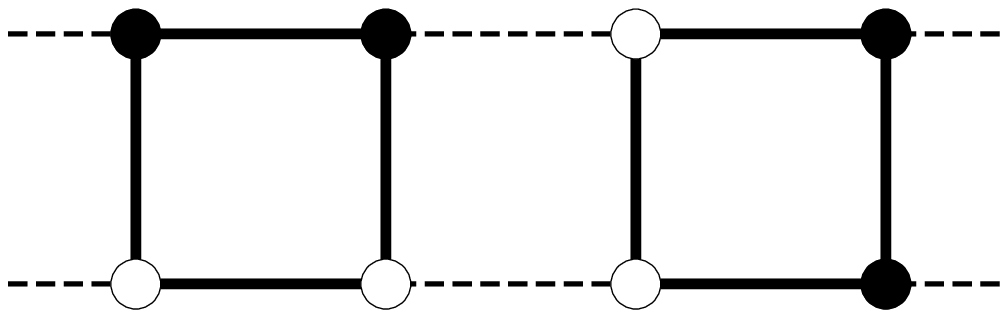}	&
\includegraphics[angle=90,scale=0.3]{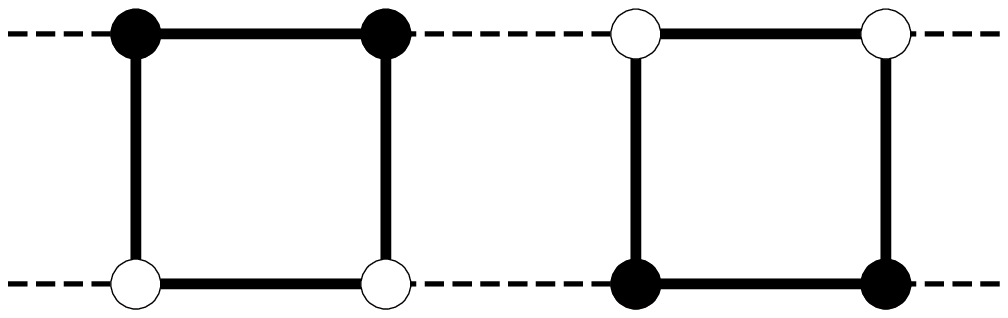}	\\
(v)	& 
(vi)   	& 
(vii) 	& 
(viii) 	\\
\includegraphics[angle=90,scale=0.3]{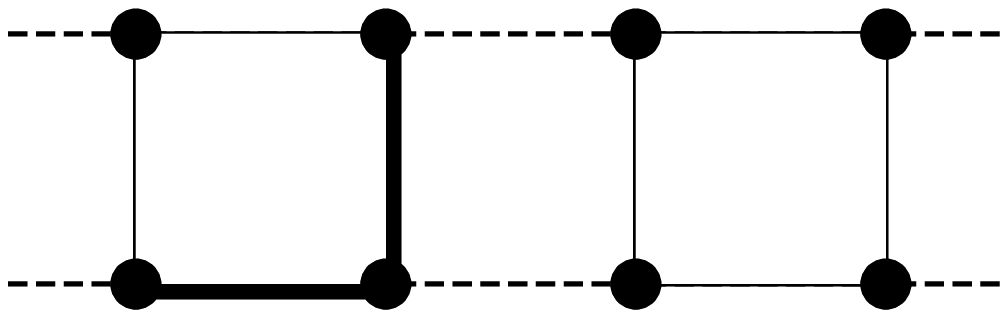}	&
\includegraphics[angle=90,scale=0.3]{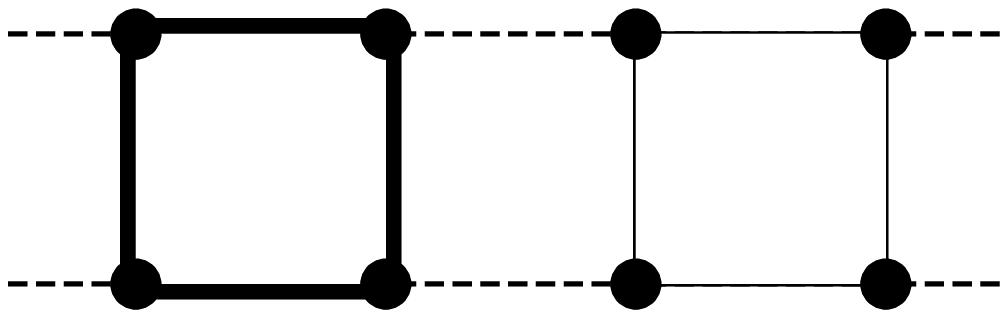}	&
\includegraphics[angle=90,scale=0.3]{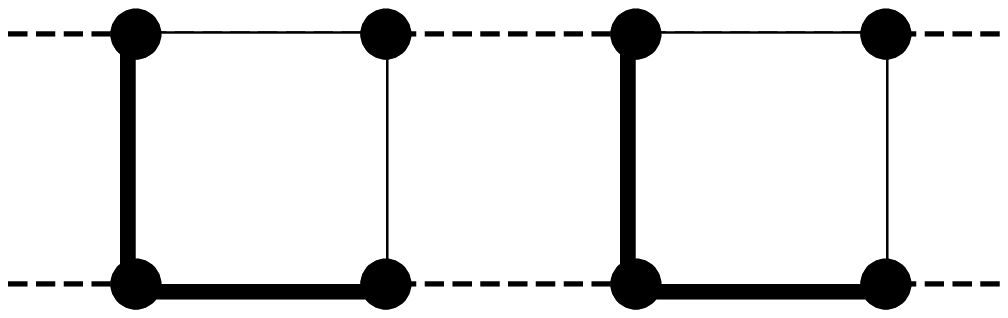}	&
\includegraphics[angle=90,scale=0.3]{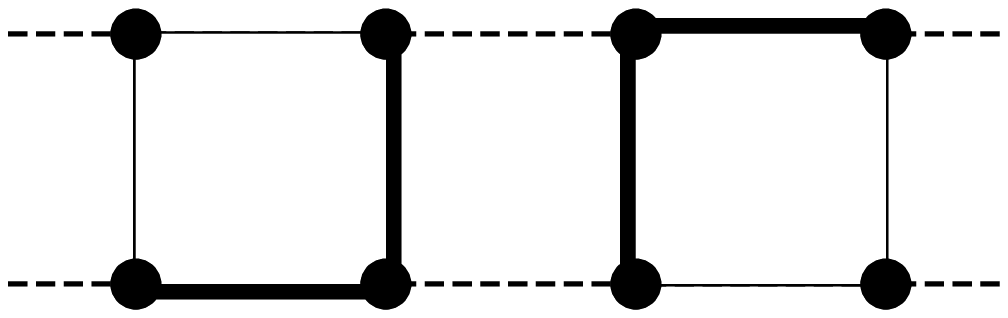}	\\
\end{array}
$
\caption{
\label{fig:StaggeredConfigurationsSmallW}
Configurations of staggered on-site ($i-iv$) and bond ($v-viii$) potentials discussed in Sec.~\ref{Sec:SmallW}. Dashed lines correspond to weak ($t'$) bonds, solid lines correspond to strong ($t$) bonds, and dots correspond to lattice sites. For configurations $(i-iv)$, white (black) dots correspond to on-site potential strengths +(-)$\frac{W}{2t}$. For configurations $(v-viii)$, solid thin (thick) lines correspond to bond strength $1+\frac{W}{2t}$ ($1-\frac{W}{2t}$). 
}
\end{figure}

For small $W/t$, the configurations $(i)$ to $(viii)$ in Fig.~\ref{fig:StaggeredConfigurationsSmallW} lead to the largest 
PBEs. With the exception of configurations $(i)$ and $(vi)$, the staggered potentials or hoppings ``pair up'' 
locally on each plaquette, resembling a dimer-like structure. The sum of all deviations from the uniform system add to 
zero except for $(i)$ and $(v)$, with $(i)$ and $(vi)$ the only configurations for which the sum of the deviations do not add to zero on each plaquette. 

\begin{figure}
$
\begin{array}{cc}
\hspA(a)	& 
\hspB(b)\vspace{-0.5cm}  	 \\

\hspA	\includegraphics[angle=-90,width=5cm]{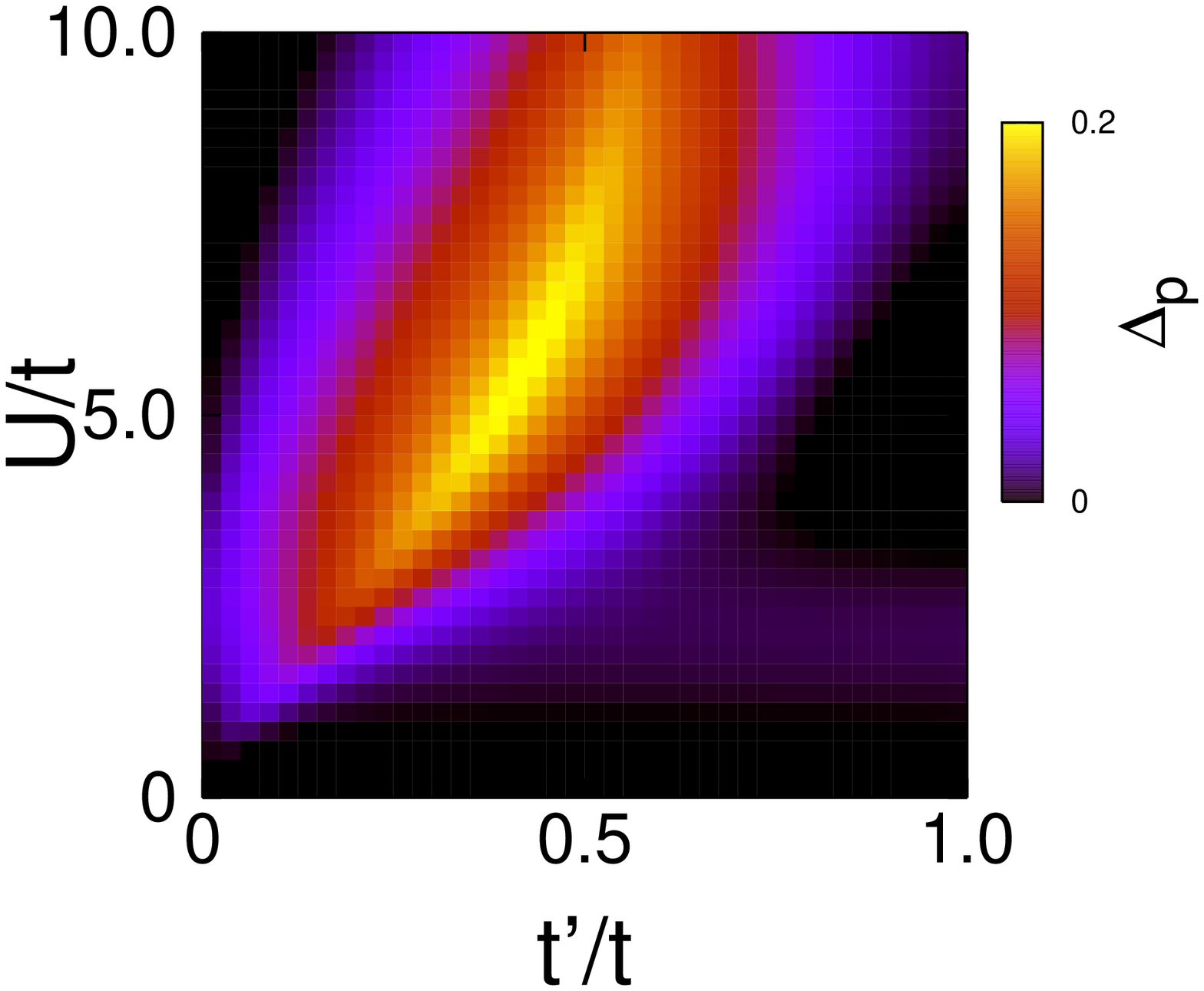} 	& 
\hspB  	\includegraphics[angle=-90,width=5cm]{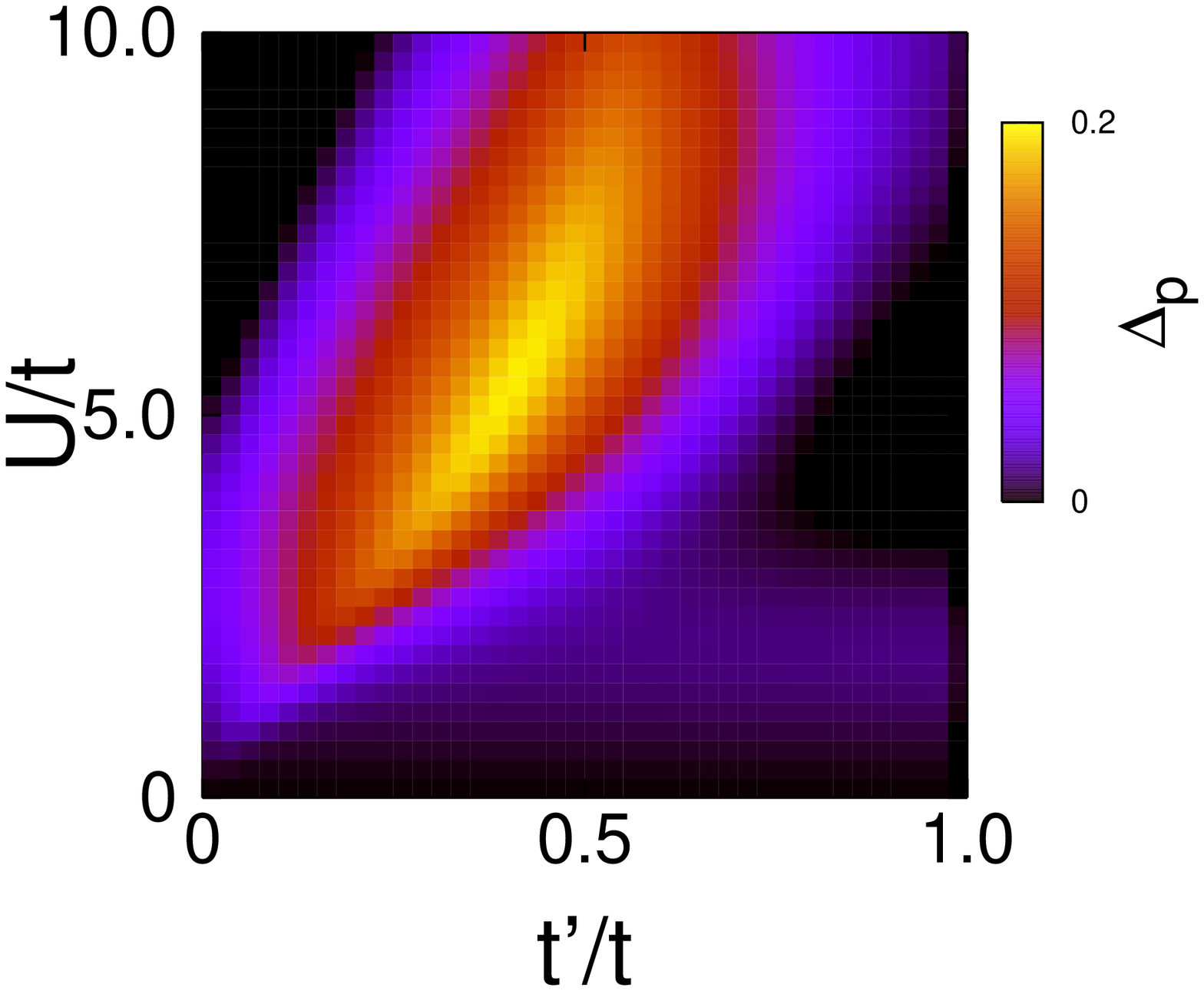}\\
\hspA(c)	& 
\hspB(d)\vspace{-0.5cm}  	 \\

\hspA	\includegraphics[angle=-90,width=5cm]{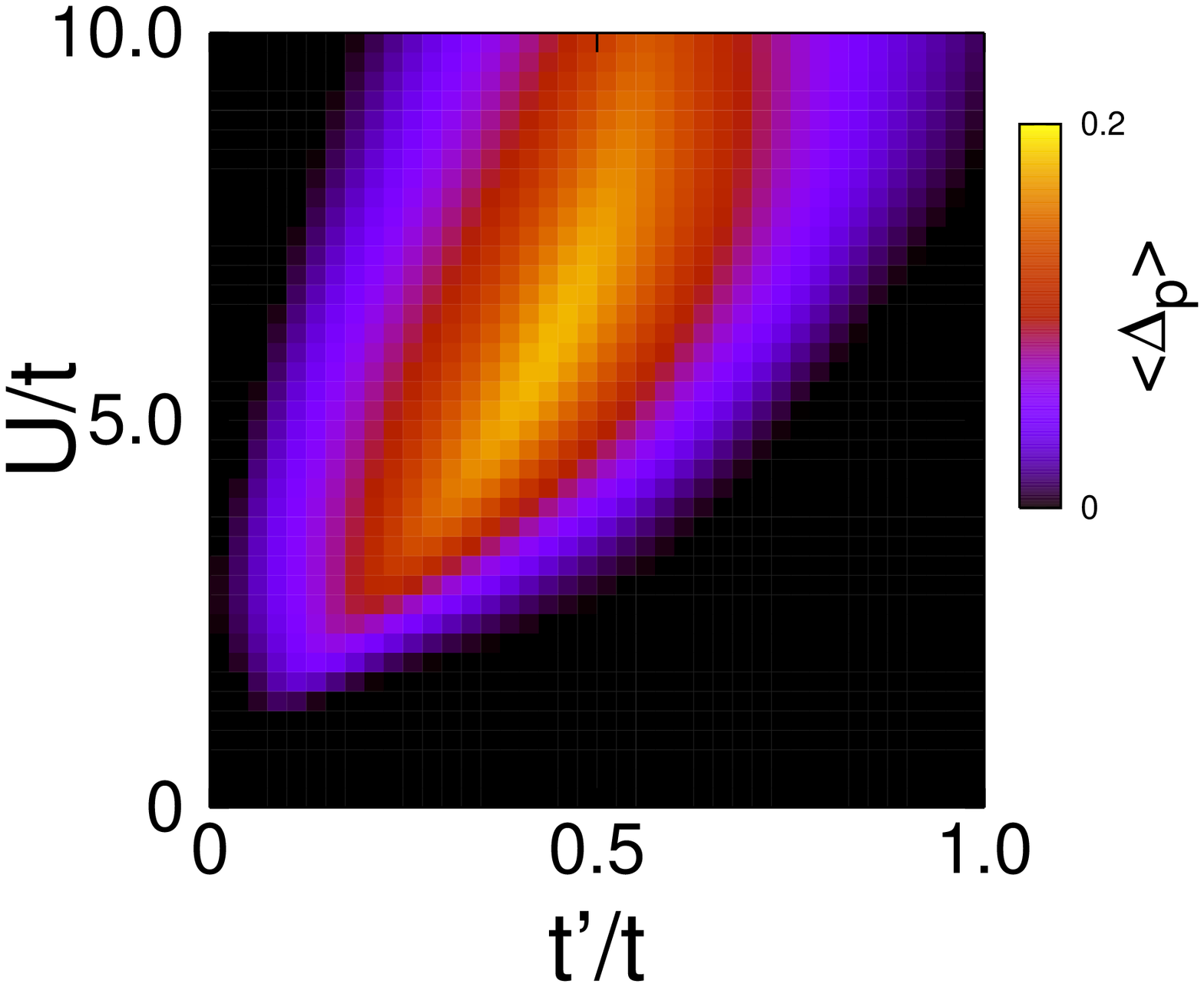} 	& 
\hspB  	\includegraphics[angle=-90,width=5cm]{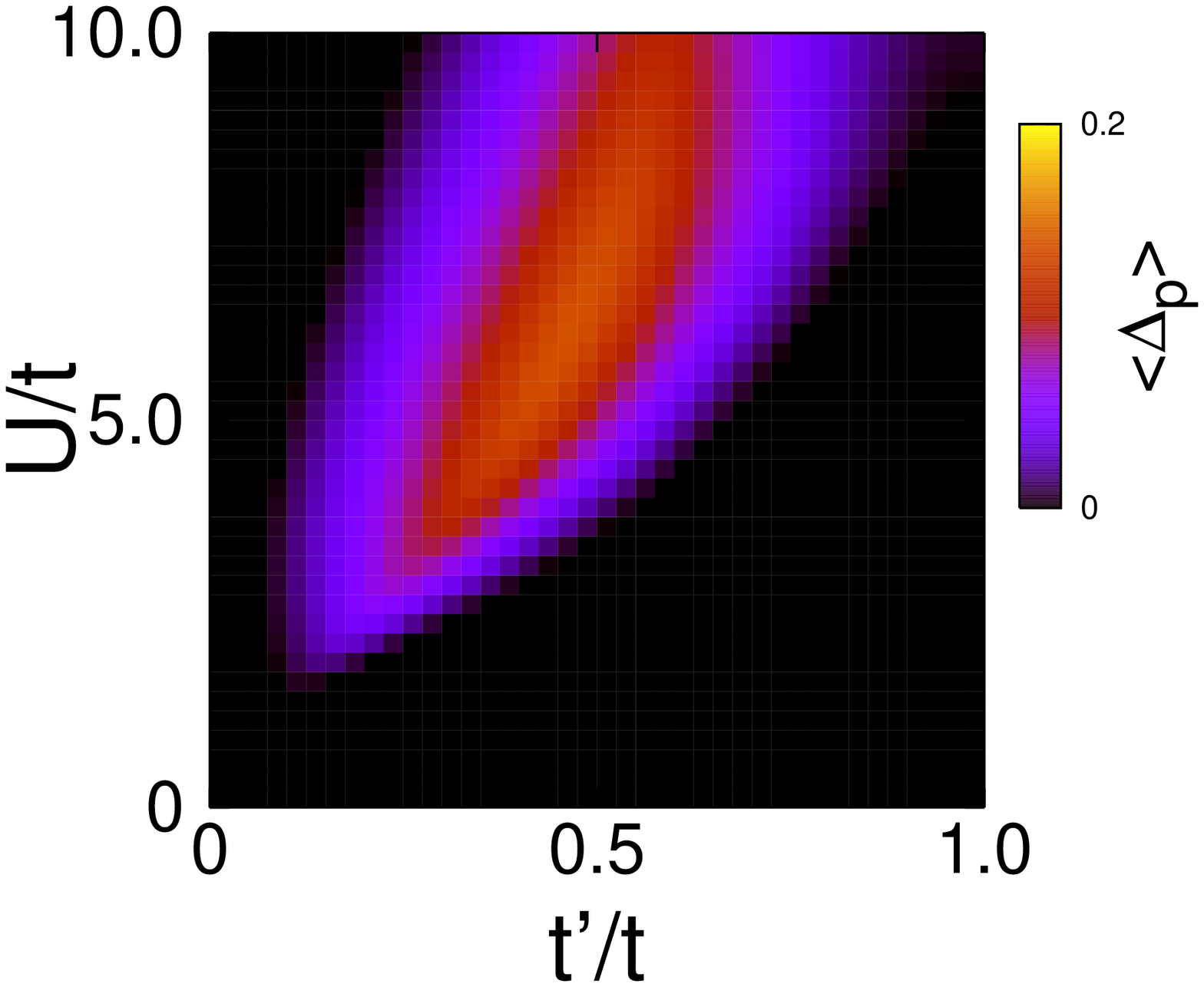}
\end{array}
$
\caption{
\label{fig:PBEStaggeredPotentialW0025}
Plots of $\Delta_{p}$ for the eight-site ladder cluster at doping $x=1/8$ and $W/t=0.25$ for $(a)$ staggered potential configuration $(ii)$ and $(b)$ staggered hopping configuration $(vi)$. For comparison, we also plot $\Dp$ at $W/t=0.25$ for $(c)$ potential disorder and $(d)$ hopping disorder.  (see Fig.~\ref{fig:StaggeredConfigurationsSmallW} for configuration details.)
}
\end{figure}

\begin{figure}
$
\begin{array}{cc}
\hspA(a)	& 
\hspB(b)\vspace{-0.5cm}  	 \\

\hspA	\includegraphics[angle=-90,width=5cm]{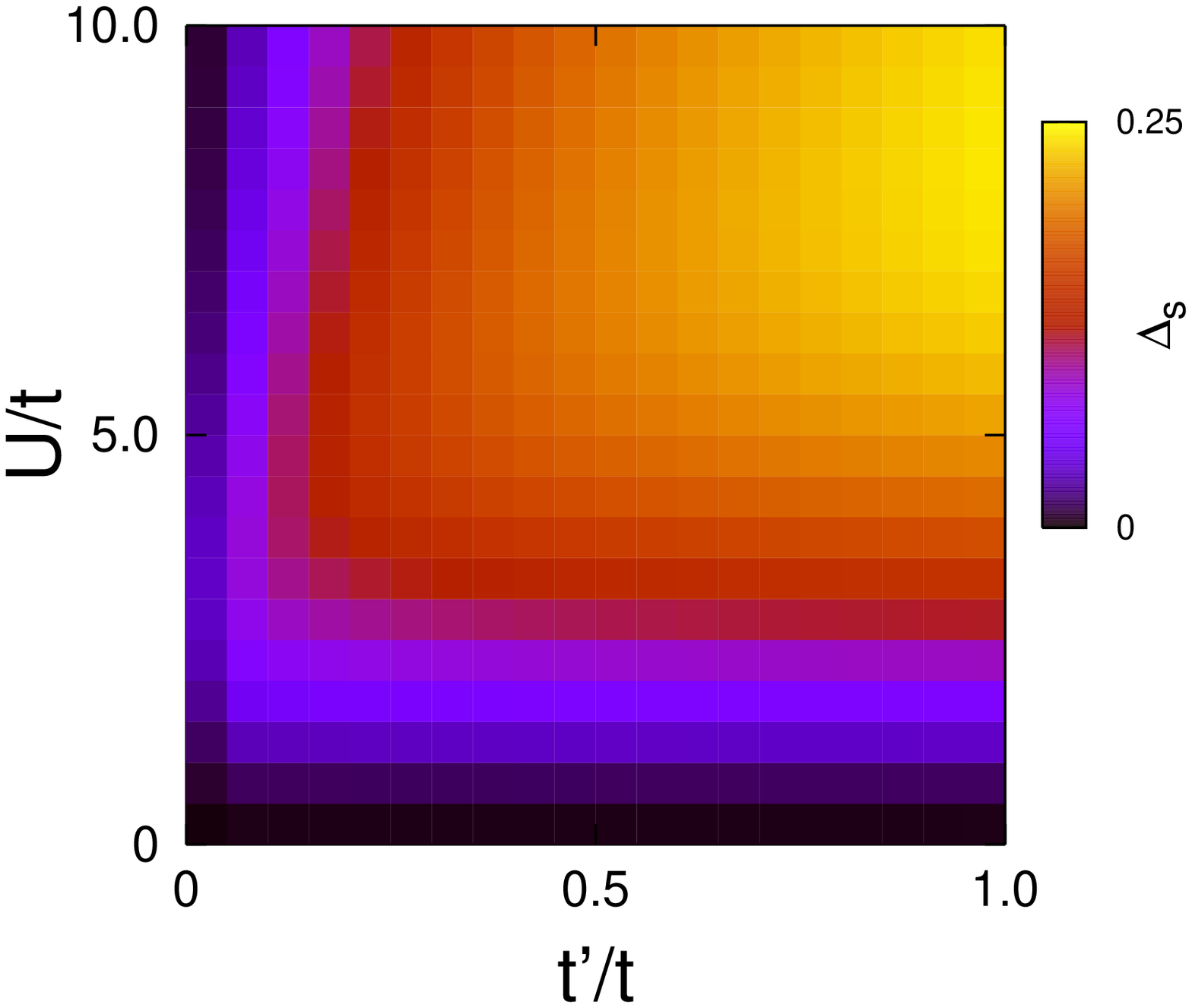} 	& 
\hspB  	\includegraphics[angle=-90,width=5cm]{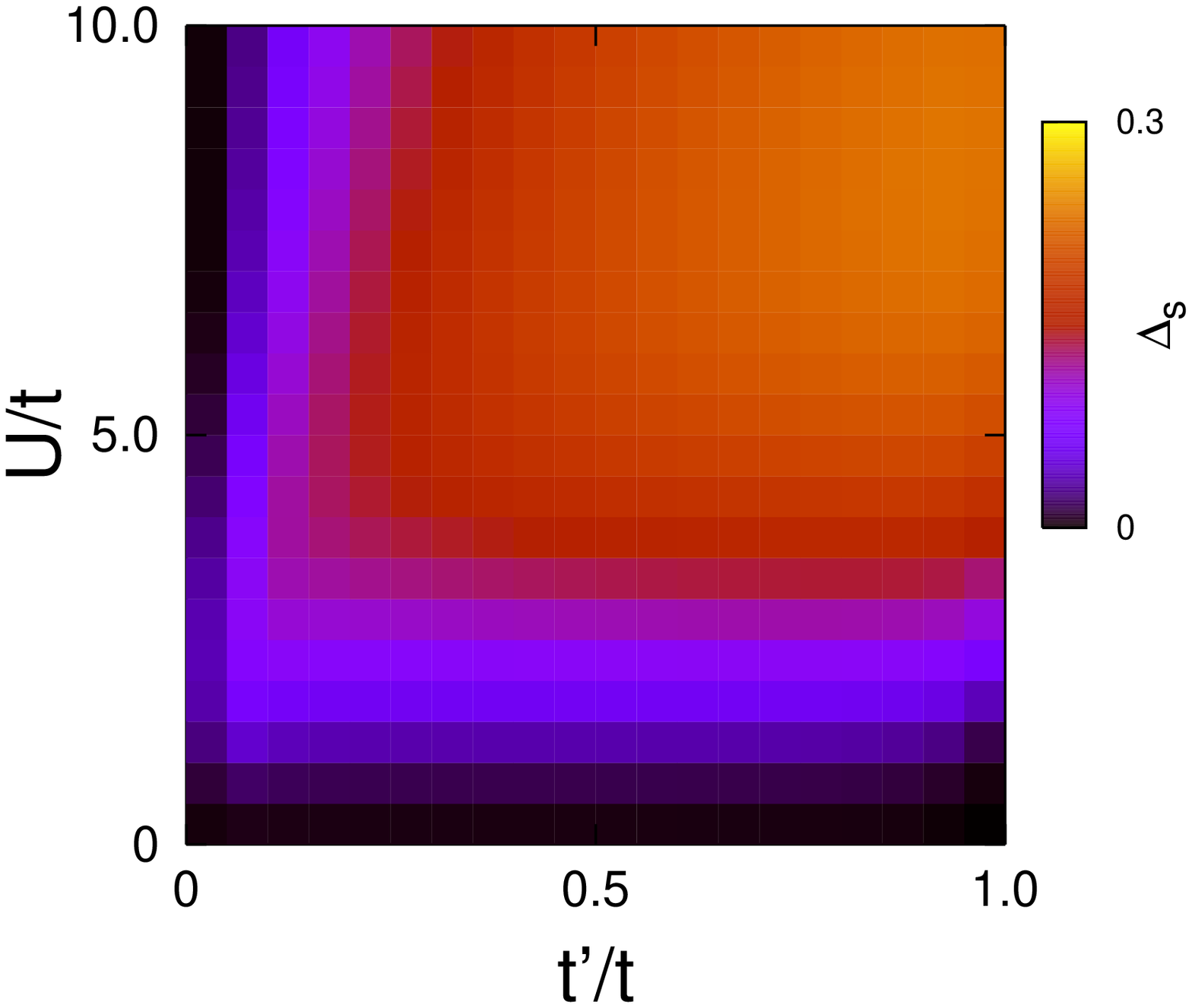}\\
\hspA(c)	& 
\hspB(d)\vspace{-0.5cm}  	 \\

\hspA	\includegraphics[angle=-90,width=5cm]{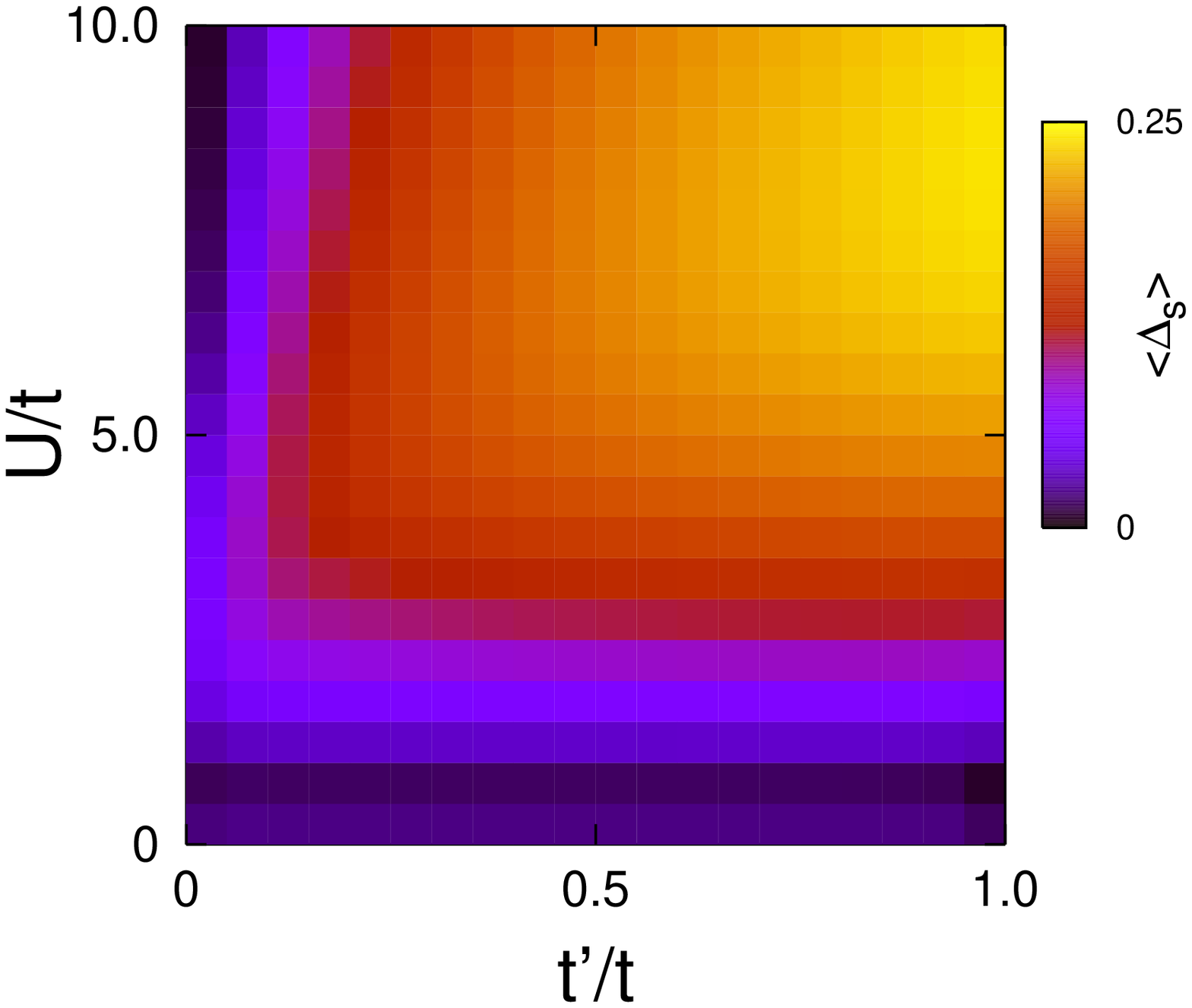} 	& 
\hspB  	\includegraphics[angle=-90,width=5cm]{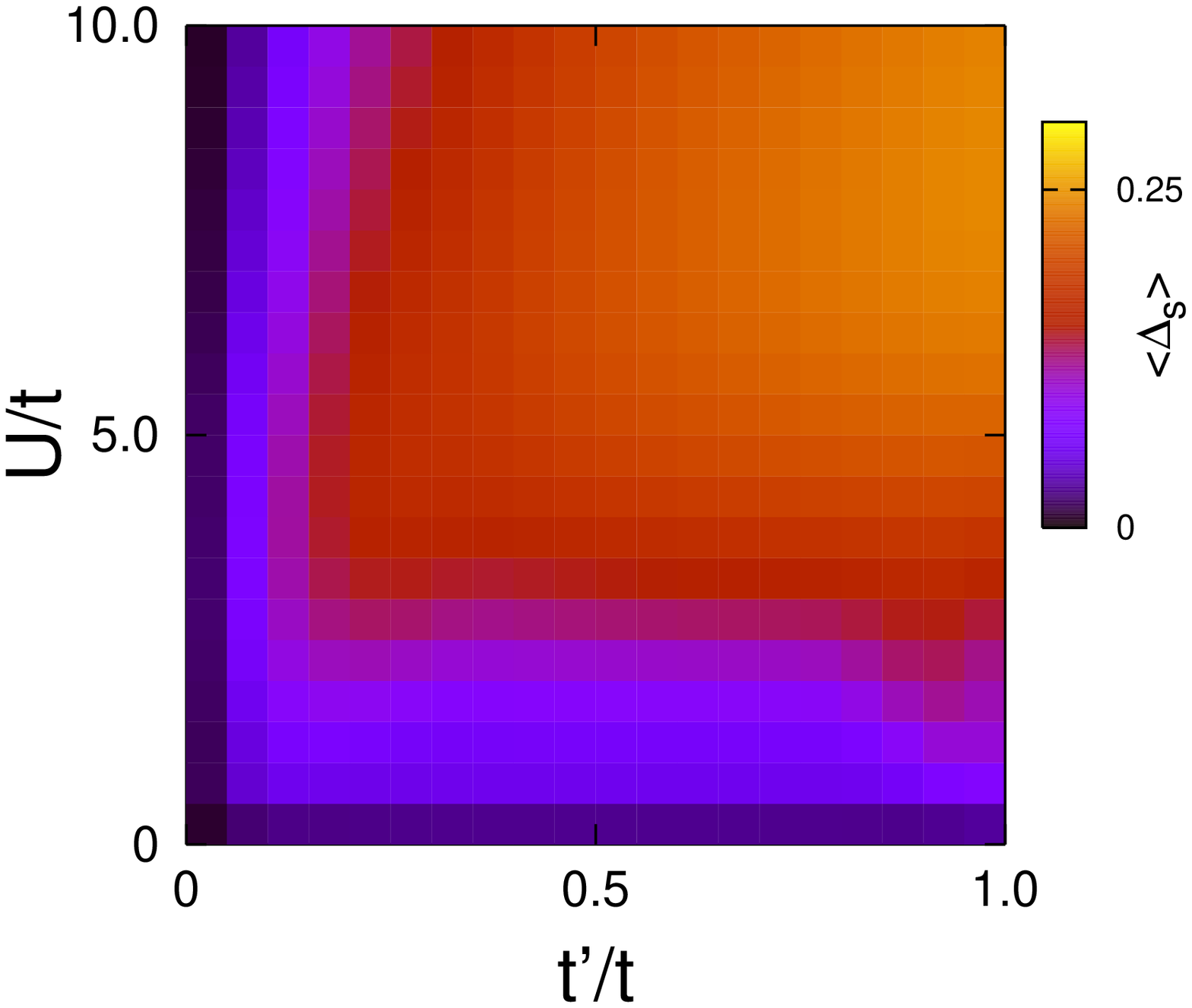}
\end{array}
$
\caption{
\label{fig:SpinGapStaggeredPotentialW0025}
Plots of $\Delta_{s}$ for the eight-site ladder cluster at doping $x=1/8$ and $W/t=0.25$ for $(a)$ staggered potential configuration $(ii)$ and $(b)$ staggered hopping configuration $(vi)$. For comparison, we also plot $\Ds$ at $W/t=0.25$ for $(c)$ potential disorder and $(d)$ hopping disorder.  (see Fig.~\ref{fig:StaggeredConfigurationsSmallW} for configuration details.)
}
\end{figure}

\begin{figure}
$
\begin{array}{cc}
\hspA(a)	& 
\hspB(b)\vspace{-0.5cm}  	 \\

\hspA	\includegraphics[angle=-90,width=5cm]{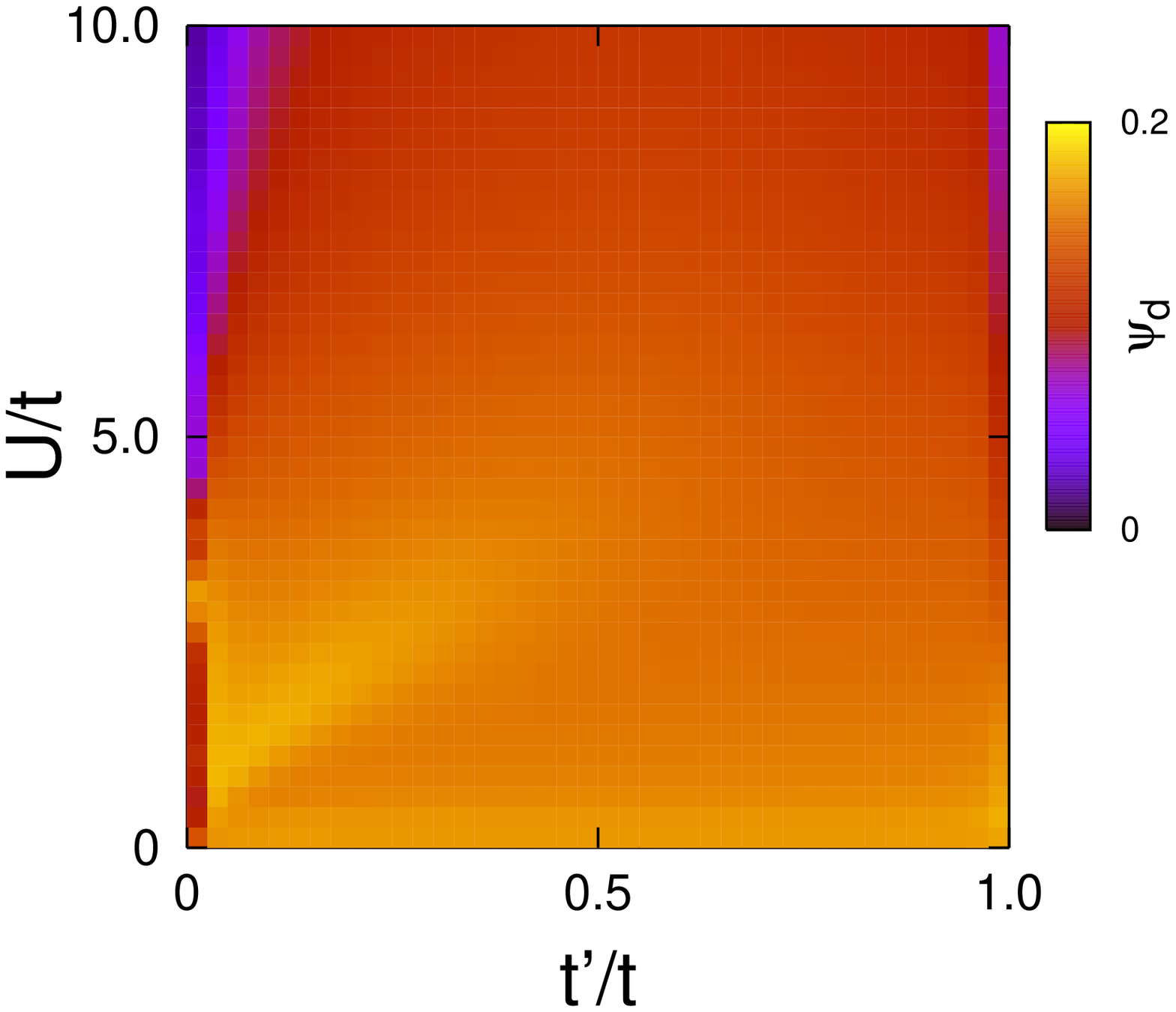} 	& 
\hspB  	\includegraphics[angle=-90,width=5cm]{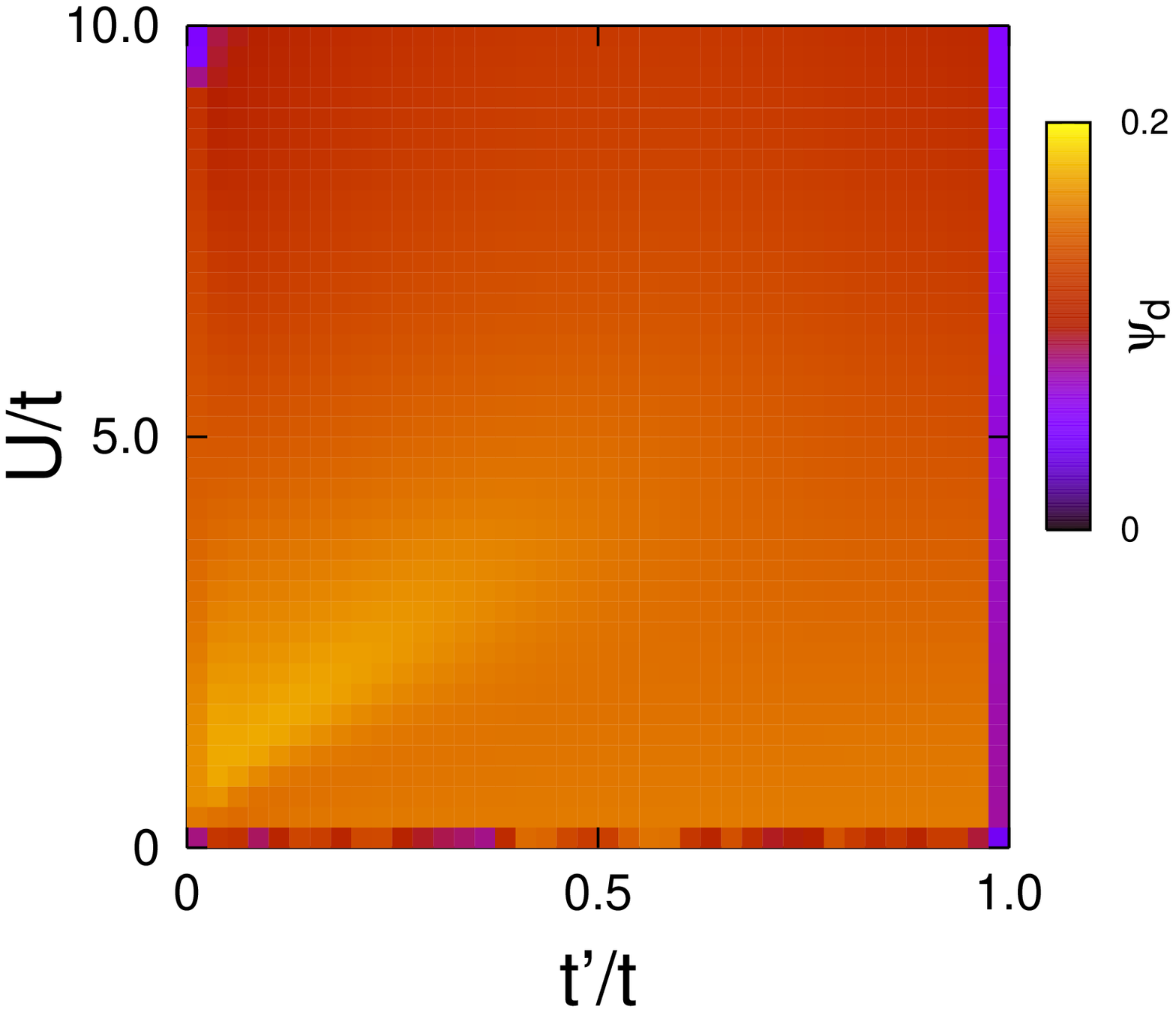}\\
\hspA(c)	& 
\hspB(d)\vspace{-0.5cm}  	 \\

\hspA	\includegraphics[angle=-90,width=5cm]{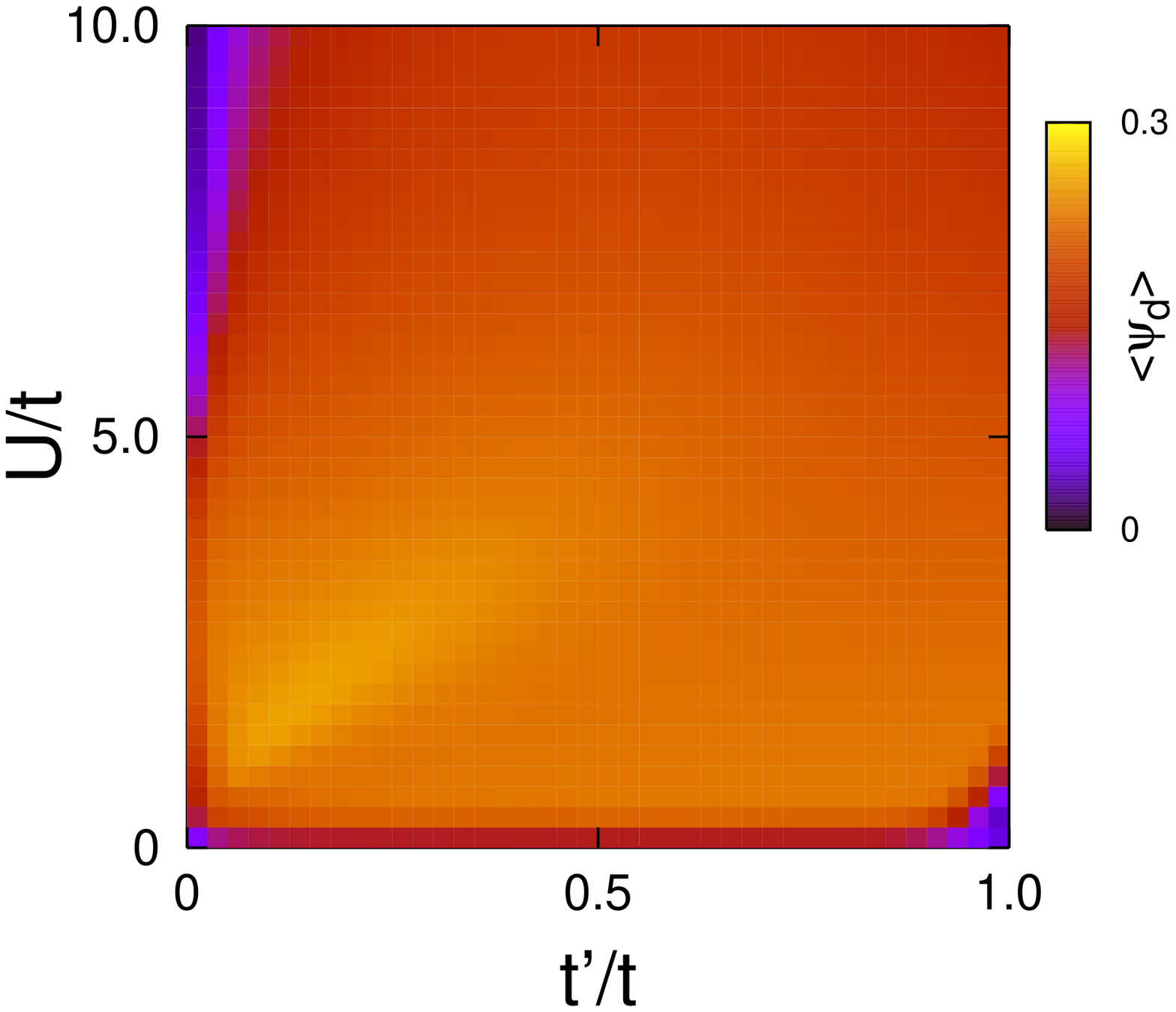} 	& 
\hspB  	\includegraphics[angle=-90,width=5cm]{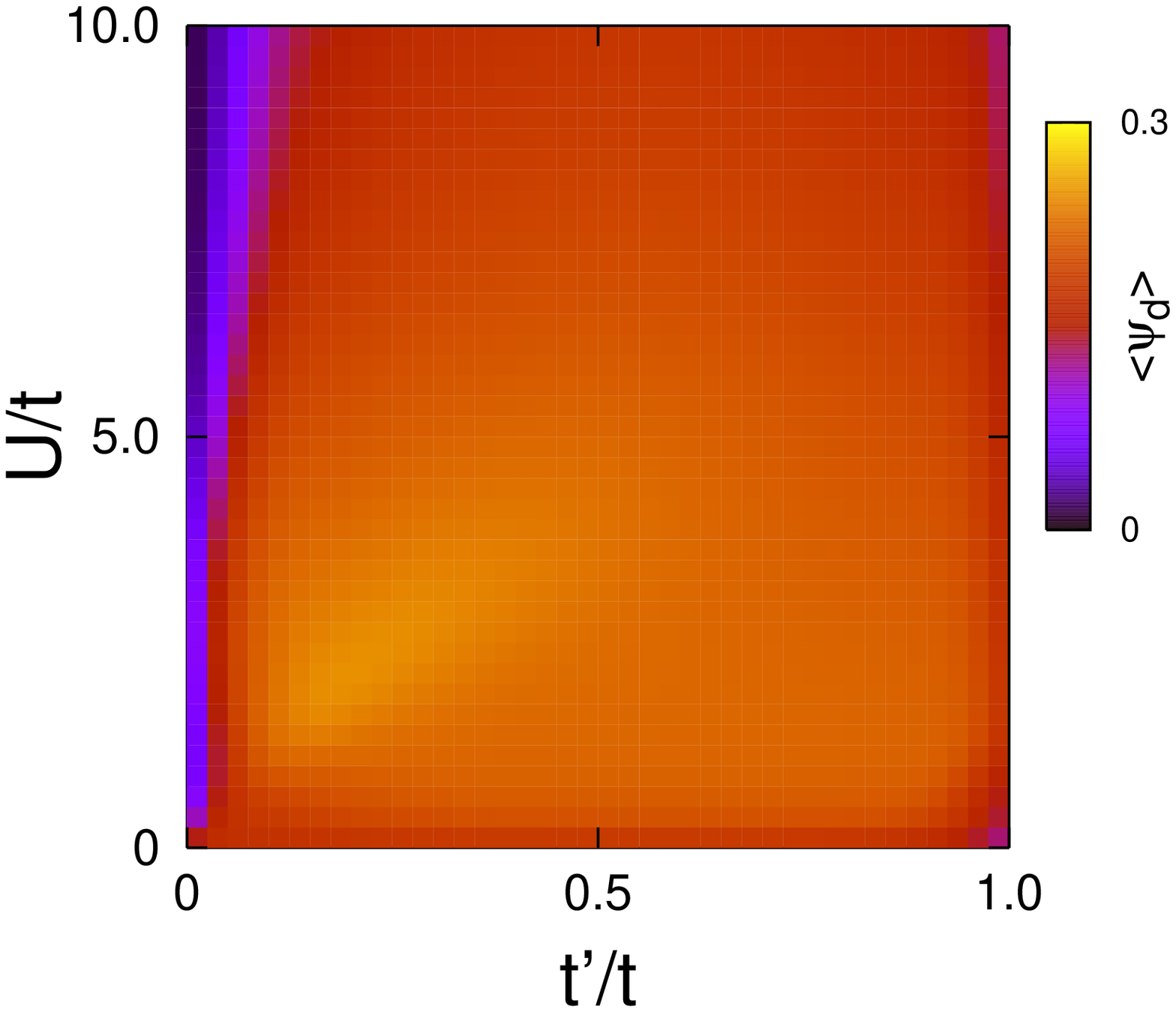}
\end{array}
$
\caption{
\label{fig:DWStaggeredPotentialW0025}
Plots of $\Psi_{d}$ for the eight-site ladder cluster at doping $x=1/8$ and $W/t=0.25$ for $(a)$ staggered potential configuration $(ii)$ and $(b)$ staggered hopping configuration $(vi)$. For comparison, we also plot $\Dd$ at $W/t=0.25$ for $(c)$ potential disorder and $(d)$ hopping disorder.  (see Fig.~\ref{fig:StaggeredConfigurationsSmallW} for configuration details.)
}
\end{figure}

We plot the values of $\Delta_{p}$, $\Delta_{s}$, and $\Psi_{d}$ as a function of $U/t$ and $t'/t$ at 
doping $x=1/8$ and $W/t=0.25$ in Figs.~\ref{fig:PBEStaggeredPotentialW0025}-\ref{fig:DWStaggeredPotentialW0025} 
for configurations $(ii)$ and $(vi)$ shown in Fig.~\ref{fig:StaggeredConfigurationsSmallW}. 
These data appear to be qualitatively similar to each other and to the results for other staggered
configurations and for random disorder for intermediate $t'/t$. 

In general, $\Delta_{p}$, $\Delta_{s}$ and $\Psi_{d}$ appear to be the most robust against small $W/t$ in configurations $(ii)-(iv)$ and $(vi)$. The local configurations of these patterns appear to favour dimerization in the case of staggered potentials and locally uniform hopping on each plaquette in the case of configuration $(vi)$.


\subsection{Large $W/t$}
\begin{figure}[htb]
$
\begin{array}{cc}
(I)     & 
(II)   \\
\includegraphics[angle=0,scale=0.3]{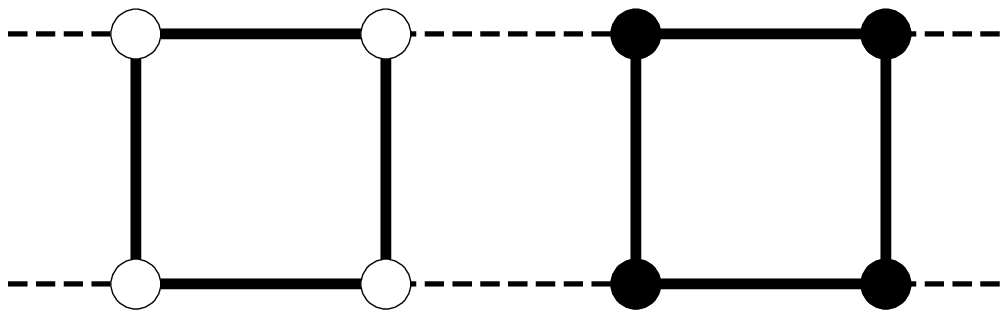}      &
\includegraphics[angle=0,scale=0.375]{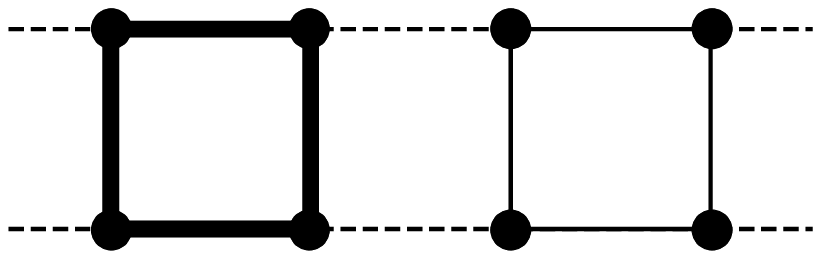}    
\end{array}
$
\caption{
\label{fig:StaggeredConfigurations}
Selected configurations of staggered on-site ($I$) and bond ($II$) potentials for which pair binding persists at large $W/t$. Dashed lines correspond to weak ($t'$) bonds, solid lines correspond to strong ($t$) bonds, and dots correspond to lattice sites. For configuration $(I)$, black (white) dots correspond to on-site potential strengths +(-)$\frac{W}{2t}$. For configuration $(II)$, solid thick (thin) lines correspond to bond strength $1+\frac{W}{2t}$ ($1-\frac{W}{2t}$).
}
\end{figure}

At large $W/t$ Fig.~\ref{fig:MaxPBEStaggered} shows that the maxima of the PBE for configurations ($I$) and ($II$) in Fig.~\ref{fig:StaggeredConfigurations} 
are more robust to disorder than other configurations. Configuration $(I)$ may be thought of as a staggered 
plaquette chemical potential $\mu=\pm 2W/t$, whereas configuration $(II)$ (which is a relabelling of 
configuration $(vi)$ shown in Fig.~\ref{fig:StaggeredConfigurationsSmallW}) may be interpreted 
as a hopping pattern staggered plaquette by plaquette, where the nearest neighbour hopping on each plaquette is either $t\pm W/2$.

\begin{figure}[h]
$
\begin{array}{cc}
\hspA(a)        & 
\hspB(b)\vspace{-0.5cm}          \\

\hspA   \includegraphics[angle=-90,width=5cm]{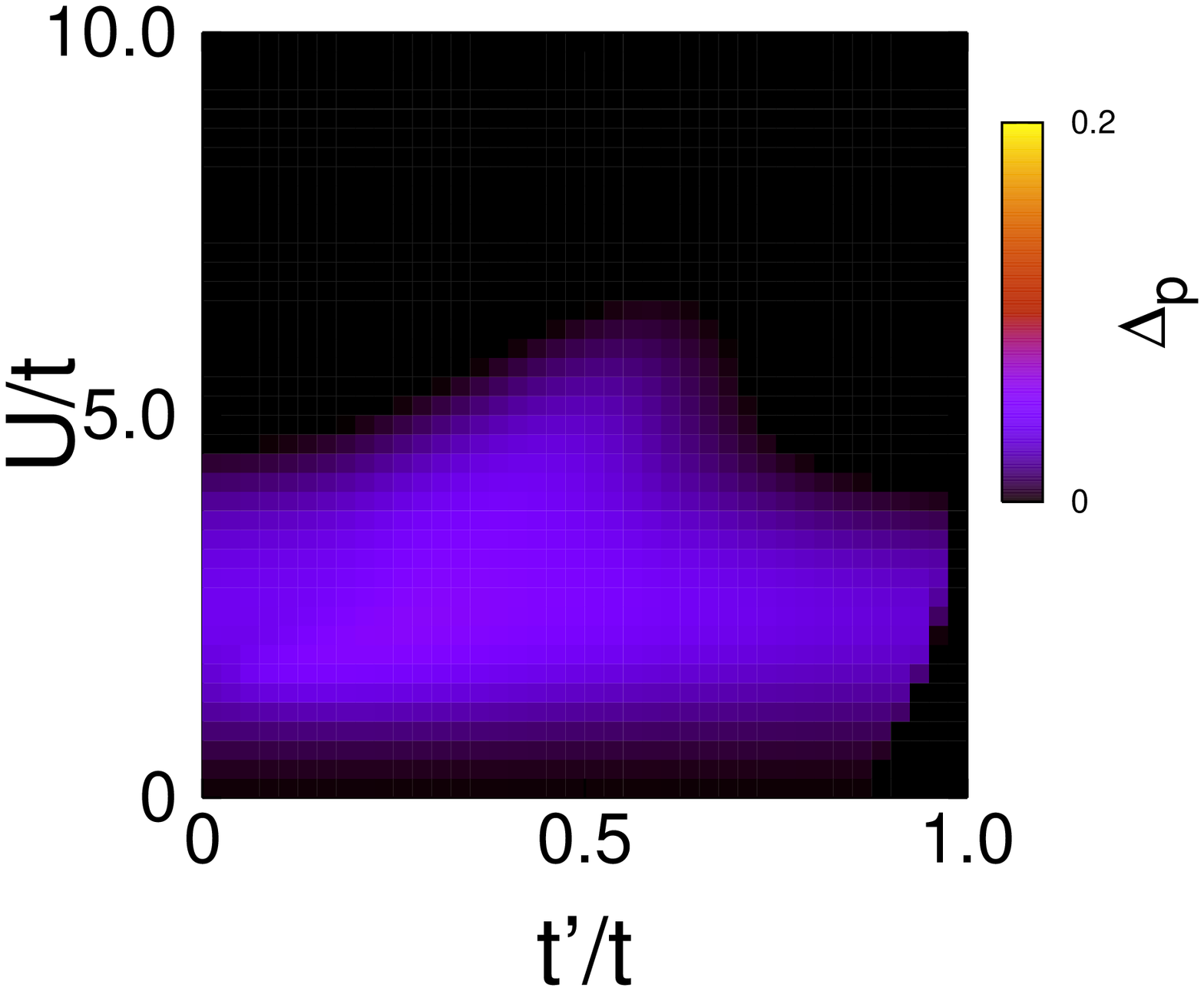}    & 
\hspB   \includegraphics[angle=-90,width=5cm]{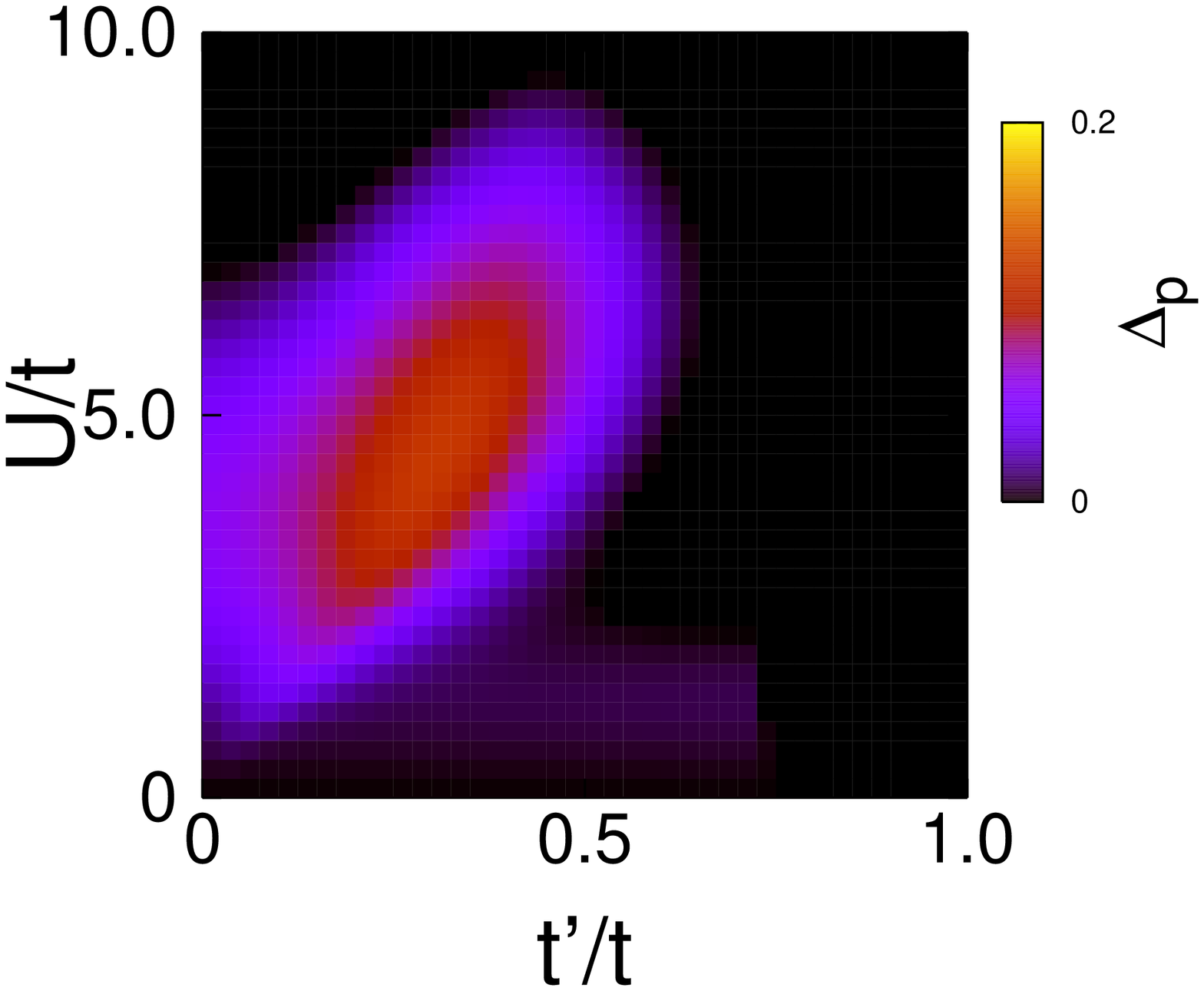} \\

\hspA(c)        & 
\hspB(d)\vspace{-0.5cm}          \\

\hspA   \includegraphics[angle=-90,width=5cm]{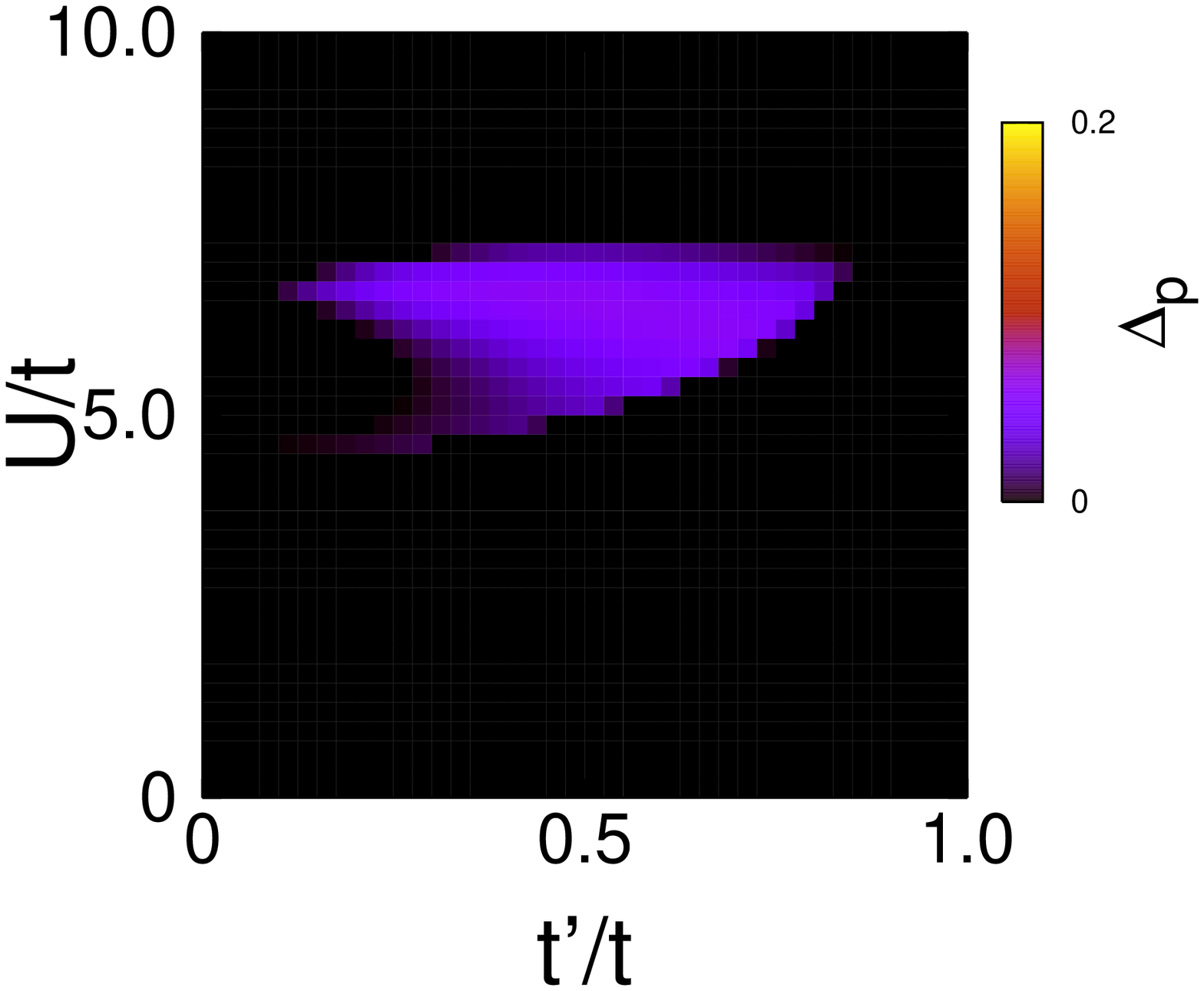}    & 
\hspB   \includegraphics[angle=-90,width=5cm]{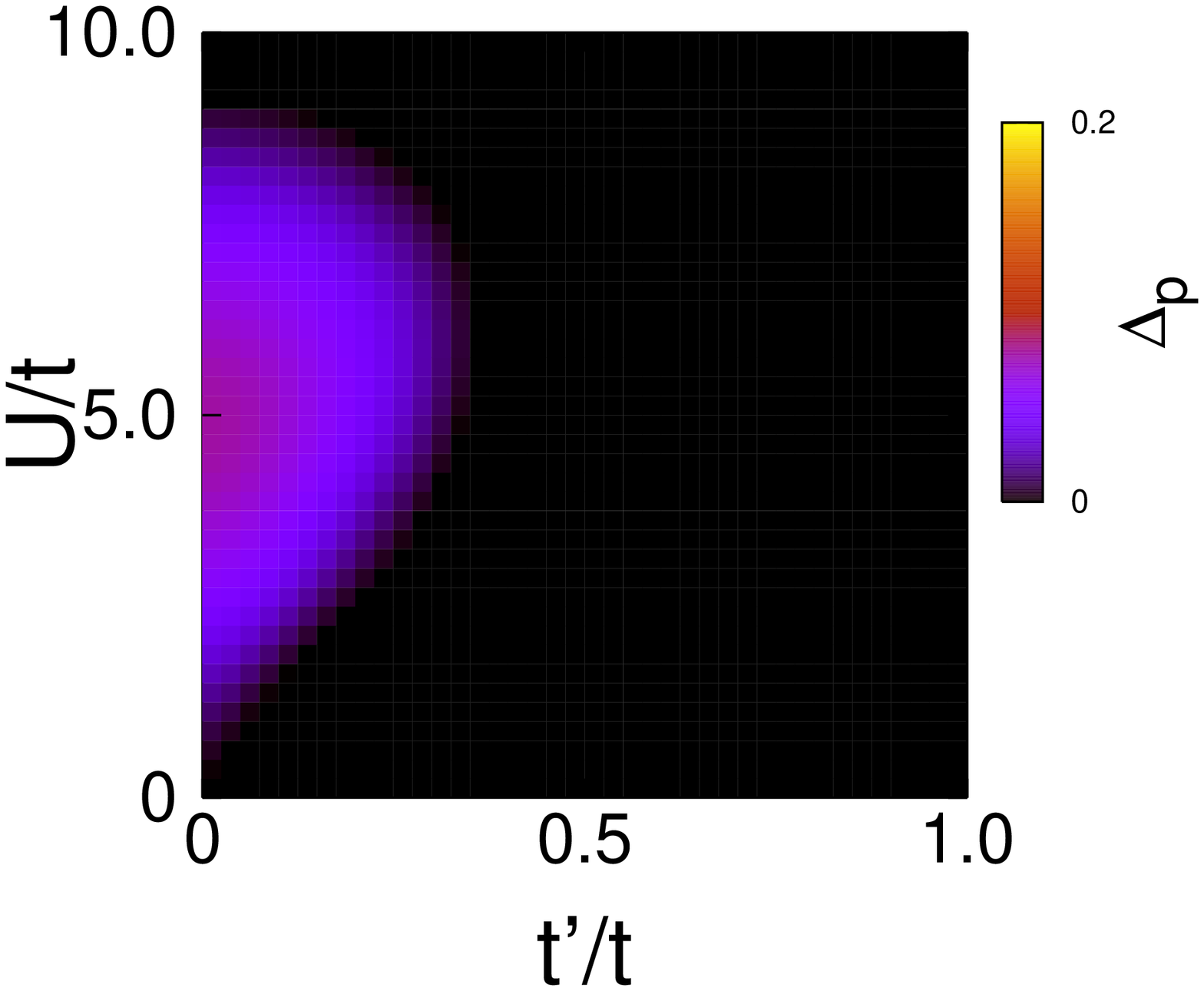}
\end{array}
$
\caption{
\label{fig:PBEStaggeredW0100}
$\Delta_{p}$ for the eight-site ladder cluster at doping $x=1/8$ for $(a)$ configuration $(I)$ at $W/t=1.00$, $(b)$ configuration $(II)$ at $W/t=1.00$.
$(c)$ configuration $(I)$ at $W/t=5.00$ and $(d)$ configuration $(II)$ at $W/t=1.98$.
(See Fig.~\ref{fig:StaggeredConfigurations} for configuration details.)
}
\end{figure}

The regions in $t'-U$ parameter space for which pair binding remains positive for large $W/t$ are quite different for 
configurations $(I)$ and $(II)$. Pair binding is favoured for intermediate $t'/t$ on configuration $(I)$ whereas pair 
binding on configuration $(II)$ is favoured primarily as $t'/t\to 0$. For all values of $W/t$ studied, there is always a region 
in parameter space where the PBE remains positive for these configurations. As shown in Fig.~\ref{fig:PBEStaggeredW0100}, 
for configuration $(I)$, $\Delta_{p}$ is maximum for $t'/t\approx0.5$ at $U/t\approx6-7$ at $W/t=5.00$, 
whilst $\Delta_{p}$ is maximum along the $t'/t=0$ axis at $W/t=1.98$ for configuration $(II)$. 

\begin{figure}[h]
$
\begin{array}{cc}
\hspA(a)        & 
\hspB(b)\vspace{-0.5cm}          \\

\hspA   \includegraphics[angle=-90,width=5cm]{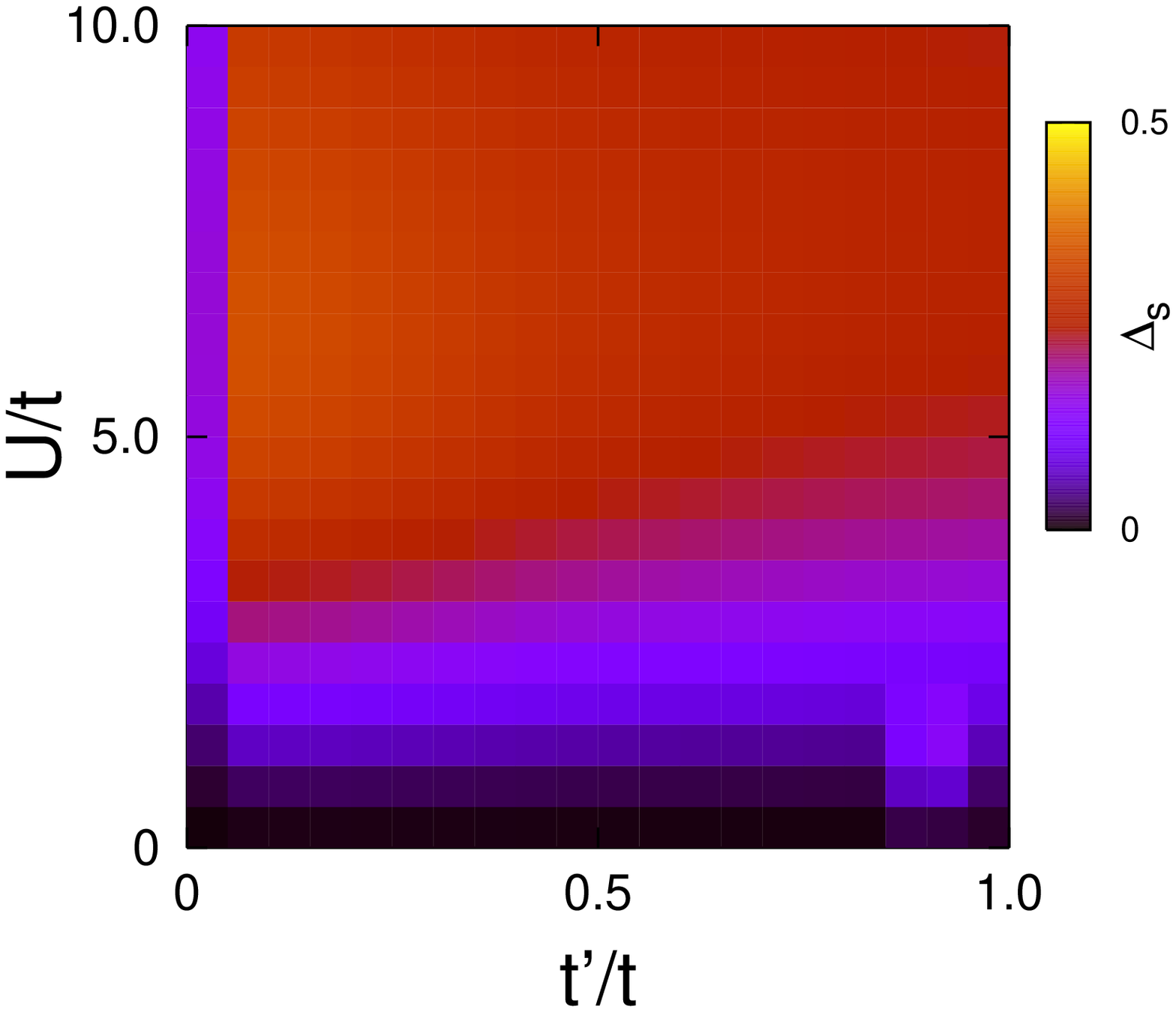}    & 
\hspB   \includegraphics[angle=-90,width=5cm]{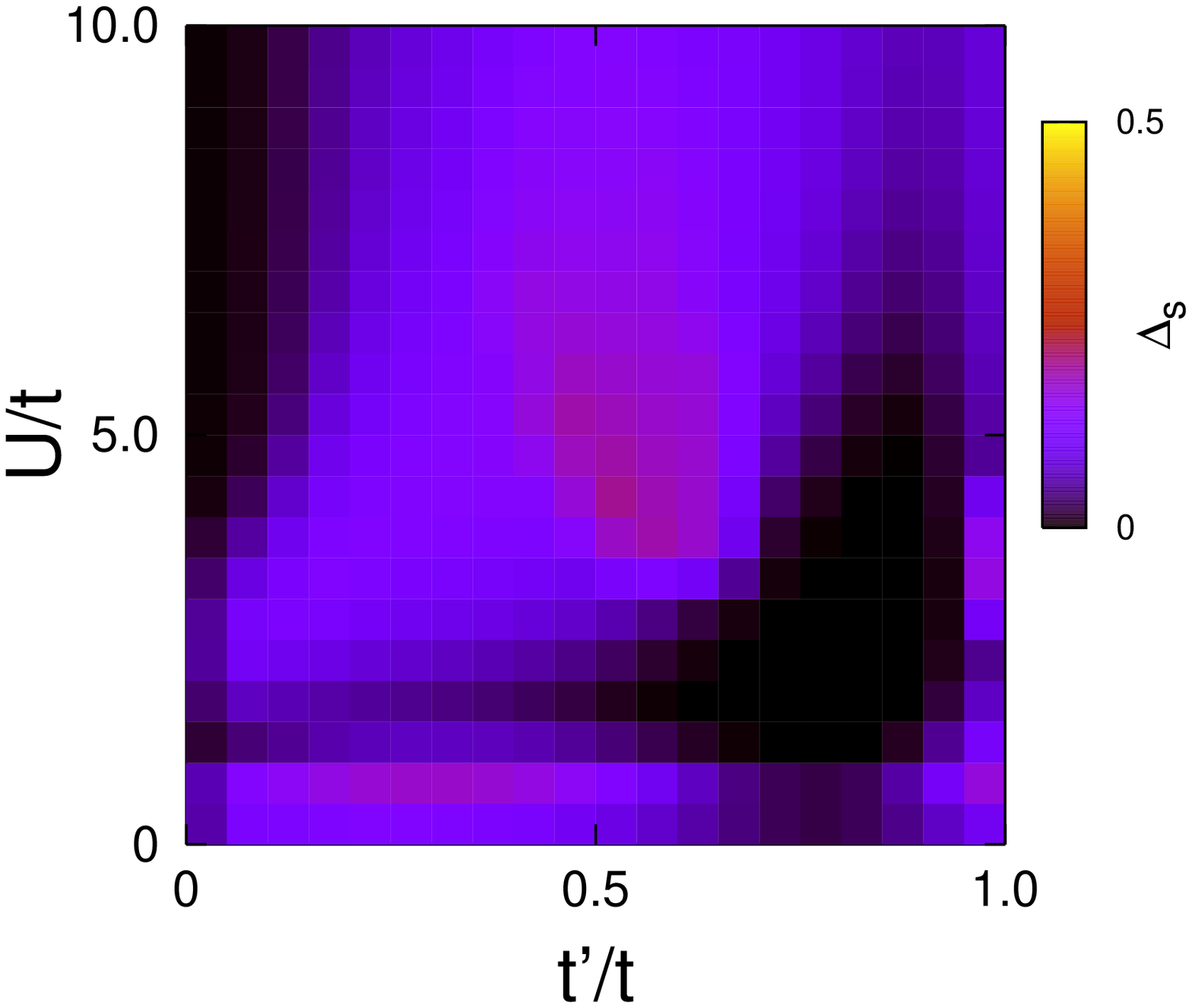}  \\
\hspA(c)        & 
\hspB(d)\vspace{-0.5cm}          \\

\hspA   \includegraphics[angle=-90,width=5cm]{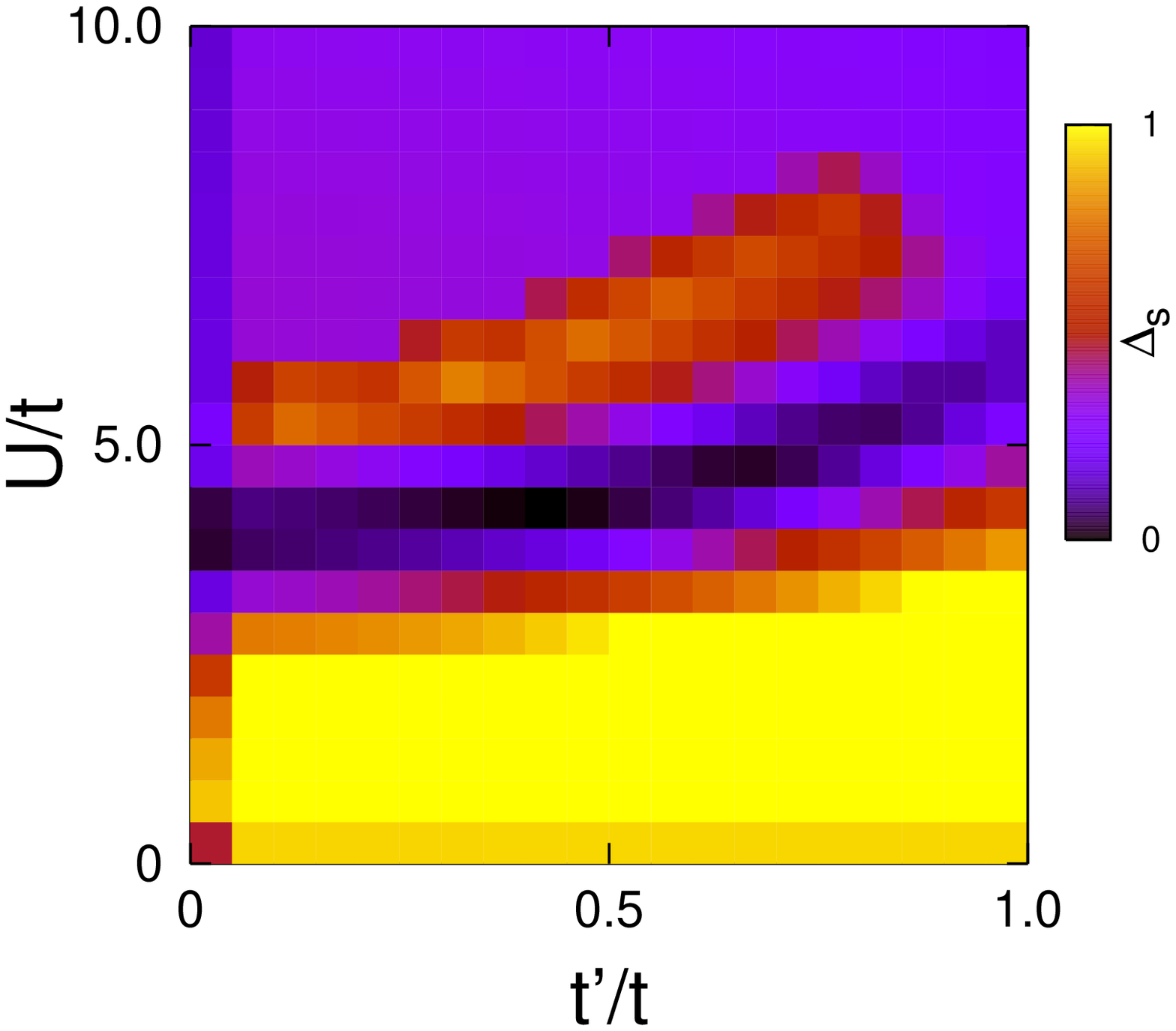}    & 
\hspB   \includegraphics[angle=-90,width=5cm]{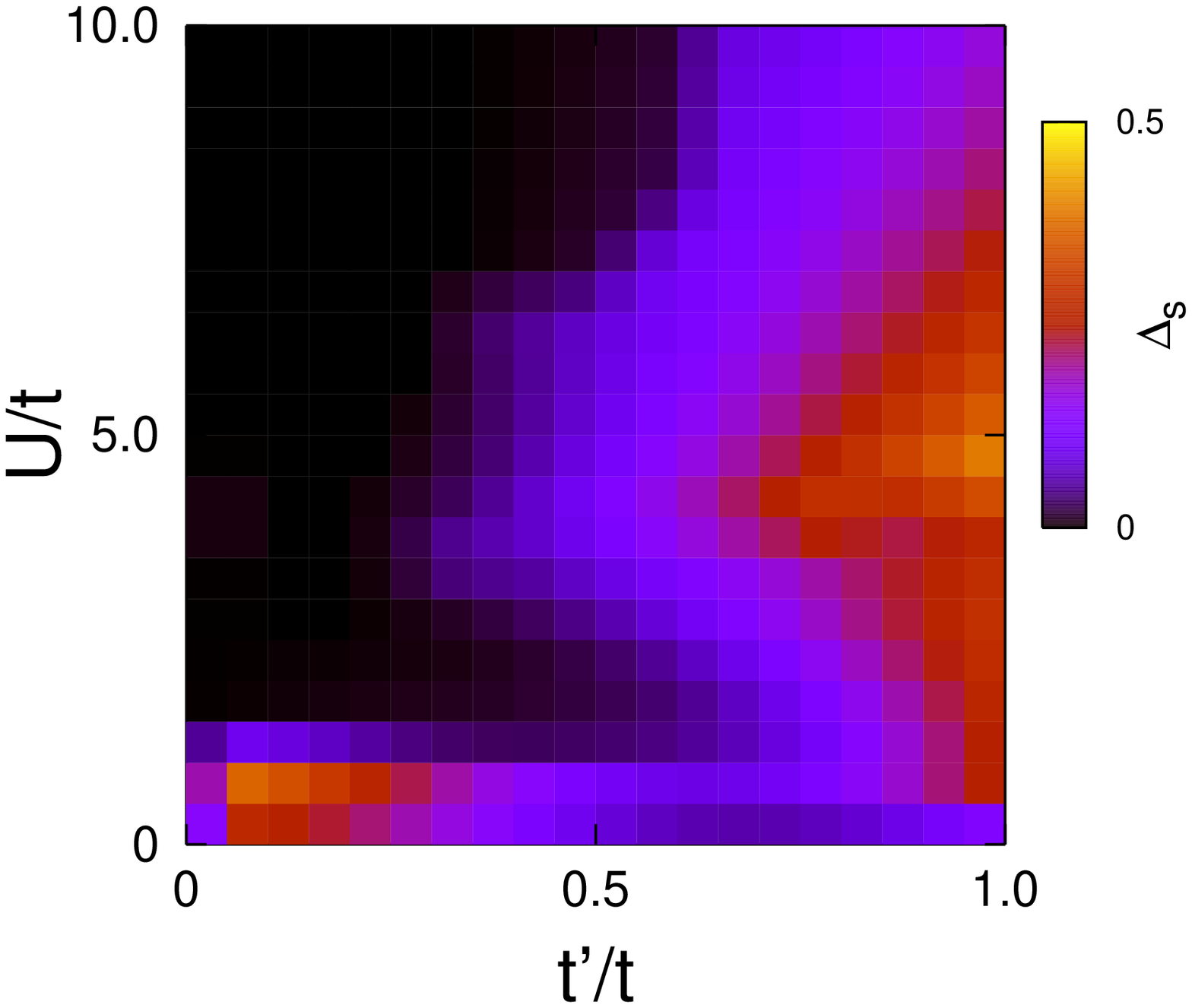}
\end{array}
$
\caption{
\label{fig:SpinGapStaggeredW0100}
$\Delta_{s}$ for the eight-site ladder cluster at doping $x=1/8$ for
$(a)$ configuration $(I)$ at $W/t=1.00$, $(b)$ configuration $(II)$ at $W/t=1.00$,
$(c)$ configuration $(I)$ at $W/t=5.00$ and $(d)$ configuration $(II)$  at $W/t=1.98$.
(See Fig.~\ref{fig:StaggeredConfigurations} for configuration details.)
}
\end{figure}

\begin{figure}[h]
$
\begin{array}{cc}
\hspA(a)	& 
\hspB(b)\vspace{-0.5cm}  	 \\

\hspA	\includegraphics[angle=-90,width=5cm]{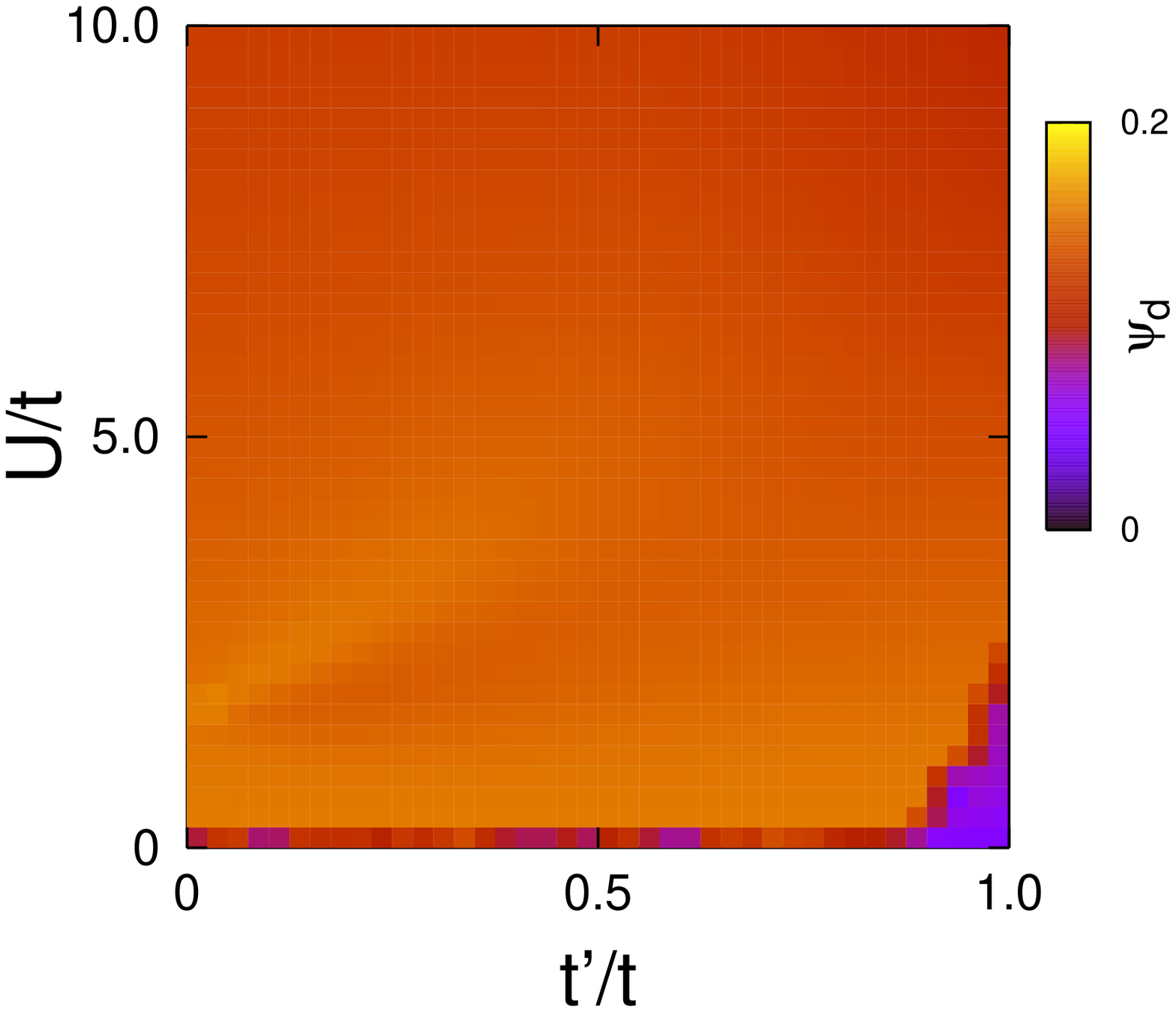} 	& 
\hspB  	\includegraphics[angle=-90,width=5cm]{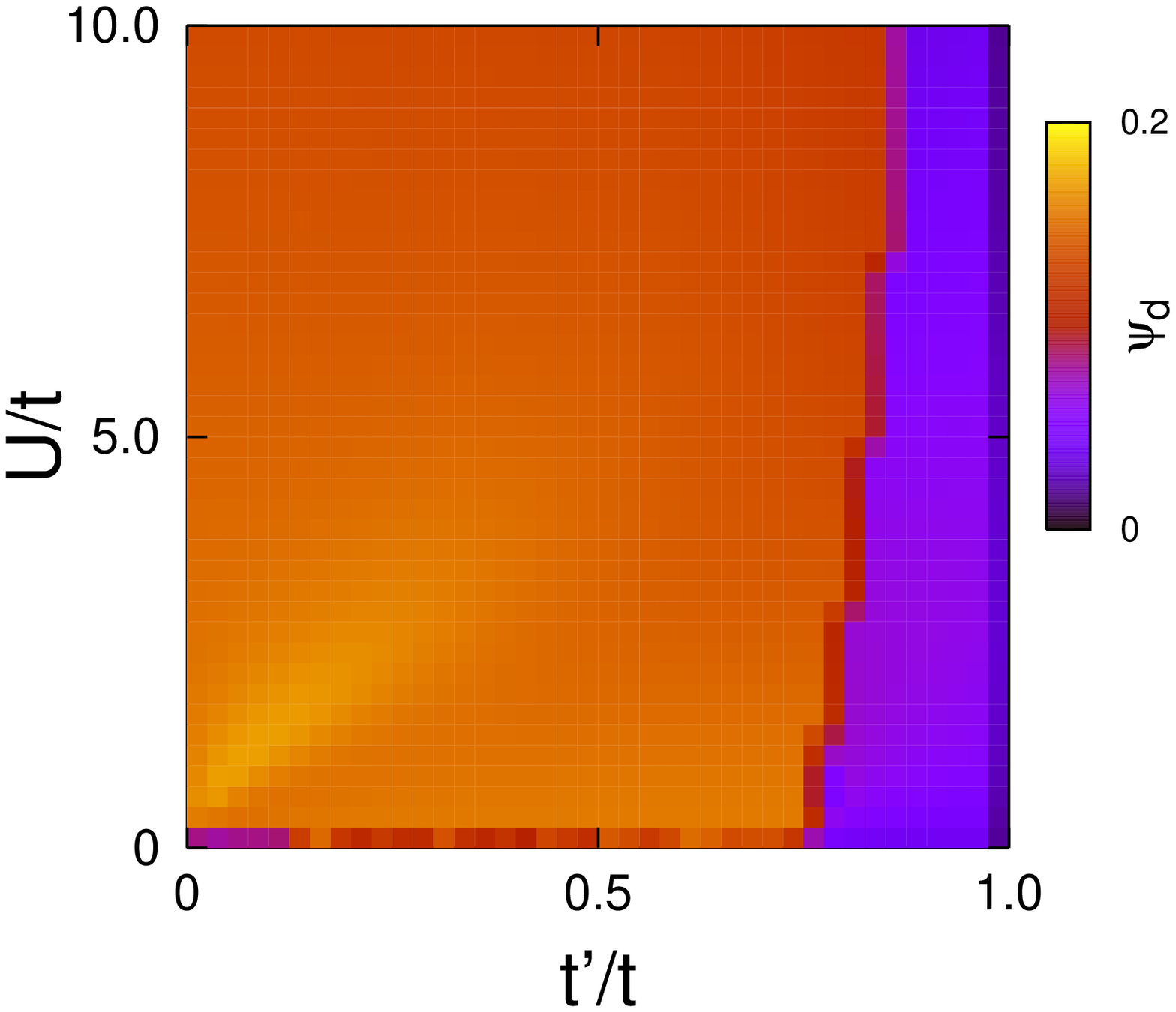} \\

\hspA(c)	& 
\hspB(d)\vspace{-0.5cm}  	 \\

\hspA	\includegraphics[angle=-90,width=5cm]{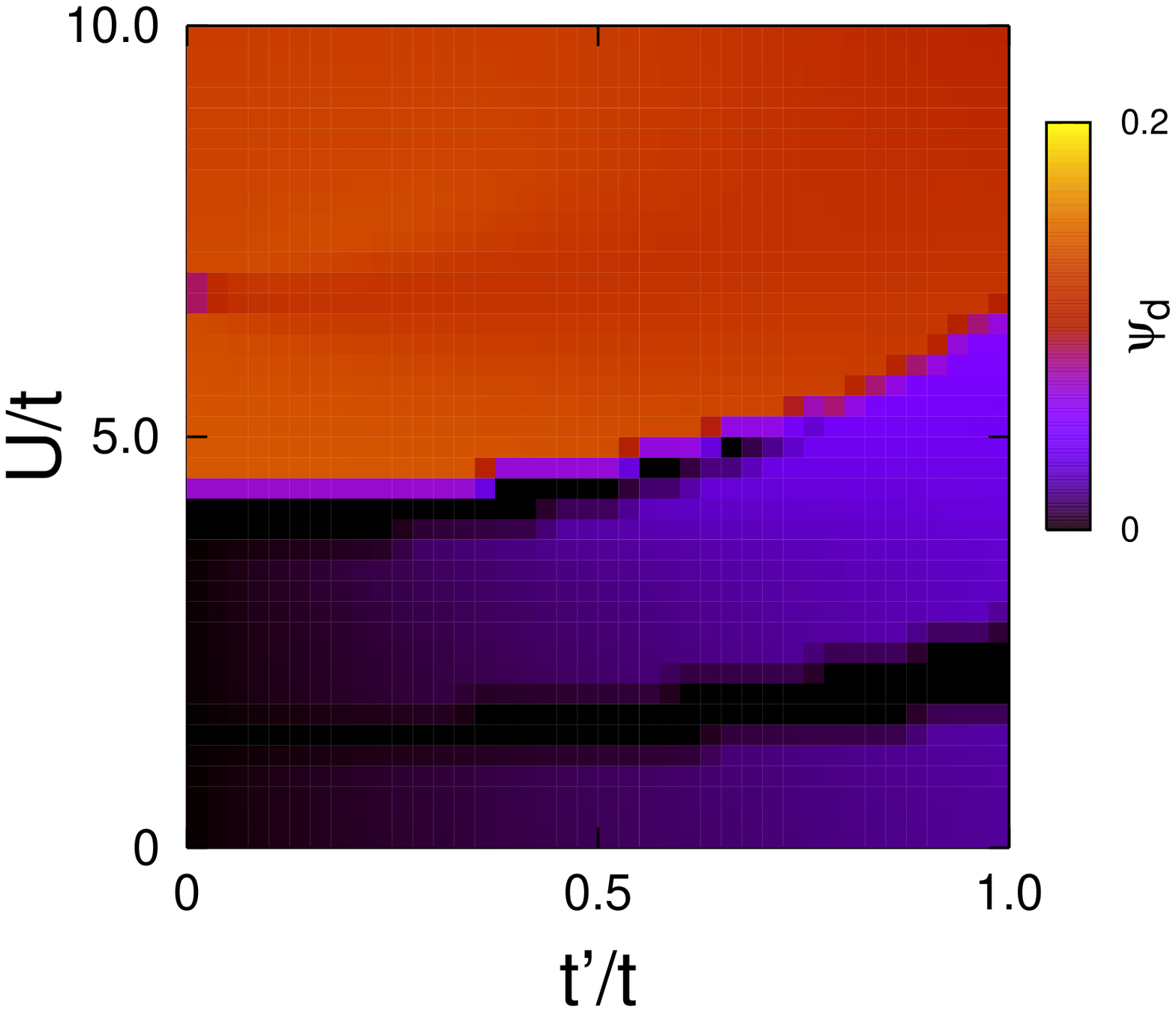} 	& 
\hspB  	\includegraphics[angle=-90,width=5cm]{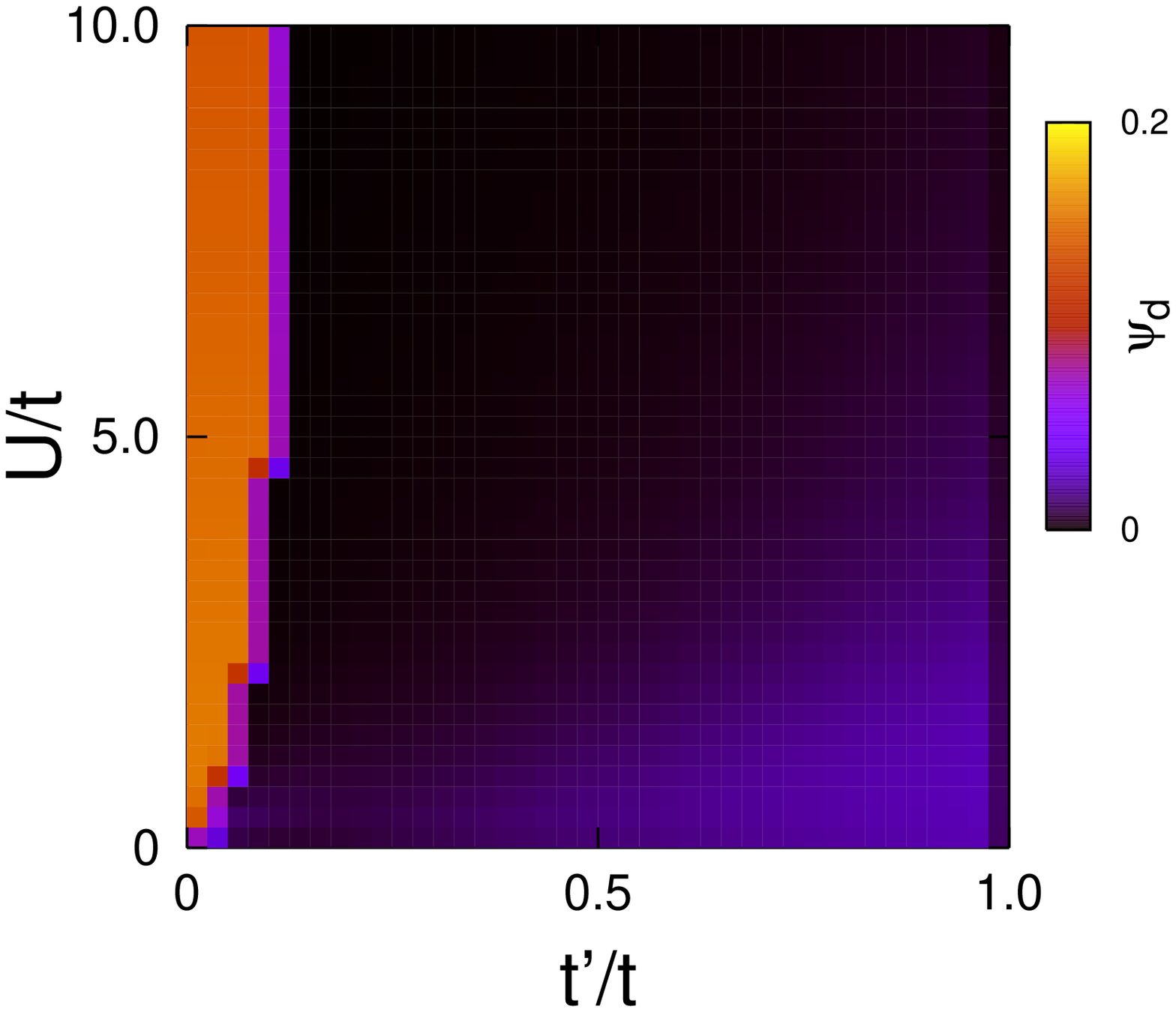}
\end{array}
$
\caption{
\label{fig:DWStaggeredW0100}
$\Psi_{d}$ for the eight-site ladder cluster at doping $x=1/8$ for $(a)$ configuration $(I)$ at $W/t=1.00$, $(b)$ configuration $(II)$ at $W/t=1.00$,
$(c)$ configuration $(I)$ at $W/t=5.00$ and $(d)$ configuration $(II)$  at $W/t=1.98$.  (See Fig.~\ref{fig:StaggeredConfigurations} for configuration details.)
}
\end{figure}

Figs.~\ref{fig:SpinGapStaggeredW0100} and \ref{fig:DWStaggeredW0100} show the effect of increasing 
$W/t$ on the spin gap and $d$-wave order parameter of configurations $(I)$ and $(II)$ at $W/t=1.00$ and beyond. 
For configuration $(I)$, the plots of $\Psi_{d}$ and $\Delta_{s}$ show a distinct crossover to a region 
where pairing is favoured at $U/t\gtrsim 4-5$ at $W/t=5.00$. 
For configuration $(II)$, increasing $W/t$ appears to shrink the spin gap in the pair 
binding region, while also suppressing $\Psi_{d}$ for $t'/t\gtrsim0.1-0.2$.

For each configuration where pair binding persists for very large disorder strengths, except for the 
configurations shown in Fig. \ref{fig:StaggeredConfigurations}, we can identify three common characteristics. First, the maximum of $\Delta_{p}$ appears for intermediate $U/t$ and $t'/t\approx0$, which is the limit of disconnected plaquettes. Second, although the sum of all potentials across the cluster is not zero for each configuration, the values of the staggered potentials add to $\pm 2W$ on one of the plaquettes. For staggered potentials, this is tantamount to a local shift in the chemical potential of $\pm 2W$, whilst in the case of staggered hoppings, this is equivalent to a local change in the interplaquette hopping from $-t\to-t-W/2$. Third, the average occupation of the plaquette for which the sum of the potentials is not $\pm 2 W$ is two electrons per plaquette, independent of the number of doped holes per cluster.  This leaves the remaining ``uniform'' plaquette with four electrons per uniform plaquette at $m=2$,  five electrons per uniform plaquette at $m=1$, and six electrons per uniform plaquette at $m=0$.

\section{Discussion and Conclusions}
\label{Sec:Discussion}

In this paper we have performed a detailed study of the effects of disorder on superconducting tendencies in
 the checkerboard Hubbard model. From exact diagonalization studies of eight- and twelve-site ladders 
at dopings $x=1/8$ and $x=1/12$, respectively, we have found that superconducting tendencies 
are much more robust to disorder at moderate $t'/t$ than for the uniform Hubbard model ($t^\prime/t = 1$).  In particular,
the disorder averaged pair binding energy $\Dp$, and the probability of non-zero pair binding $P(\Dp >0)$
are peaked for intermediate $U$ and $t^\prime/t$ and decay more slowly with disorder than for the uniform case.  
We observed similar behaviour for $\Ds$ and $\Dd$.  This implies a real space picture
in which disorder leads to patches of superconductivity, some of which persist even to strong disorder.  
Such a picture emerges from studying the full distribution of pair-binding energies in the presence of disorder, 
not just the mean value.  This additional robustness to disorder is reminiscent of the observation of stabilization
of the pseudogap by disorder in Lanczos and Quantum Monte Carlo simulations.\cite{Chiesa}

Examining fluctuations in the pair binding energy, we find that these have a cusp-like minimum
in the region of strongest superconducting tendencies, which appears to correspond to a cross-over between
single-plaquette and more delocalized physics, reminiscent of the phase transition to the d-Mott state at half-filling.
To gain further insight into disorder effects on the CHM, we studied all eight site configurations in which 
a staggered potential or staggered hopping was superimposed on the underlying CHM.  We found that the 
configurations with the greatest robustness to
increasing disorder strength generally appeared to be those with a pattern of dimerization in either the
staggered potential or hopping.
A caveat to our results is the issue of finite size effects, which are always present in numerical calculations.  In 
disordered systems, one generically expects shorter correlation lengths than the corresponding ordered system, 
which is encouraging.  However, it would be desirable to have our results confirmed via other techniques, 
such as the contractor renormalization approach\cite{BaruchOrgad} or the Dynamical Cluster Approximation.\cite{Tremblay}

Beyond the focus on the two dimensional Hubbard model from the perspective of high temperature superconductivity, the
interest in ``designer Hamiltonians'' \cite{Jaksch1,Bloch,Jaksch2} and the checkerboard Hubbard model in 
particular \cite{Peterson,Demler} in the context of cold atom systems gives an additional area to which 
our results may be of interest.  The crossover between single plaquette and multi-plaquette physics we see away
from half-filling is reminiscent of the transition to the $d$-Mott insulator phase \cite{YaoKivelson1,YaoKivelson2,Peterson,Lauchli} 
which has been argued to be favoured at half filling and is not adiabatically connected to any band insulators.
In this phase each plaquette on the lattice has a local $d$-wave symmetry: rotation of a single plaquette 
by 90$^{\circ}$ leads to a change in sign of the wave function of the system.  Possible experimental
signatures to identify the presence of this state were suggested by Peterson \etal.\cite{Peterson}
Cold atom systems lack disorder, but efforts to introduce disorder using incommensurate lattices \cite{Diener}
and optical speckle fields \cite{DeMarco} might allow for experimental realization of the disordered CHM.


\section{Acknowledgements}
The authors thank Bill Atkinson, Igor Herbut, Wei-Fang Tsai, Xin Wan and Rachel Wortis 
for helpful discussions, Martin Siegert for technical support, and Westgrid for computer resources.
This work was supported by NSERC.


\begin{thebibliography}{99}
\bibitem{footnote01} Other authors have also recently studied models with hopping modulations other than the checkerboard 
model, e.g. Ref.~\onlinecite{Su}.
\end{thebibliography}
\end{document}